\documentclass[11pt, a4paper]{article}
\usepackage[utf8]{inputenc}
\usepackage[T1]{fontenc}
\usepackage[english]{babel}
\usepackage[a4paper, margin=2.5cm]{geometry}
\usepackage{lmodern}
\usepackage{amsmath}
\usepackage{amssymb}
\usepackage{upgreek}
\usepackage{graphicx}
\usepackage{float}
\usepackage{xcolor}
\usepackage{tabularx}
\usepackage{booktabs}
\usepackage{array}
\usepackage{multirow}
\usepackage{subcaption}
\usepackage{changepage}
\usepackage{hyperref}

\hypersetup{
    colorlinks=true,
    linkcolor=blue,
    citecolor=blue,
    urlcolor=blue
}

\newcolumntype{C}{>{\centering\arraybackslash}X}

\newlength{\extralength}
\newlength{\fulllength}
\makeatletter
\newcommand{\Title}[1]{\gdef\@Title{#1}}
\newcommand{\Author}[1]{\gdef\@Author{#1}}
\newcommand{\address}[1]{\gdef\@address{#1}}
\newcommand{\corres}[1]{\gdef\@corres{#1}}
\renewcommand{\abstract}[1]{\gdef\@abstract{#1}}
\newcommand{\keyword}[1]{\gdef\@keyword{#1}}
\newcommand{\featuredapplication}[1]{\gdef\@featuredapplication{#1}}

\newcommand{\makeMDPIheader}{%
  \begin{center}
    {\LARGE \bfseries \@Title \par}
    \vspace{1.5em}
    {\large \@Author \par}
    \vspace{1em}
    {\small \@address \par}
    \vspace{0.5em}
    {\small \@corres \par}
  \end{center}
  \vspace{1em}
  \ifx\@abstract\@undefined\else
    \begin{quote}
      \small \textbf{Abstract:} \@abstract
    \end{quote}
  \fi
  \ifx\@keyword\@undefined\else
    \vspace{0.5em}
    \noindent\small \textbf{Keywords:} \@keyword \par
  \fi
  \ifx\@featuredapplication\@undefined\else
    \vspace{0.5em}
    \noindent\small \textbf{Featured Application:} \@featuredapplication \par
  \fi
  \vspace{1.5em}
}
\makeatother

\newcommand{\firstpage}[1]{}
\newcommand{\pubvolume}[1]{}
\newcommand{\issuenum}[1]{}
\newcommand{\articlenumber}[1]{}
\newcommand{\pubyear}[1]{}
\newcommand{\copyrightyear}[1]{}
\newcommand{\externaleditor}[1]{}
\newcommand{\datereceived}[1]{}
\newcommand{\daterevised}[1]{}
\newcommand{\dateaccepted}[1]{}
\newcommand{\datepublished}[1]{}
\newcommand{\CorrStatement}[1]{}
\newcommand{\longauthorlist}[1]{}
\newcommand{\IsAssociation}[1]{}

\newcommand{\xdeleted}[1]{}

\newcommand{\orcidauthorA}[1]{}
\newcommand{\orcidauthorB}[1]{}
\newcommand{\orcidauthorC}[1]{}
\newcommand{\orcidauthorD}[1]{}
\newcommand{\orcidA}{}

\newcommand{\orcidC}{}
\newcommand{\orcidD}{}

\newcommand{\appendixtitles}[1]{}

\newcommand{\authorcontributions}[1]{\section*{Author Contributions}\noindent #1}
\newcommand{\funding}[1]{\section*{Funding}\noindent #1}
\newcommand{\dataavailability}[1]{\section*{Data Availability Statement}\noindent #1}
\newcommand{\acknowledgments}[1]{\section*{Acknowledgments}\noindent #1}
\newcommand{\conflictsofinterest}[1]{\section*{Conflicts of Interest}\noindent #1}
\newcommand{\abbreviations}[2]{\section*{#1}\noindent #2}
\newcommand{\reftitle}[1]{\renewcommand{\refname}{#1}}

\newcommand{\entrylink}[1]{}
\newcommand{\reviewreports}[1]{}

\Title{RPC Telescope Tests for Muon Detection at\linebreak   Laser-Plasma Accelerators}

\Author{Sumaira Ikram $^{1}$, Maxime Lagrange $^{1}$, Donya Ahmadi $^{2}$, Uddhab Chaulagain $^{3}$\orcidA{}, Anna Cimmino $^{3}$\orcidA{}, Eduardo Cortina Gil $^{1}$, Pavel Demin $^{1}$, Gabriele M. Grittani $^{3}$, Sebastian Lorenz $^{3}$, Michal Nevrkla $^{3,4}$, Ela Rockafellow $^{5}$, Jaron E. Shrock $^{5}$, Jiri Sisma $^{3,4}$, Ari J. Sloss $^{5}$, Michael Tytgat $^{2}$, Roberto Versaci $^{3}$\orcidC{}, Filip Vitha $^{3,4}$, Illia Zymak $^{3}$, Howard M. Milchberg $^{5}$ and Andrea Giammanco $^{1,}$* \orcidD}

\address{
$^{1}$ \quad Centre for Cosmology, Particle Physics and Phenomenology (CP3), UCLouvain, B-1348 Louvain-la-Neuve, Belgium; sumaira.ikram@uclouvain.be (S.I.); maximelagrange98@gmail.com (M.L.);\linebreak    eduardo.cortinagil@uclouvain.be (E.C.G.); pavel.demin@uclouvain.be (P.D.); andrea.giammanco@cern.ch (A.G.) 
\\
$^{2}$ \quad Inter-University Institute for High Energies (IIHE), Vrije Universiteit Brussel, B-1050 Brussels, Belgium; donya.ahmadi@vub.be (D.A.); michael.tytgat@vub.be (M.T.)
\\
$^{3}$ \quad ELI Beamlines Facility, The Extreme Light Infrastructure ERIC, 252 41 Dolní Břežany, Czech Republic; uddhab.chaulagain@eli-laser.eu (U.C.); anna.cimmino@eli-laser.eu (A.C.); gabriele.grittani@eli-laser.eu (G.M.G.);  sebastian.lorenz@eli-laser.eu (S.L.); michal.nevrkla@eli-laser.eu (M.N.);  jiri.sisma@eli-laser.eu (J.S.); roberto.versaci@eli-laser.eu (R.V.);  filip.vitha@eli-laser.eu (F.V.); illia.zymak@eli-laser.eu (I.Z.)
\\
$^{4}$ \quad Faculty of Nuclear Sciences and Physical Engineering, Czech Technical University in Prague, 115 19 Prague, Czech Republic
\\
$^{5}$ \quad Institute for Research in Electronics and Applied Physics, University of Maryland, College Park, MD 20742, USA; erock22@terpmail.umd.edu (E.R.); jshrock@umd.edu (J.E.S.); asloss@umd.edu (A.J.S.);\linebreak    milch@umd.edu (H.M.M.)
}

\corres{Correspondence: andrea.giammanco@cern.ch}

\abstract{We report on a feasibility study conducted at the ELBA station at the ELI Beamlines facility to investigate the possible production of muons from high-energy electron beams generated by extended laser–plasma interactions in optically generated plasma waveguides. 
Our team operated a portable, autonomous, and compact telescope based on Resistive Plate Chamber (RPC) detectors, positioned to detect high-penetration charged particles originating from the beam dump.
The campaign demonstrated the stable operation of the RPC telescope in the challenging radiation and electromagnetic environment of ELBA. The collected datasets, though statistically limited and acquired during parasitic operation of the accelerator, allow detailed characterization of the background and validated the detectors' stability and tracking performance. These results confirm the feasibility of the approach and provide the foundation for a dedicated future run under optimized beam conditions, where muon detection sensitivity will be substantially improved.}

\keyword{laser-plasma acceleration; muon production; muon detectors}

\featuredapplication{This work establishes the feasibility of using compact, portable gaseous detector arrays to characterize laser-driven secondary particle beams, paving the way for mobile, man-made muon sources designed for on-site muographic inspection of critical civil infrastructure, industrial systems, and geophysical surveys.}

\begin{document}

\makeMDPIheader

\section{Introduction}

Muon production driven by high-power lasers represents an emerging frontier that connects plasma physics, particle physics, and applied imaging science. When high-intensity laser pulses accelerate electrons to GeV energies, their interaction with dense targets can initiate photonuclear reactions leading to muon pair production. This process has been recently verified by groups at SULF~\cite{zhang2025proof}, Berkeley~\cite{terzani2025measurement} and ELI-NP~\cite{Calvin2026}. 

Cosmic-ray muons already provide a natural source for muographic imaging~\cite{tanaka2023muography}, where their high penetration depth enables the study of large or shielded structures. Applications include geophysical surveys, civil engineering, and cultural heritage studies~\cite{IAEA2022,procureur2018muon, Kaiser2018muography,bonomi2020applications,bae2026muography}. However, muography with cosmic rays is intrinsically limited by the fixed and relatively low flux, which constrains the achievable spatial resolution and requires long acquisition times. We commented on these limitations, and on the comparative advantages of controllable muon sources, in Ref.~\cite{muonsCH-Giammanco2024}. Controlled, man-made muon sources would therefore constitute a useful complement, allowing systematic calibration studies, validation of reconstruction methods, and measurements in environments where long-term detector deployment is impractical. 
Such muon sources would have wide-ranging scientific and technological implications. Today, muon beams are available only at a few large accelerator complexes worldwide, requiring extensive infrastructure and resources~\cite{MakingMuons2021}. 
Facilities that provide muon beams have demonstrated their value across a broad range of disciplines~\cite{ninomiya2015nondestructive,reidy1978use,cataldo2022novel,biswas2023non,Hofmann2023}, underscoring the scientific demand for wider access to this tool.
A laser-driven alternative could drastically lower the threshold for access to muon science, enabling smaller laboratories and interdisciplinary research groups to explore muon-based applications.

In the longer term, the advent of next-generation, high-intensity lasers---especially systems that are increasingly compact and transportable---will make the realization of mobile muon sources conceivable. A portable laser system capable of generating muons would open unprecedented opportunities: on-site muographic inspection of critical infrastructure, flexible experimental campaigns, and the possibility of extending muon-based methods beyond dedicated accelerator facilities. 

Resistive Plate Chambers (RPCs) are gaseous particle detectors consisting of two parallel resistive electrodes separated by a thin gas gap~\cite{santonico1981development, SANTONICO20041,fonte2002review}. When a charged particle traverses the chamber, it ionizes the gas; the resulting electrons trigger an avalanche under a high electric field, producing a fast electrical signal. Pickup strips placed outside the resistive plates record the position and timing of the event. 
These detectors combine excellent timing resolution, good spatial granularity, and robustness in variable environmental conditions. The resulting instruments are a popular choice as muon detectors in large particle physics experiments at particle colliders~\cite{CMSmuonTDR,ATLASmuonTDR} and for cosmic-ray studies~\cite{aielli2006layout, abreu2018marta}, and have been deployed in a variety of configurations for cosmic-ray muography applications~\cite{carloganu2013towards,RPC-MuographyBook2022}. The portability and modularity of RPC telescopes make them suitable candidates for characterizing potential laser-driven muon sources, as they can be positioned flexibly around the interaction region while maintaining stable operation.

\begin{figure}[H]
    \centering
    \includegraphics[width=0.84\linewidth]{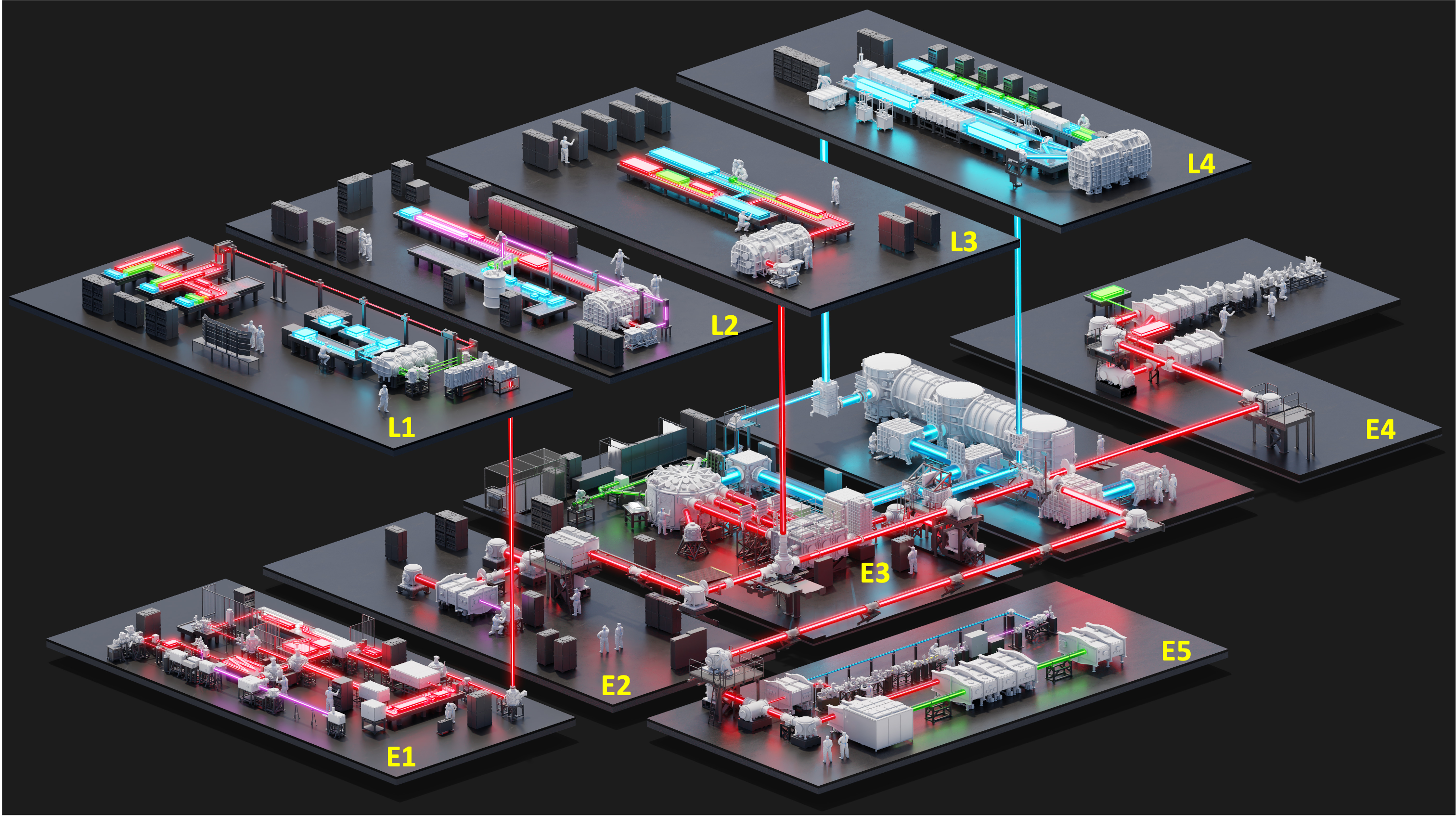}
    \caption{Schematic layout of the ELI Beamlines facility, showing the main experimental areas. {\it (Courtesy from ELI ERIC)}.}
    \label{fig:laser_building}
\end{figure}

This context motivates the present study, in which RPC-based telescopes are employed to assess the feasibility of detecting muons produced by high-energy laser-accelerated electrons interacting with a beam dump. 
The tests reported here were performed at ELBA (ELectron Beam Accelerator for fundamental science), a laser–electron collider hosted at ELI Beamlines {\cite{eliwhitebook}} and powered by the high-repetition-rate L3-HAPLS laser system {\cite{l3hapls1, l3hapls2}}. ELBA now delivers multi-GeV electron beams produced via Self-Waveguided Laser Wakefield Acceleration (SW-LWFA)~\cite{Feder2021,Miao2022,Sisma2025} for secondary interactions including the ability to interact them with additional, ultra-intense laser pulses. ELBA is operated either in single shot mode or at a repetition rate of up to 3.3 Hz. 
LWFA \cite{Tajima1979,Esarey_review} utilizing the strong (100 MeV/mm) electrostatic fields in plasma waves are excited by ultra-high intensity ($\gtrsim$10$^{19}$ W/cm$^{2}$) laser pulses to produce GeV-class, femtosecond electron bunches over short distances. Use of a plasma waveguide (a plasma analogue to glass optical fibers \cite{Durfee1993}) to confine the LWFA-drive pulse enables extension of the interaction to meter-scale distances without reliance on non-linear effects. Elsewhere, this has enabled the production of electron bunches with energies up to $\approx$10 GeV \cite{Rockafellow2025,Picksley2024} and charge above 1~nC~\cite{Rockafellow2025} in 30~cm accelerators. 
This environment provides a unique opportunity to investigate secondary particle production, including muons generated in beam-dump interactions, with high shot statistics. Figure~\ref{fig:laser_building} shows the layout of the laser building.

The present work should be regarded as a proof-of-principle demonstration of the deployment of an RPC-based muon detection system in a laser-plasma accelerator environment. These detectors were originally developed for autonomous cosmic-ray muography and were operated here for the first time in the challenging electromagnetic and radiation environment of a high-power laser facility. Furthermore, the measurements were performed parasitically alongside the commissioning and optimization of the primary ELBA experiment, without control over the accelerator settings or beam parameters. 

This paper is organized as follows. Section \ref{sec:beam-conditions} describes the ELBA experimental environment and the beam conditions under which the measurements were performed. Section~\ref{sec:detectors} presents the RPC detector system, the data acquisition chain, and its deployment at ELI Beamlines. Section~\ref{sec:results} discusses the detector performance, including the tracking algorithm, timing analysis, and a principal component analysis used to characterize the beam-induced background. Finally, Section~\ref{sec:summary} summarizes the main findings and the lessons learned from this proof-of-principle campaign, which will guide future dedicated measurements under controlled beam conditions. Additional technical details are provided in the appendices: Appendix~\ref{sec:thresh_scan} describes the validation of the DAQ thresholds, Appendix ~\ref{sec:cosmics_validation} presents the validation of the detector performance and analysis procedures using cosmic-ray data collected at CP3, and Appendix~\ref{sec:pca} provides further details of the principal component analysis.

\section{Electron Beam Conditions}
\label{sec:beam-conditions}

\begin{figure}[H]
    \centering
    \includegraphics[width=0.93\linewidth]{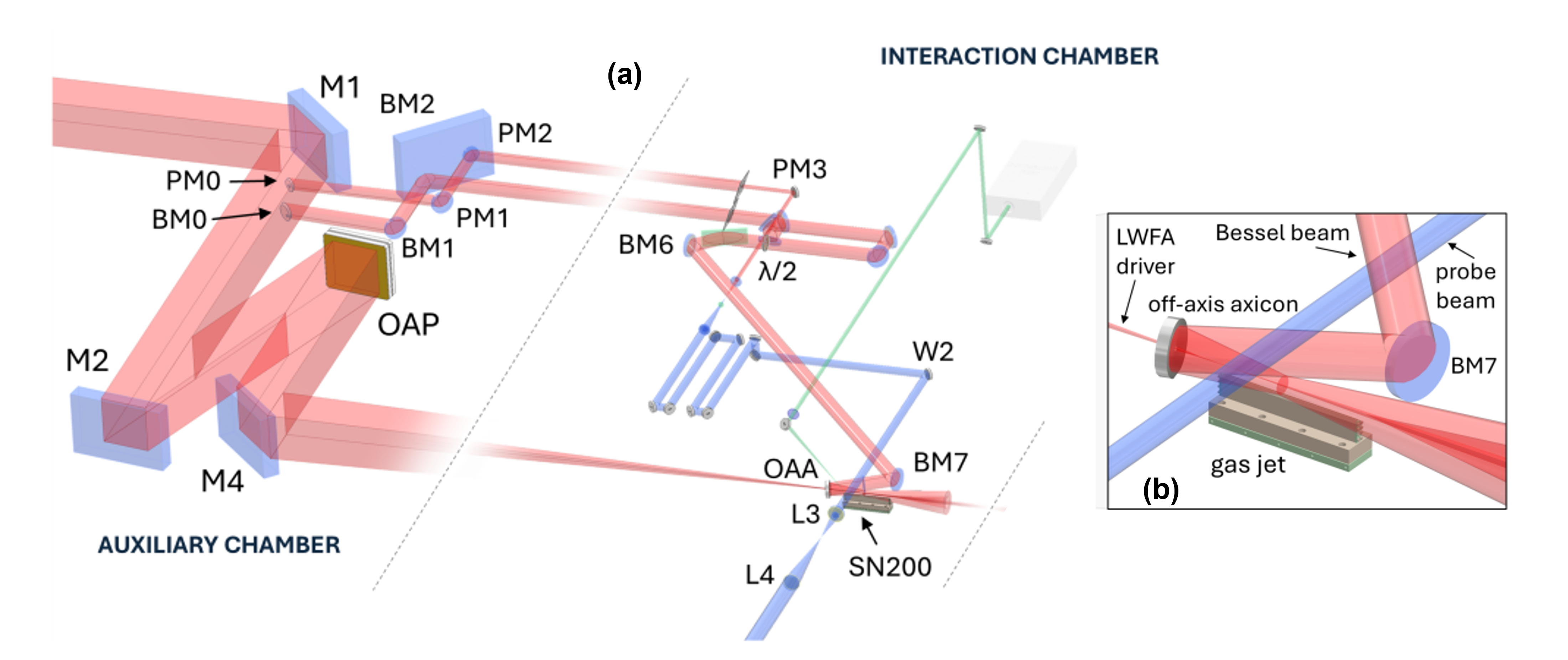}
    \caption{Schematic layout of the SW-LWFA beamline at ELBA (adapted from Figures 1 and 2 in \cite{Sisma2025}): (\textbf{a}) Beamlines for all-reflective SW-LWFA over a 20 cm gas jet. (\textbf{b}) Zoomed in schematic of 20~cm interaction region.}
    \label{fig:beam_schematics}
\end{figure}

 The ELBA setup, designed for experiments based on laser wakefield acceleration, has recently adopted a new platform for SW-LWFA \cite{Miao2022,Sisma2025}, enabling the production of Multi-GeV electron beams with O(100) pC of charge in optically generated plasma waveguides. A schematic of this new geometry, which uses masking and beam routing techniques to deliver multiple beams to the interaction area at precise time delays, is shown in Figure~\ref{fig:beam_schematics}. The 13~J, 30~fs L3 laser pulse is directed to mirror M1 after which small pick-off mirrors (PM0 and BM0) sample sub-apertures of the beam and direct them down separate beamlines. PM0 samples a few mJ portion of the beam which can be used for probing interactions. The optics in the channel-forming beamline, BM0-7 and the off-axis axicon (OAA), are used to generate a ~1.1 J $J_0$ Bessel beam above a 20 cm supersonic gas jet \cite{Miao2025} flowing Helium gas with small percentage dopants of Nitrogen or Argon to inject electrons into the accelerator \cite{Pak2010}. The $J_0$ pulse ionizes a narrow column of plasma, which expands into the surrounding neutral gas on nanosecond time-scales, forming a low-density ($1-8 \times 10^{17}$ cm$^{-3}$) plasma waveguide core surrounded by a much higher density neutral gas shock \cite{Feder2021,miao2020}. 

 Mirrors M2, M4, and the f = 10 m off-axis parabola direct and focus the remainder of the energy in the SW-LWFA drive pulse to a high-intensity focal spot with a 35 $\upmu$m FWHM. The leading wings of this pulse are sufficiently intense to ionize the neutral gas shock, enabling self-waveguiding \cite{Feder2021,Miao2022} as shock-ionization generates the high density plasma cladding of waveguide, confining the bulk of the drive pulse energy through the 20 cm accelerator. In commissioning experiments, this platform was able to produce multi-GeV electron bunches with up to 100's of pC and energies as high as 5 GeV~\cite{Sisma2025}.  

A range of diagnostics, including focal spot cameras, a top-view camera, a diagnostic wedge, and an interferometer are used to monitor the laser quality and characterize plasma conditions. The accelerated electron beam then propagates downstream through magnetic elements, where particles are deflected according to their energy via the Lorentz force, enabling energy measurement. In some diagnostic configurations, a ~0.2 mrad acceptance tungsten collimator is used before the magnetic field to improve spectral resolution of the diagnostic. Lanex screens and cameras record the electron distribution, while alignment systems ensure precise beam steering. Ultimately, the electrons are directed into a beam dump, which serves as a converter target for secondary particle production. When the collimator is in place, it typically samples less than 50\% of the total beam charge, and so also serves as a converter target.

\begin{figure}[H]
    \centering
    \includegraphics[width=0.85\linewidth]{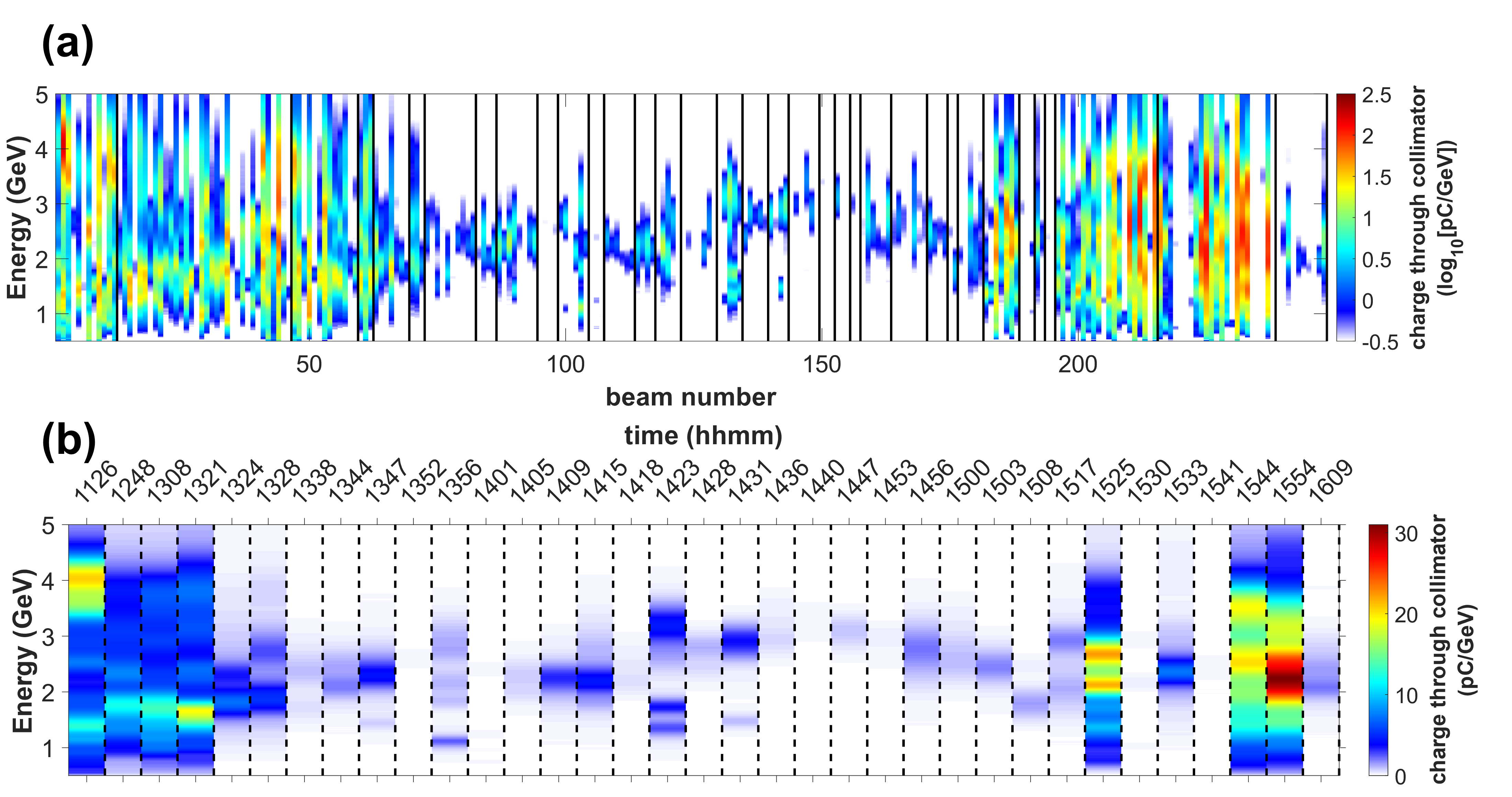}
    \caption{Example of beams produced during a typical day of parasitic operation. (\textbf{a}) All angle-integrated spectra collected during a day of operation at 0.2 Hz repetition rate. The solid lines denote periods of operation when the operating conditions were fixed.  (\textbf{b}) Mean angle-integrated collected for fixed operating conditions noted in panel (\textbf{a}). The top axis labels show the time each data set was collected at a 0.2 Hz repetition rate.}
    \label{fig:beams_example}
\end{figure}

During the typical parasitic operation employed in these tests, the accelerator underwent cycles of parametric tuning, misalignment, and attempts at spectral optimization. This means that beam parameters shifted frequently throughout a day of operation at <1 Hz repetition rates, with multiple, longer periods of downtime. An example of the resulting electron beam fluctuations during a typical day of operation are shown in Figure~\ref{fig:beams_example}. Panel (a) plots integrated spectra from each electron beam produced during a ~5 h experimental runtime. The black lines denote intervals for which the operating conditions (laser and plasma parameters, but not alignment) of the accelerator were fixed. The colormap is plotted in log-scale due to large shot to shot fluctuation in charge. These are primarily due to shot-to-shot alignment jitter between the SW-LWFA drive pulse and channel. Panel (b) below shows the mean spectrum for each of the 35 different sets of accelerator conditions. For these measurements, a collimating slit was used with the electron spectrometer, and so the measured charge represents only 5--50\% of the total beam charge. The range of charge and bunch energies shown here are typical for the data collected: electron energies of 500 MeV-5 GeV and charge from 10's--100's of pC.

In this configuration, muons are generated through the interaction of high-energy electrons with the dense converter target. The bremsstrahlung radiation produced in the target induces photonuclear reactions, leading to the formation of secondary mesons, primarily pions and, to a lesser extent, kaons, which subsequently decay into muons. Direct muon pair production from high-energy photons may also occur, although it remains subdominant at ELBA energies.
The resulting muon yield depends strongly on the properties of the electron beam, particularly its energy and charge, as well as on the material and geometry of the converter target.

The shot-to-shot variability of the beam conditions reduces the effective statistical power of the dataset. Because secondary muon photoproduction exhibits a steep threshold dependence on electron energy (suppressing production sharply below $\sim$2~GeV), shots that fluctuate toward lower energies contribute negligible signal yield. Based on the spectral time-series in Figure~\ref{fig:beams_example}, stable high-energy ($\ge$2.5~GeV) and high-charge ($\ge$50~pC) bunches were present in only approximately 20--30\% of the delivered shots.

\begin{figure}[H]
    \centering
    \begin{tabular}{cc}
    \includegraphics[width=0.49\linewidth]{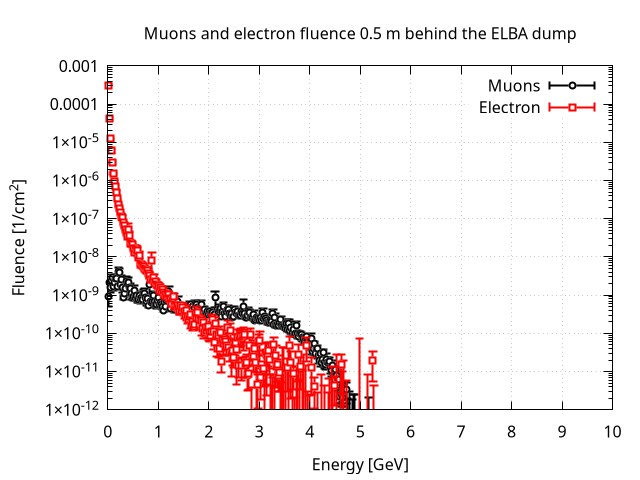} & 
    \includegraphics[width=0.49\linewidth]{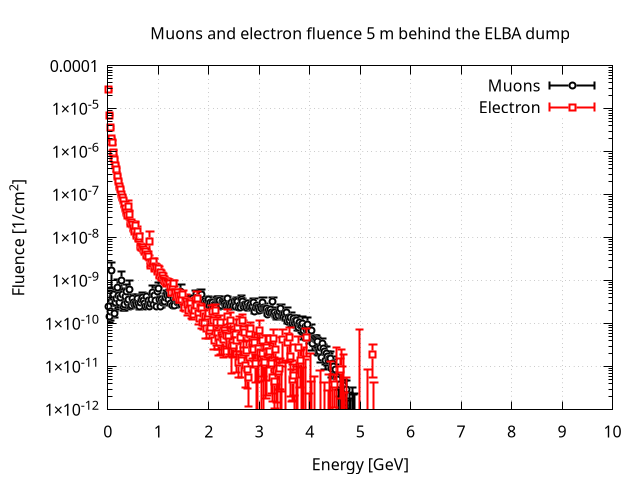} 
    \end{tabular}
    \caption{Muon and electron fluences as a function of energy estimate 0.5 m (left) and 5 m (right) downstream of the ELBA beam dump.} 
    \label{fig:Fluence} 
\end{figure}

\begin{figure}[H]
    \centering
    \begin{tabular}{cc}
    \includegraphics[width=0.99\linewidth]{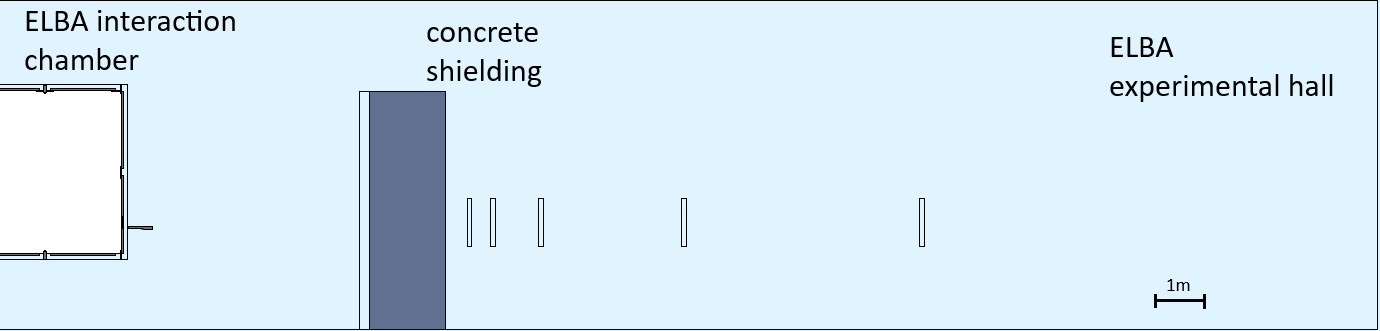} 
    \end{tabular}
    \caption{Horizontal cut at beam height of the FLUKA geometry of the ELBA experimental hall. The interaction vacuum chamber and the downstream shielding are visible. Behind the shielding the locations where the fluences of muons and electrons have been estimated are visible (thin rectangles).} 
    \label{fig:geometry} 
\end{figure}

Two data taking campaigns have been performed labeled as D1 and D2 (see Section \ref{sec:deployement}). The electron beam conditions do not substantially differ between the two campaigns.
The energy-dependent fluence of electrons and muons at distances of 0.5 m and 5 m downstream of the electron beam dump is shown in Figure~\ref{fig:Fluence}. The electron fluence is dominant at low energies (below ~1 GeV), but it decreases rapidly with increasing energy due to strong electromagnetic interactions and energy losses within the dump material. In contrast, the muon fluence is lower at low energies but decreases much more gradually with energy, extending up to approximately 3–5 GeV. This behavior reflects the greater penetrating power of muons, which interact less strongly with matter than electrons.
As a consequence, the beam dump effectively attenuates the electron component, while muons become the dominant contribution at higher energies. This distinction is important for radiation shielding considerations and for evaluating background contributions in downstream detectors.
FLUKA Monte Carlo simulations, normalized per primary electron, estimates the muon fluence at the detector location under realistic ELBA beam conditions. For electron beams in the 3-5 GeV energy range and charges on the order of 100 pC corresponding to approximately $10^{8}$--$10^{9}$ electrons per shot, the simulated muon fluence in the GeV range is found to be of the order of $10^{-10}$--$10^{-9}\ \mathrm{cm}^{-2}$ per primary electron. 
The simulations also indicate a large presence of secondary neutrons that are produced in photo- and electro-nuclear interactions induced by the electromagnetic shower generated by the interaction of the electron beam.
The FLUKA geometry of the experimental area is shown in Figure~\ref{fig:geometry}. 

\section{Detectors}
\label{sec:detectors}

Since 2017, the muography team at UCLouvain and VUB has been developing a portable, compact, and fully autonomous muon detector based on Resistive Plate Chambers (RPC)~\cite{RPC-Ikram2025}. The project has produced a detector system optimized for reliability, low-maintenance operation, and flexible deployment in challenging environments. Over the years, we have demonstrated stable long-term operation, field deployments in various configurations, and the successful integration of data acquisition, power management, and environmental monitoring subsystems, validating the viability of RPC technology in mobile muon applications. This experience is the foundation on which we now propose to bring the detector into a high-energy laser-accelerator context, leveraging years of expertise in portable muon instrumentation.

\begin{figure}[H]
    \centering
    \includegraphics[width=0.9\linewidth]{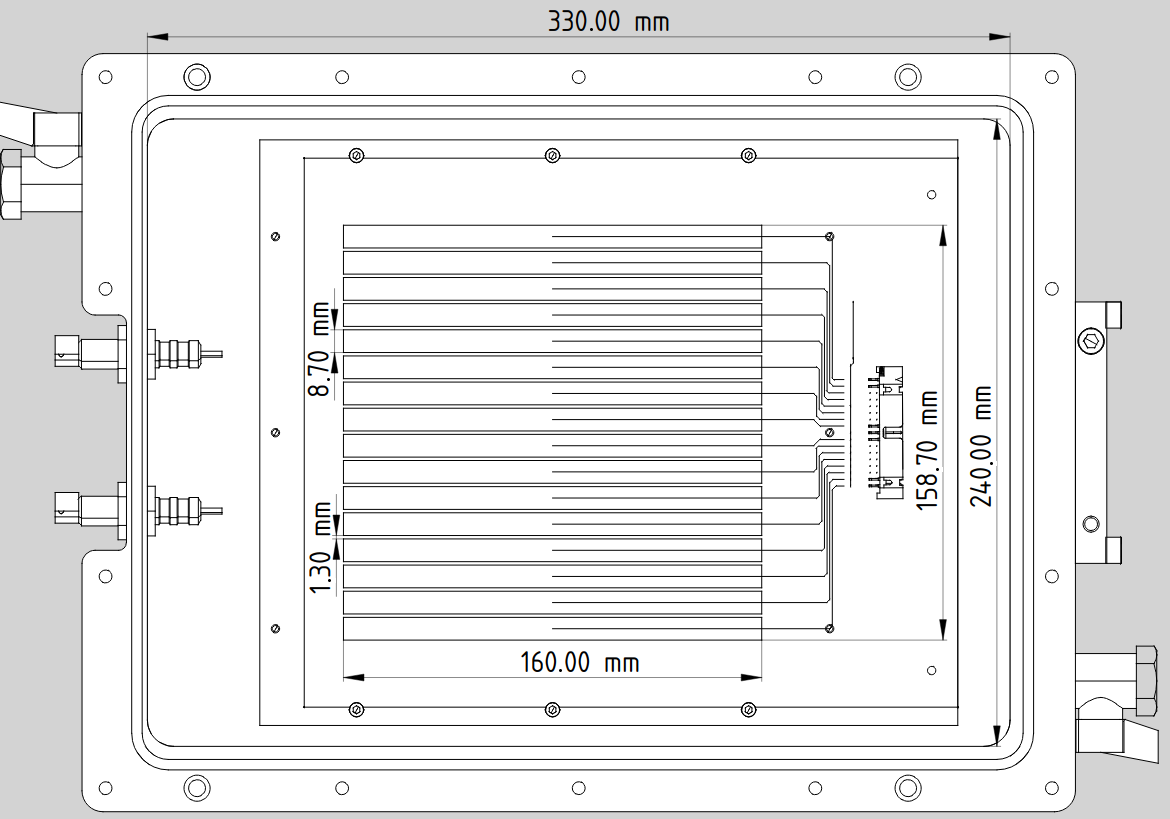}
    \caption{Sliced view of one of the RPC chambers.}
    \label{fig:rpc_slice}
\end{figure}

Our setup includes multiple RPC detectors, typically between two and four depending on the application; three RPCs have been employed in the work reported in this document. 
Each RPC is fabricated from $20 \times 20~\text{cm}^2$ glass plates with a thickness of $1.1~\text{mm}$. The plates feature an active area of $16 \times 16~\text{cm}^2$ and are coated with a resistive paint layer, with a surface resistivity ranging from $0.5$ to $1.0~\text{M}\Omega/\square$. A uniform gas gap of $1~\text{mm}$ is maintained between the plates. Glass electrodes and readout board enclosed in an aluminum box filled with a standard gas mixture composed of 95.2\% Freon, 0.3\% SF6, and 4.5\% isobutane at atmospheric pressure.  The enclosure has a total gas volume of $1.9~\text{L}$ and internal dimensions of $32.5 \times 24.0 \times 2.45~\text{cm}^3$. A technical drawing of the RPC is presented in Figure~\ref{fig:rpc_slice}.
The gas volume is flushed 4–5 times at a flow rate of $0.2~\text{L/min}$ to ensure complete removal of residual air. After flushing, both the inlet and outlet valves are closed. The readout system comprises 16 strips, each 160.0 mm long and 8.7 mm wide, separated by a 1.3 mm gap. 
The RPCs are positioned between two plastic scintillators, whose coincident signals act as both a trigger for the DAQ system and a reference for measuring efficiency. The signals from the scintillators are first digitized using a Constant Fraction Discriminator (CFD) before being sent to the DAQ. 
More information can be found in Ref.~\cite{RPC-Ikram2025}. 

\subsection{Data Acquisition System}
The Data Acquisition (DAQ) system comprises three integrated components, each fulfilling a distinct role within the experimental setup. These modules supply high voltage to the RPC, collect signals from the readout strips via a 34-pin ribbon cable—consisting of 16 signal pins and 18 ground pins—and process these signals to produce interpretable data for subsequent analysis. The system incorporates front-end electronics originally developed for the large-area RPCs used in the CMS experiment at CERN~\cite{CMSmuonTDR}, featuring an ASIC based on 0.8 $\upmu$m BiCMOS technology with a charge dynamic range of 20 fC to 20 pC. Each ASIC contains eight identical channels consisting of an amplifier, zero-crossing discriminator, monostabilizer, and differential line driver, and the present configuration employs four such chips to support 64 channels. The DAQ also includes the Trenz Electronic TE0720-03-1CF FPGA + CPU module, built around the Xilinx Zynq XC7Z020 system-on-chip, which integrates a dual-core ARM Cortex-A9 processor with an FPGA. Providing 75 differential LVDS input/output interfaces with a 5 ns clock cycle, this module interfaces directly with the front-end electronics. It operates autonomously as a self-contained unit that stores FPGA configuration files and data-acquisition software, and enables external communication via Ethernet and USB interfaces~\cite{RPC-Kumar2025}.

\begin{figure}[H]
    \centering
    \includegraphics[width=0.6\linewidth]{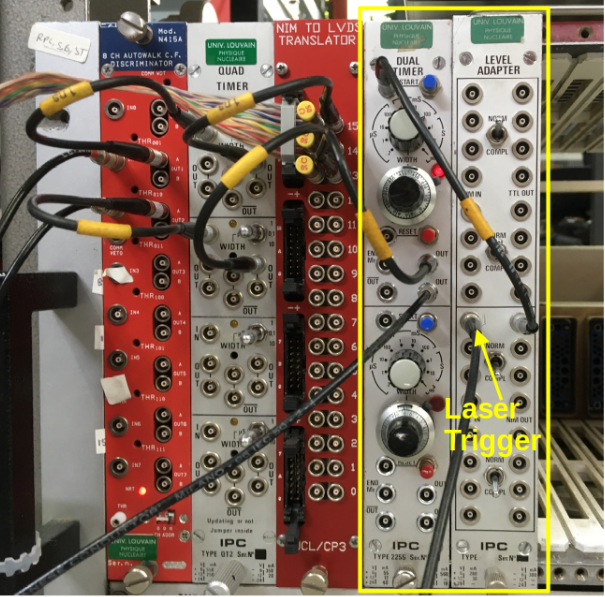}
    \caption{Modules highlighted in yellow correspond to TTL-to-NIM signal conversion and pulse-duration extension.}
    \label{fig:laser_trigger}
\end{figure}

\begin{figure}[H]
    \centering
    \includegraphics[width=0.7\linewidth]{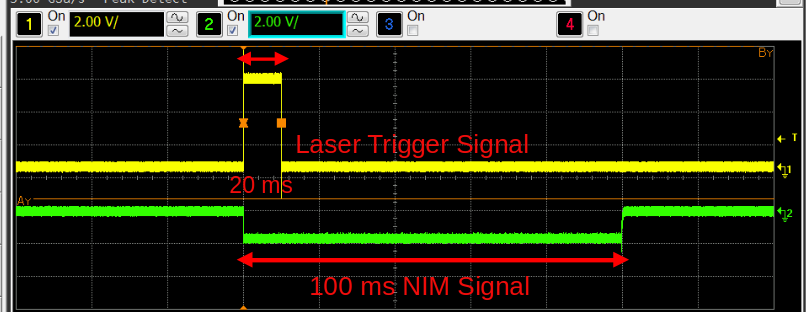}
    \caption{Oscillogram showing the approximately 20 ms laser-trigger pulse (yellow) and the corresponding 100 ms NIM pulse (green). The NIM pulse defines the tagging window for associating detector events with the laser trigger.}
    \label{fig:laser_signal}
\end{figure}

After the initial data-taking campaign (see Section \ref{sec:deployement}), several improvements were implemented. It was established that the laser trigger consists of a 5 V TTL signal with a custom pulse length, arriving 80 ms prior to the laser pulse. To accommodate this, a dedicated laser trigger signal was incorporated, necessitating updates to the FPGA configuration and data format, including the addition of a specific bit to register the trigger. On the hardware side, grounding wires were replaced to enhance signal stability, and new NIM modules were introduced: one for TTL-to-NIM signal conversion and another to extend the pulse duration to 100 ms, further optimizing the experimental setup as can be seen in Figure \ref{fig:laser_trigger} and real time laser trigger TTL signal converted to NIM pulse with width of 100 ms is shown in Figure \ref{fig:laser_signal}.

The RPC data-acquisition system is triggered by the coincidence of signals from two external plastic scintillators. The coincidence is formed through a logical AND, with an effective coincidence window of approximately 40 ns, determined from the pulse width at the output of the quad-timer NIM module. The rising edge of this coincidence signal defines the event trigger time. RPC strip hits are recorded relative to this trigger with a time-bin width of 5 ns. 

The ELBA timing system provides a TTL signal approximately 80 ms before each laser shot, which is converted to NIM level and stretched to a 100 ms pulse to tag beam-related events. The laser trigger is used only for event tagging, while the RPC hit timing is measured relative to the scintillator trigger with 5 ns resolution. The shot-to-shot jitter of the laser trigger relative to the laser pulse is approximately 30 ps, which is negligible compared with both the 100 ms tagging window and the 5 ns timing bins used throughout this work.

\subsection{Deployment at ELI Beamlines}
\label{sec:deployement}

The RPC telescope was successfully transported to ELI Beamlines. Data were collected during both beam-on and beam-off (cosmic muon data-taking) periods. The different datasets used in this work are categorized as follows:
\begin{itemize}
    \item D1: The first beam test setup, conducted in the ELBA experimental hall, is illustrated in Figure \ref{fig:detectorsetup_may}. Data collection was carried out over a two‑week period. The experimental arrangement includes a polyethylene block measuring 25 cm in height, 14 cm in thickness, and 12 cm in length, positioned directly in front of the detector system. As part of the shielding configuration, lead blocks with dimensions of 5 cm thickness, 28 cm height, and 33 cm length were also installed later. The distance from the polyethylene block to the first scintillator was 24 cm and to the last scintillator was 53 cm. 
    Vertically, the center of the detector system was placed at beam height.
    Layer 1 was oriented orthogonally to layers 0 and 2. The vertical distance between hits in layer 0 and layer 2 is used as a discriminant between cosmic muons (large distance) and beam-induced muons (small distance), while the distribution of hits in layer 1 is only used to verify whether the detector is at the core or at the edge of the cone of particles downstream of the beam. Two different data takings were conducted: D1 \textit{cosmic}, where the beam was off, and D1 \textit{beam} where the beam was on, with different shielding configurations. 
    
\begin{figure}[H]
    \centering
    \begin{tabular}{cc}
        \includegraphics[width=0.47\linewidth]{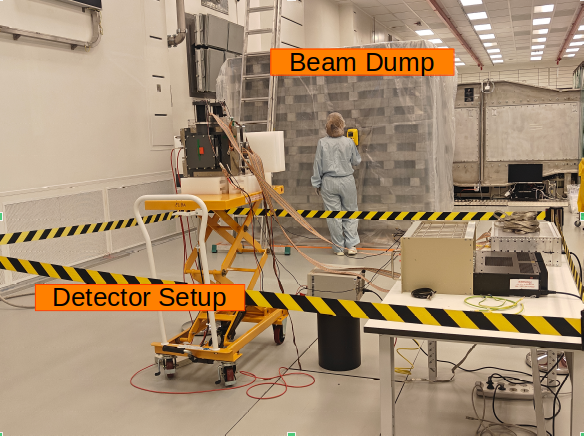} &
        \includegraphics[width=0.41\linewidth]{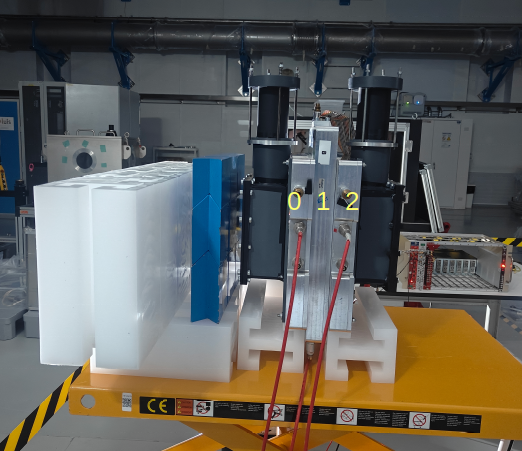} \\
        (\textbf{a}) & (\textbf{b})
    \end{tabular}
    \caption{(\textbf{a}) Detector setup in the experimental hall during the D1 beam test, positioned 335 cm from the beam dump; (\textbf{b}) three detectors with polyethylene and lead shielding. The numbers indicate the order assigned to the detectors in the analysis code.}
    \label{fig:detectorsetup_may}
\end{figure}
    
    \item D2: During the second data-taking period, the detector setup remained largely unchanged; however, its position was moved to 18 m from the beam dump, as shown in Figure~\ref{fig:detectorsetup_august}. In this configuration, all three chambers were aligned in the same orientation, enabling tracking analyses, as illustrated in Figure~\ref{fig:setup}. The detector geometry is shown in Figure~\ref{fig:detector_schematics}, where the setup is placed at an angle and located 18 m downstream of the beam dump for the D2 data-taking campaign. Three distinct data-taking modes were performed: D2 \textit{cosmic}, with the beam off and no shielding; D2 \textit{beam}, with the beam on in two configurations: first without shielding, and then with shielding consisting of 5 cm of lead and 14 cm of polyethylene; and D2 \textit{tag}, with the beam on and recording the trigger signal (\textit{laser tag}).
    \\ In the experimental configuration, a layered shielding system composed of lead (Pb) and polyethylene (PE) was installed upstream of the detector stack, with all elements aligned along the same axis such that the incoming particles traversed the shielding before reaching the detectors. The most upstream layer consisted of a 5 cm thick Pb shield, intended to attenuate gamma radiation prior to the detector system. Immediately downstream of the lead, an air gap of 6 cm was maintained, after which the particles encountered a horizontal PE shielding block placed in front of the detector system. This horizontal PE shielding had dimensions of 60 cm in length, 14 cm in width, and 25 cm in thickness. A further air gap of 16.5 cm was present between the horizontal PE shielding and the vertical PE shielding, which was positioned above the horizontal PE. The vertical PE shielding had dimensions of 73 cm in length, 25 cm in width, and 14 cm in thickness, and was installed upright relative to the table, laterally aligned to fully cover the active detector region. Downstream of the PE shielding, the distance from the back face of the PE to the scintillator system was 3 cm, while the distance to the RPC detector system was 11 cm. The RPC detector system occupied a total thickness of 12 cm along the particle direction, with transverse dimensions of 38 cm × 29 cm, and was positioned such that its active area was fully covered by the shielded region defined by the Pb and PE blocks. The trigger system consisted of two plastic scintillators with dimensions of 8 cm in thickness, 19 cm in length, and 17 cm in width, positioned upstream and downstream of the RPC stack. Finally, an additional 14 cm thick PE block was placed at the back of the detector setup to further suppress back-scattered secondary particles, particularly neutrons.
 
\begin{figure}[H]
    \centering
    \begin{tabular}{cc}
    \includegraphics[width=0.55\linewidth]{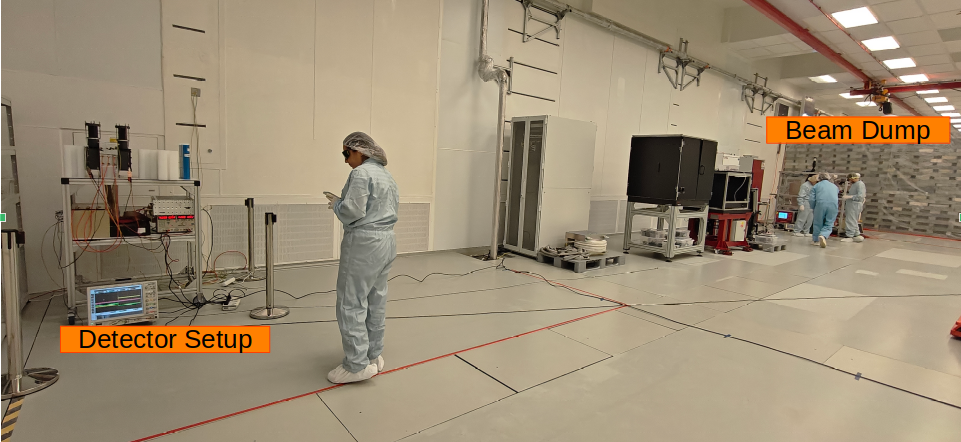} &
    \includegraphics[width=0.4\linewidth]{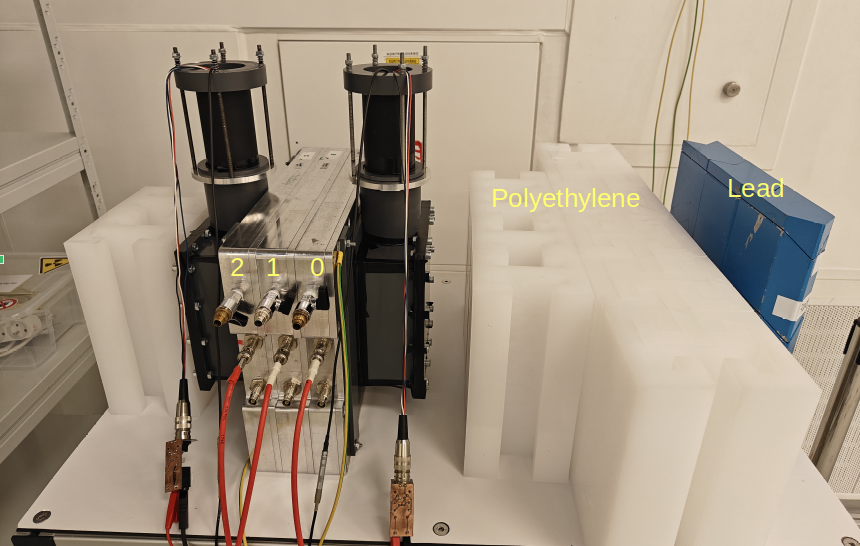} \\
    (\textbf{a}) & (\textbf{b})
    \end{tabular}
    \caption{(\textbf{a}) Detector setup in the experimental hall during the D2 beam test, positioned 18 m from the beam dump; (\textbf{b}) three detectors aligned in the same direction with polyethylene and lead shielding.}
    \label{fig:detectorsetup_august}
\end{figure}
 
    \item D3: A final campaign was conducted at the Gammatron\cite{bib:gammatron} beamline, a separate experimental station at ELI Beamlines delivering electron beams with energies up to 1 GeV. At these energies, muon production is expected to be negligible, allowing the collection of a control dataset for the characterization of beam-induced backgrounds. The apparatus was located approximately 9 m behind a concrete shielding wall. Several lead and polyethylene (PE) shielding configurations were investigated to study their effect on the detector response and to aid the background classification discussed in Section~\ref{sec:shielding}.

\begin{figure}[H]
    \centering
    \includegraphics[width=0.7\linewidth, angle=270]{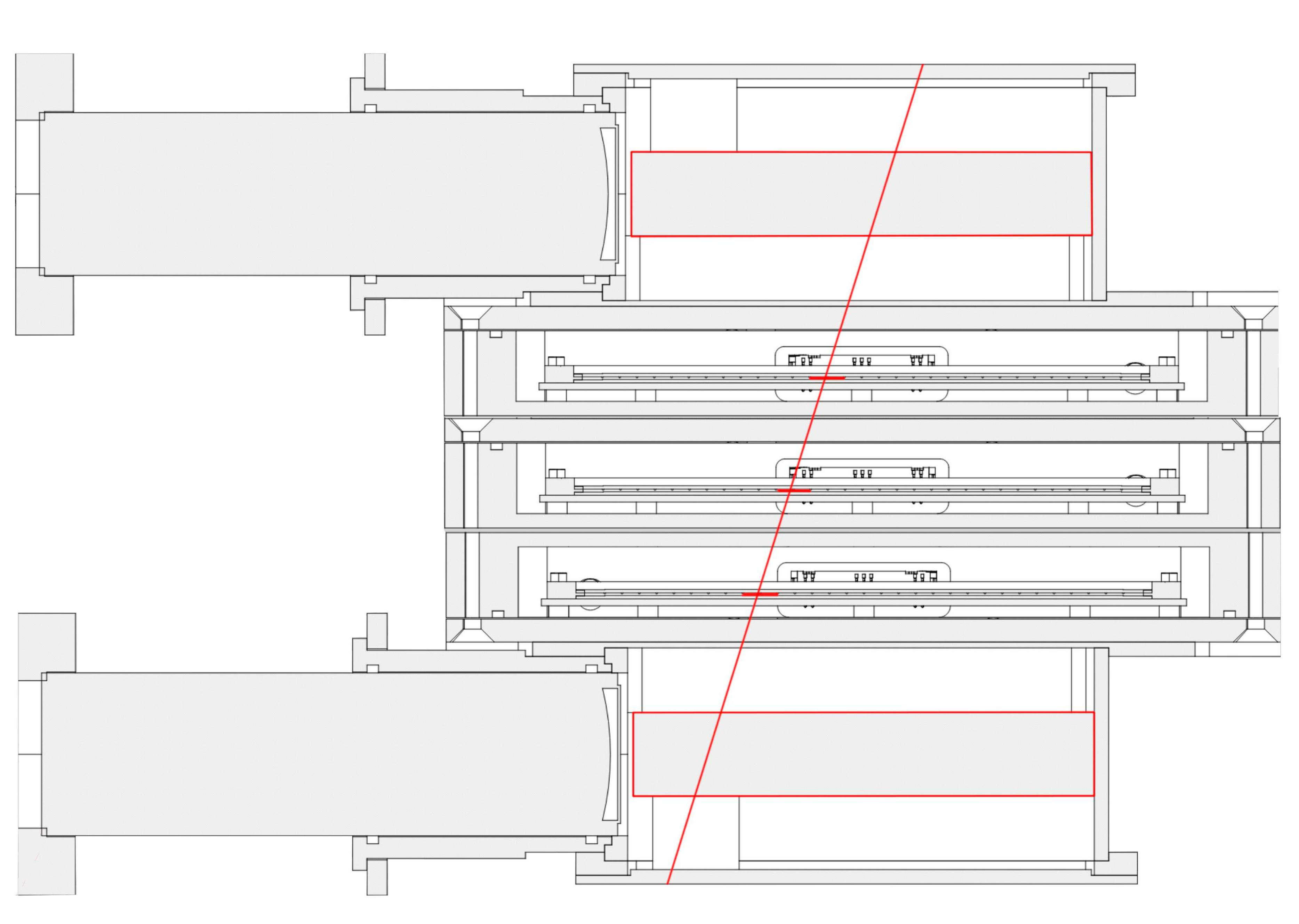}
    \caption{Side sliced view of the three RPCs and the two scintillators detection setup deployed during the D2 and D3 data taking. Scintillator active area and RPCs fired strips are highlighted in red.}
    \label{fig:setup}
\end{figure}

\begin{figure}[H]
    \centering
    \includegraphics[width=\textwidth]{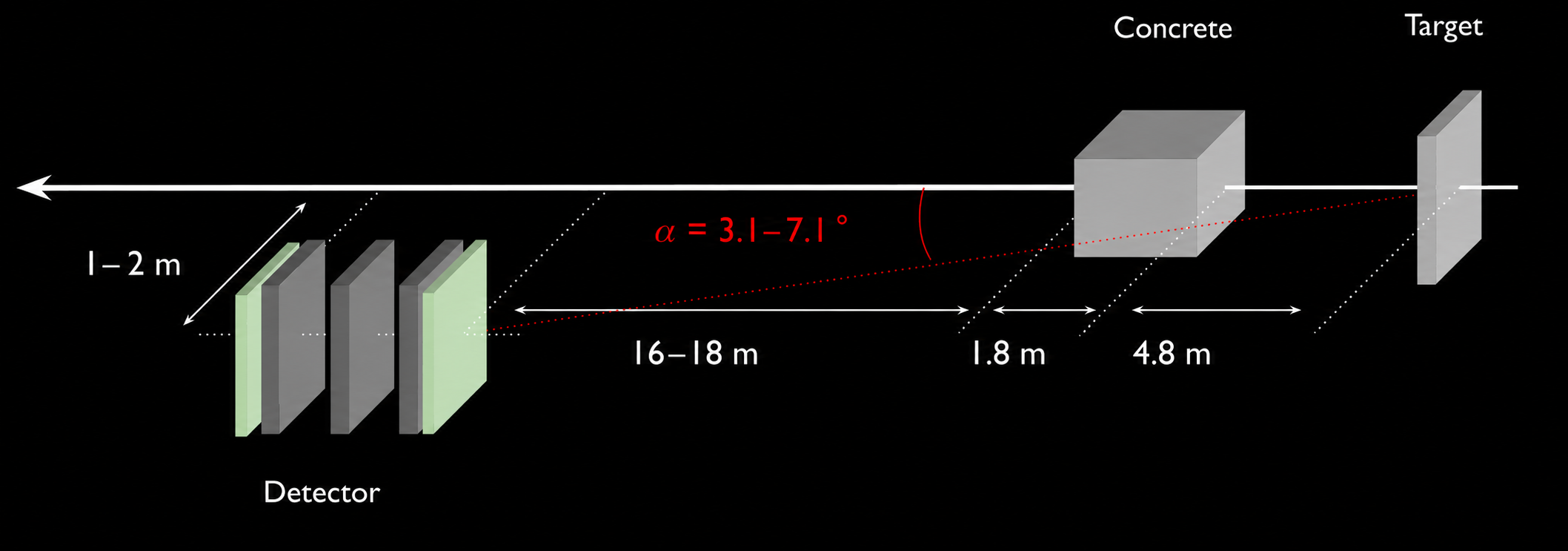}
    \caption{Schematic of the experimental configuration and geometry used to define the detector system in the experimental hall.}
    \label{fig:detector_schematics}
\end{figure}
\end{itemize}

\subsection{Event Selection and Track Identification}
\label{sec:event-selection-and-track-id}

A simple reconstruction algorithm was developed to characterize muon tracks and determine their direction relative to the horizontal axis. When an event is triggered by the external scintillators coincidence, the identifiers of RPC strips registering induced charge from traversing particles are recorded. 

In the majority of cosmic muon events, only one strip per detector layer is activated, which allows for straightforward track reconstruction. Nevertheless, additional strip activations may occur due to electronic noise or secondary particle production. These contributions must be handled using a more robust selection strategy, through clustering of adjacent strips and by rejecting inconsistent signals. 
This becomes even more important in beam events, where several particles induce simultaneous signals in the same time bins. 

A cluster is defined as a set of at least two spatially contiguous strips registered within the same 5~ns time bin. The position of a cluster is computed as the centroid of the strip positions involved. 

\begin{figure}[H]
    \centering
    \includegraphics[width=0.9\linewidth]{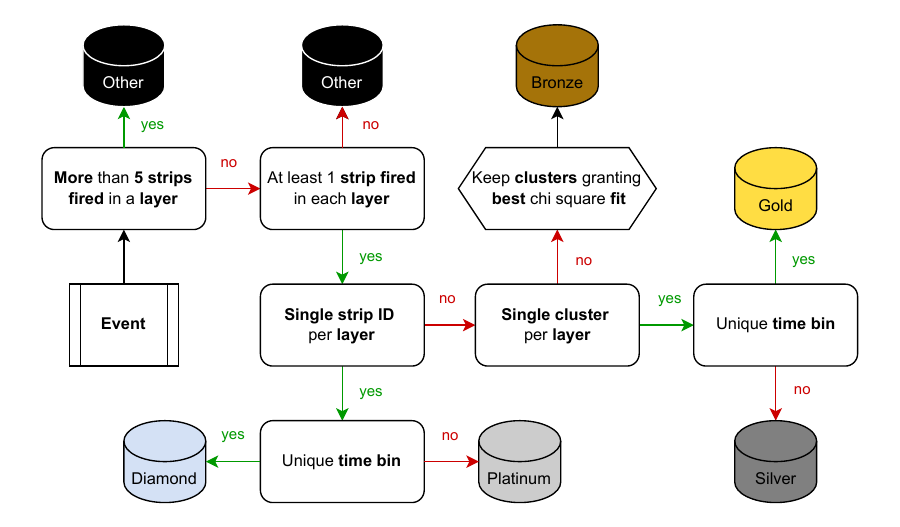}
    \caption{Decision Tree of the tracking algorithm used to process data from the D2 and D3 data takings. In the D1 case, $\chi^2$ is not used in the Bronze category.}
    \label{fig:decision_tree}
\end{figure}

A classification scheme was introduced to assess the reliability of the reconstructed tracks. This procedure is implemented using a decision tree to ensure a uniform treatment of all events, regardless of whether they belong to cosmic runs or beam shots; the corresponding workflow is shown in Figure \ref{fig:decision_tree}. The classification is based on the number of fired strips and/or clusters per layer, as well as their temporal distribution within 5 ns time bins. Events are classified into five categories based on the number of fired strips and/or clusters per layer and their temporal distribution, as well as a final exclusion class:
\begin{itemize}
\item \textbf{Diamond} events correspond to the highest reconstruction quality, in which exactly one strip is fired per layer and all strips are contained within the same time bin. 
\item In \textbf{Platinum} events, exactly one strip per layer is observed, but the strips are distributed across different time bins. 
\item \textbf{Gold} events are defined by exactly one cluster per layer, with all clusters occurring within the same time bin; these typically arise when muons traverse near strip boundaries, leading to charge sharing between adjacent strips. 
\item In \textbf{Silver} events, exactly one cluster per layer is present, but clusters are distributed across different time bins. 
\item \textbf{Bronze} events contain more than one cluster per layer, with clusters potentially spanning multiple time bins. In the D2 and D3 datasets, additionally, $\chi^2$ is used to select the clusters belonging to the best fitting track, and discard the others from further analysis. 
\item Finally, \textbf{Other} events correspond to cases in which only two of the three RPC chambers register hits and are therefore excluded from the tracking analysis. 
\end{itemize}

This classification procedure is implemented through a decision tree to ensure consistent treatment of all events; the corresponding scheme is shown in Figure \ref{fig:decision_tree}. 

\begin{figure}[H]
    \centering
    \includegraphics[width=0.99\linewidth]{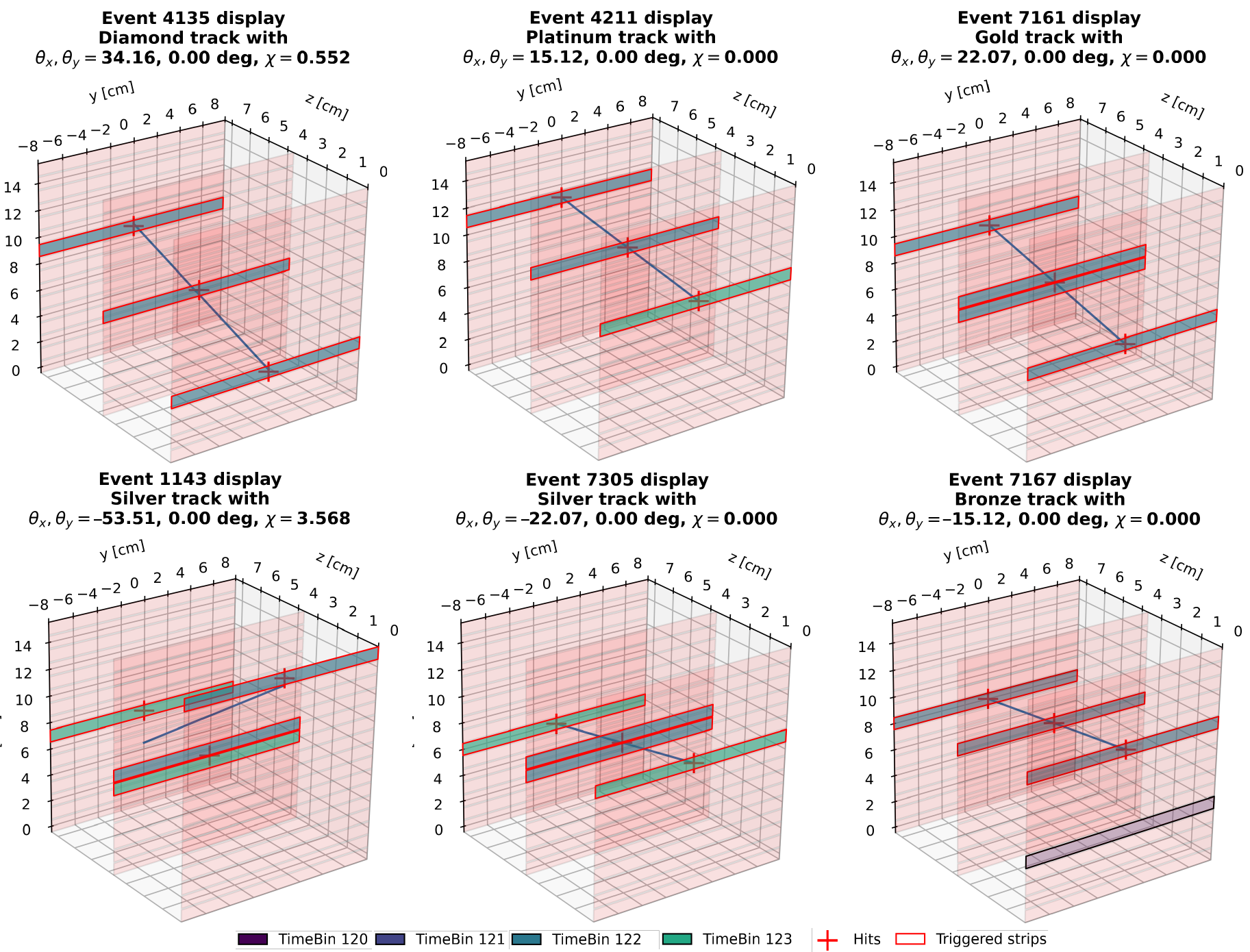}
    \caption{Event display of diamond, Platinum and Gold tracks (above), silver and bronze tracks (bottom). The track angle with respect to the horizontal plane $\theta_x$ in $^\circ$ and the residuals of the track linear fit $\chi$ in cm are given for each event.}
    \label{fig:event_display_class}
\end{figure}

Representative event displays for each category are shown, for dataset D2, in Figure \ref{fig:event_display_class}. After classification, the relevant strip identifiers are converted into spatial hit positions using the known geometry of the RPC chambers, as illustrated in Figure \ref{fig:setup}.

\section{Results}
\label{sec:results}

\subsection{D1 Campaign}
This section describes the data collected during the D1 data-taking campaign, with key details provided in Table~\ref{table:summary1}. The operating high voltage of 7 kV was selected by optimizing for maximum efficiency during cosmic-ray measurements in laboratory conditions, while maintaining a leakage current below 60 nA at thresholds of 30 and 35. During this period, layer 1 was oriented orthogonally to layers 0 and 2. The search for beam-induced muons was conducted by an angular distribution analysis. 
The rationale for orienting layer 1 orthogonally was to use the distribution of its signals to confirm that our detectors were horizontally positioned at the core, and not at the edge, of the cone of particles produced downstream of the beam dump. It was implicitly assumed, at the time of planning the campaign, that the backgrounds would have been sufficiently low that a 3-point tracking (which would have required the three detectors to be oriented all parallel) was not necessary. 

\begin{table}[H]
\caption{Summary of the D1 Cosmic, and D1 Beam data takings. The DAQ threshold settings were validated through a threshold scan, as described in Section~\ref{sec:thresh_scan}.\label{table:summary1}}

\begin{tabularx}{\textwidth}{CcCC}
 \toprule
 \textbf{Data Taking} & \textbf{Shielding} & \textbf{Duration} & \textbf{Threshold [a.u]} \\ 
 \hline
 D1 Cosmic & Polyethylene & 4 h & [30, 35] \\ 
 D1 Beam & Polyethylene & 28 h & [30, 35]  \\
 D1 Beam &Lead + Polyethylene & 4 h & 35  \\
 D1 Beam &Polyethylene + Lead & 10 h & 30 \\
\bottomrule
\end{tabularx}
\end{table}

\subsubsection{Strip Occupancy and Multiplicity in D1}

\begin{figure}[H]
    \begin{adjustwidth}{-\extralength}{0cm}
    \centering
    \subfloat[Cosmic-ray data.]{\includegraphics[width=\linewidth]{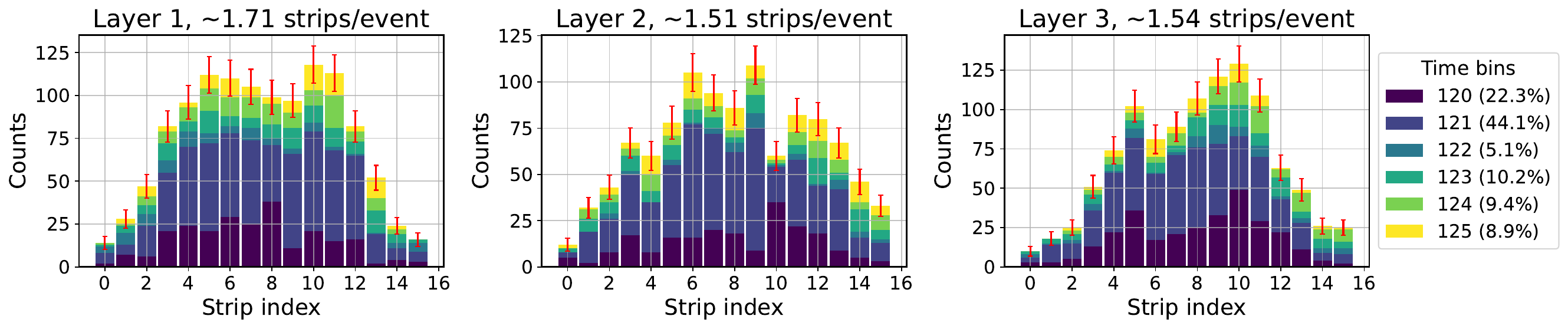}}\\
    \subfloat[Beam data (PE–Pb).]{\includegraphics[width=\linewidth]{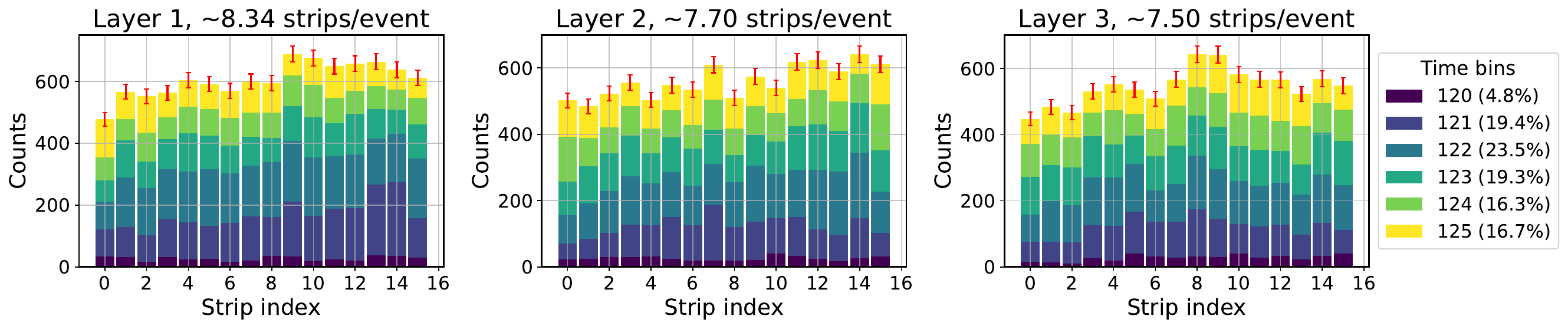}}\\
    \subfloat[Beam data (Pb–PE).]{\includegraphics[width=\linewidth]{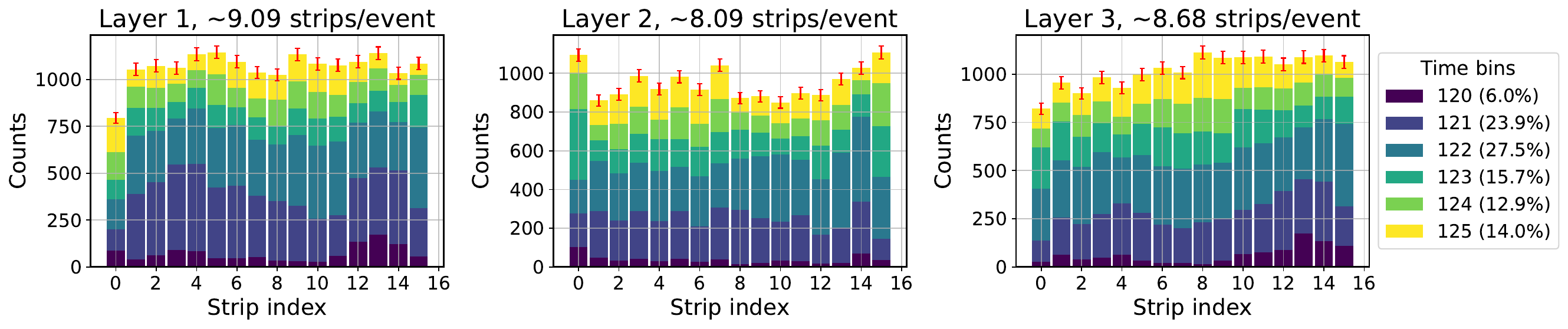}}\\
    \subfloat[Beam data (PE only).]{\includegraphics[width=\linewidth]{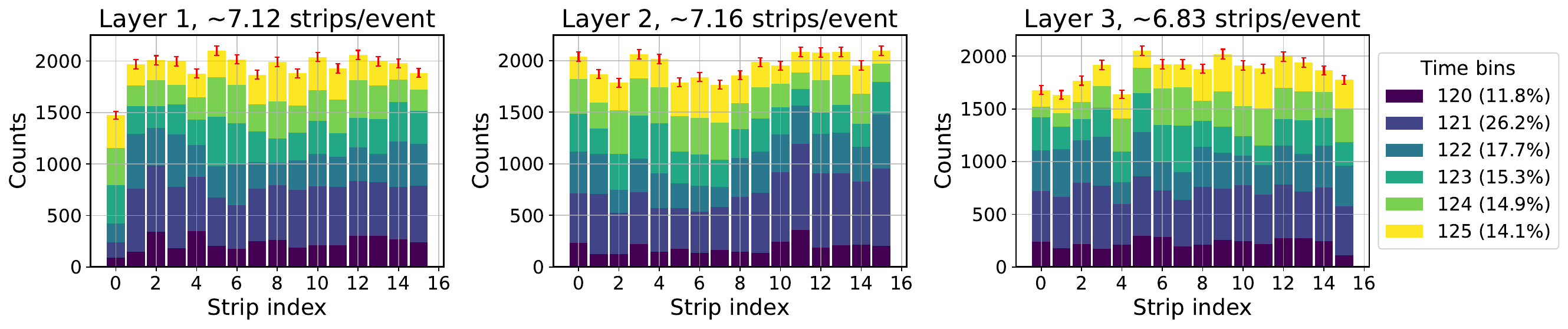}}
    \end{adjustwidth}
    \caption{Strip occupancy distributions for the D1 dataset under different configurations. The top panel shows cosmic-ray data, while the lower panels correspond to beam data with different material arrangements: polyethylene followed by lead (PE–Pb), lead followed by polyethylene (Pb–PE), and polyethylene only. The occupancy patterns reflect variations in particle flux and secondary production induced by the different absorber configurations. Poissonian uncertainties are shown in red.\label{fig:D1_occupancies}}
\end{figure}

During the D1 data-taking campaign, multiple shielding configurations were studied, including lead facing the beam followed by polyethylene, polyethylene facing the beam followed by lead, and a configuration consisting solely of polyethylene. Detectors were placed only 335 cm away from the beam dump. Strip occupancy distributions corresponding to the data takings are shown in Figure~\ref{fig:D1_occupancies}.
The D1 cosmic run is used as a baseline to evaluate the detector response to single muons. With the beam turned off, cosmic muons constitute the dominant component of the charged-particle flux. The strip-occupancy distribution for the cosmic dataset shows that the majority of fired strips are concentrated in the 5 ns-wide time bin 121, corresponding to 605 ns after the scintillator trigger.

For the D1 beam data, the activity extends over multiple time bins, namely 121, 122, and 123 (corresponding to 605, 610, and 615 ns after the scintillator trigger). While the D1 cosmic run yields approximately one fired strip per layer per event, the D1 beam dataset exhibits significantly higher multiplicities, reaching up to nine strips in the first layer (closest to the beam and the polyethylene slab) and decreasing to approximately 7–8 strips in the third layer.

\begin{figure}[H]
    \centering
    \subfloat[D1 cosmic]{\includegraphics[width=.48\linewidth]{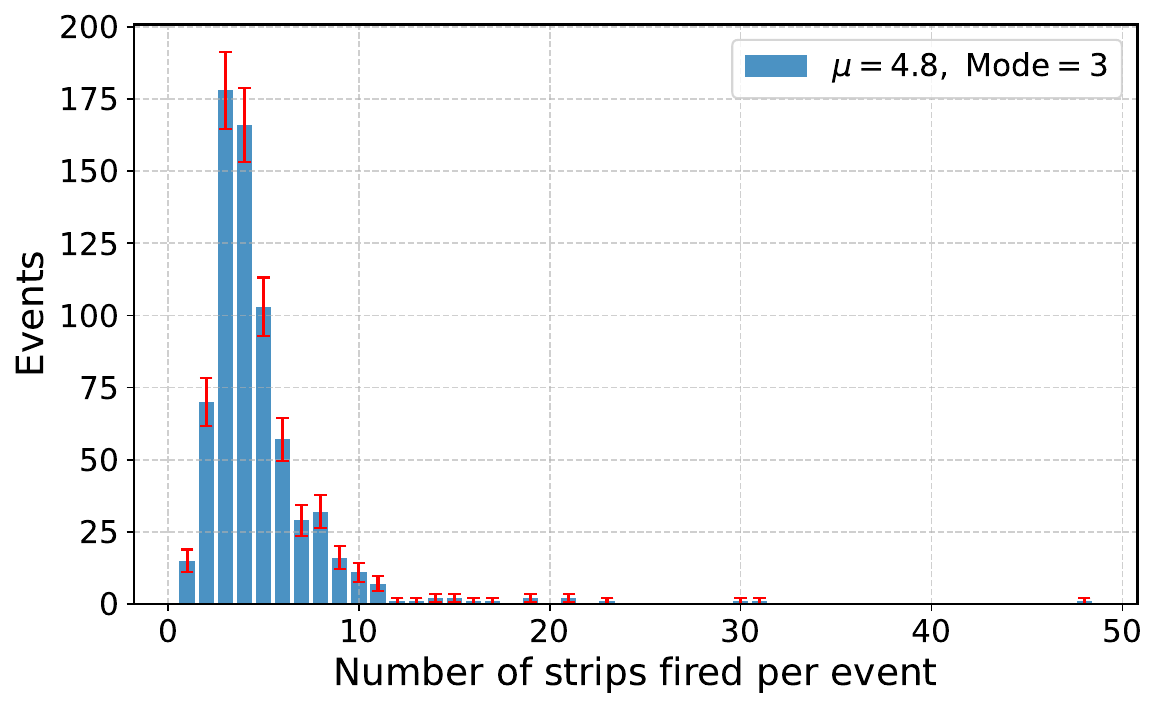}}  
    \hfill
    \subfloat[Beam (PE–Pb)]{\includegraphics[width=.48\linewidth]{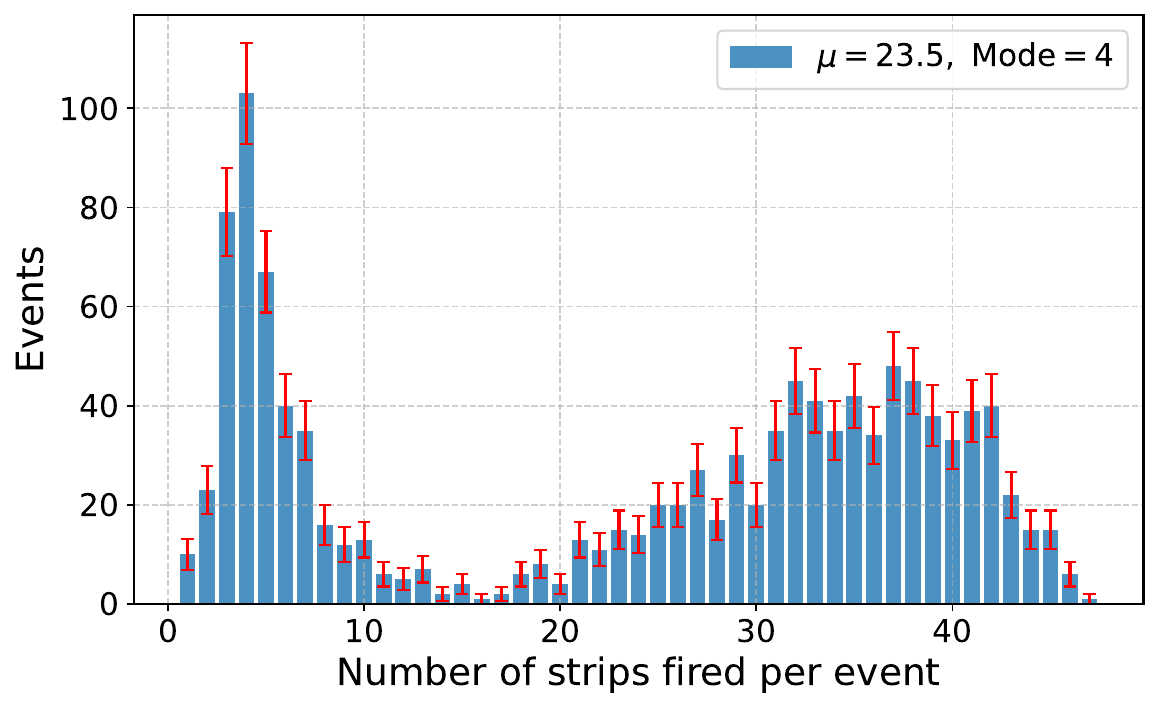}}\\
    \vspace{0.2cm} 
    \subfloat[Beam (Pb–PE)]{\includegraphics[width=.46\linewidth]{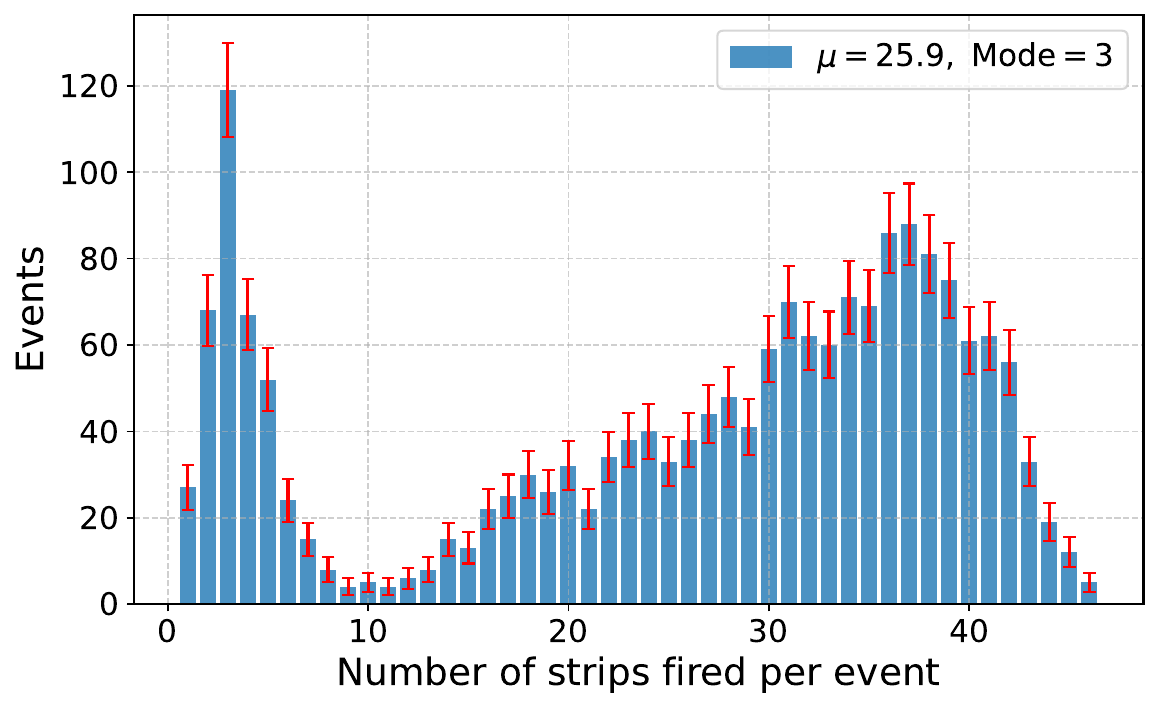}}    
    \hfill
    \subfloat[Beam (PE only)]{\includegraphics[width=.48\linewidth]{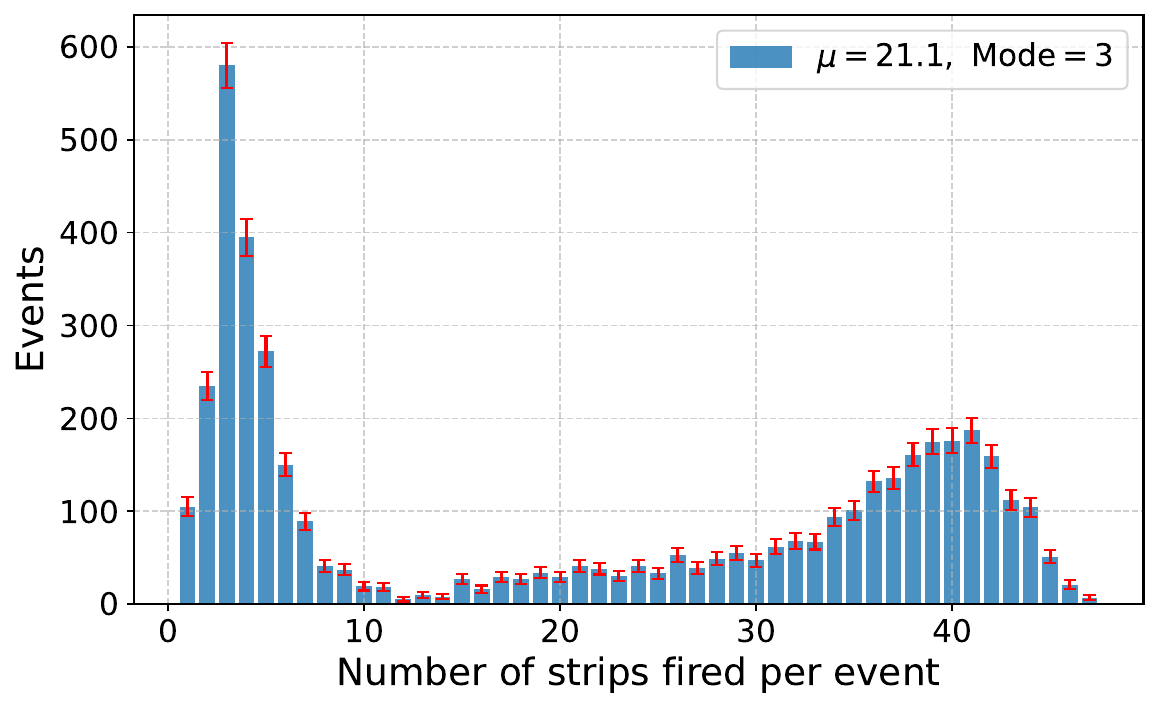}}
    \caption{Strip multiplicity distributions for the D1 dataset for cosmic and beam configurations. Beam data are shown for different absorber arrangements (PE–Pb, Pb–PE, and PE only), highlighting the increase in multiplicity relative to cosmic-ray events. Poissonian uncertainties are shown in red.}
    \label{fig:D1_multiplicities}
\end{figure}

This behaviour indicates an increased charged-particle multiplicity in regions closer to the polyethylene blocks, consistent with secondary particle production induced by the beam.
The distributions of the number of strips fired per event shown in Figure~\ref{fig:D1_multiplicities} corroborate the hypothesis of a higher number of charged particles entering the detector acceptance for the D1 beam data taking, for which the average number of strips fired per event reaches 26 compared to 4.8 for D1 cosmic. While the distribution means differ significantly, the mode of the distribution remains the same at 3.0 and 4.0, suggesting that the cosmic muon contribution is dominant. The distribution exhibits a pronounced peak at low multiplicity (mode = 3), characteristic of cosmic-like events, together with a long high-multiplicity tail extending up to ~50 strips, indicating the presence of beam-induced secondary particle showers for different shielding configurations.

\subsubsection{Angular Distribution}

Beam-induced muons are expected to exhibit a strong alignment due to their common production direction, and to propagate predominantly horizontally, leading to a narrow distribution around $\theta_x \approx 0$. Cosmic muons are expected to cross the detector at relatively large angles with respect to the beam axis, with a distribution shaped by the acceptance of the detector. This distribution is well characterized using the D1 cosmic control dataset. Combinatorial background arises from random associations of detector hits induced by the beam, which are unrelated to true muon tracks.

\begin{figure}[H]
    \centering
    \subfloat[D1 cosmic]{\includegraphics[width=.48\linewidth]{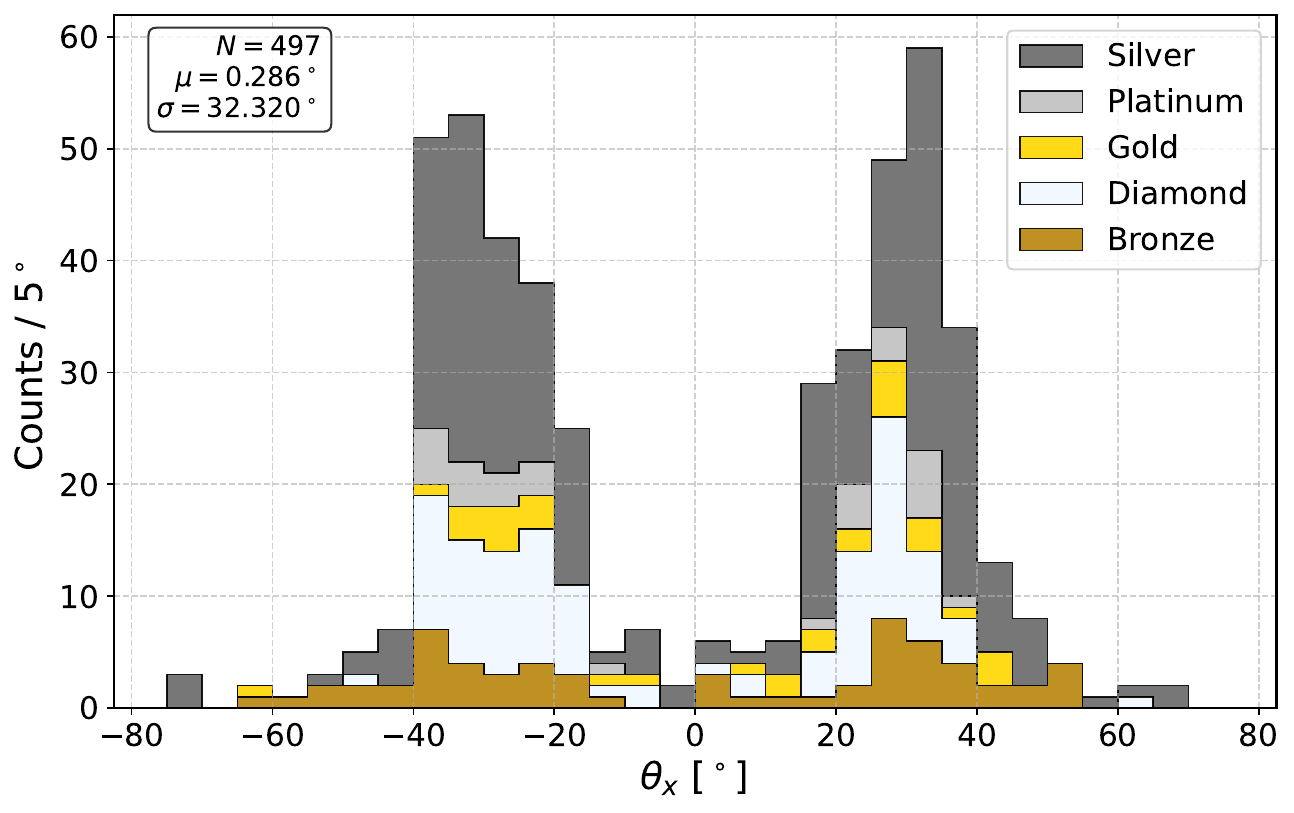}}       
    \hfill
    \subfloat[Beam (PE–Pb)]{\includegraphics[width=.48\linewidth]{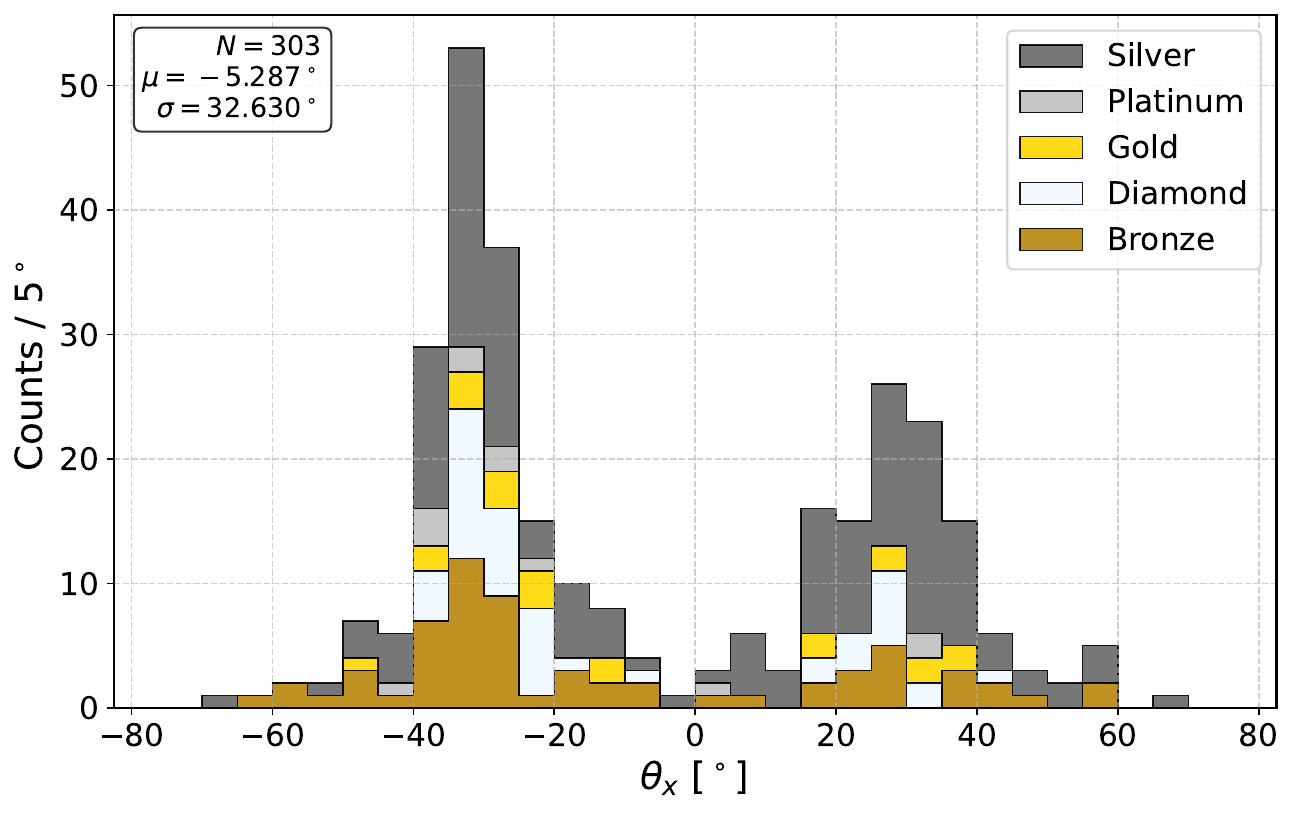}}\\
    \vspace{0.2cm}
    \subfloat[Beam (Pb–PE)]{\includegraphics[width=.48\linewidth]{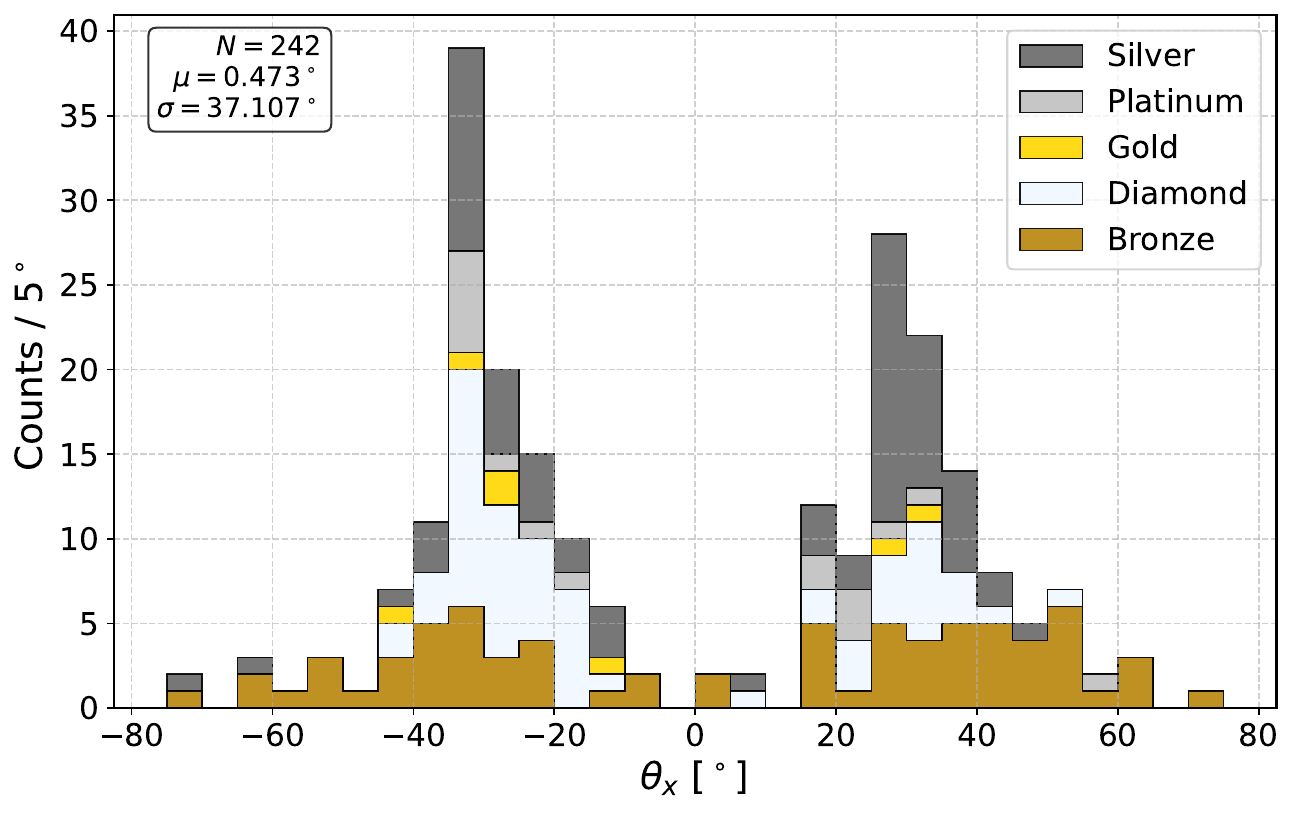}}       
    \hfill
    \subfloat[Beam (PE only)]{\includegraphics[width=.48\linewidth]{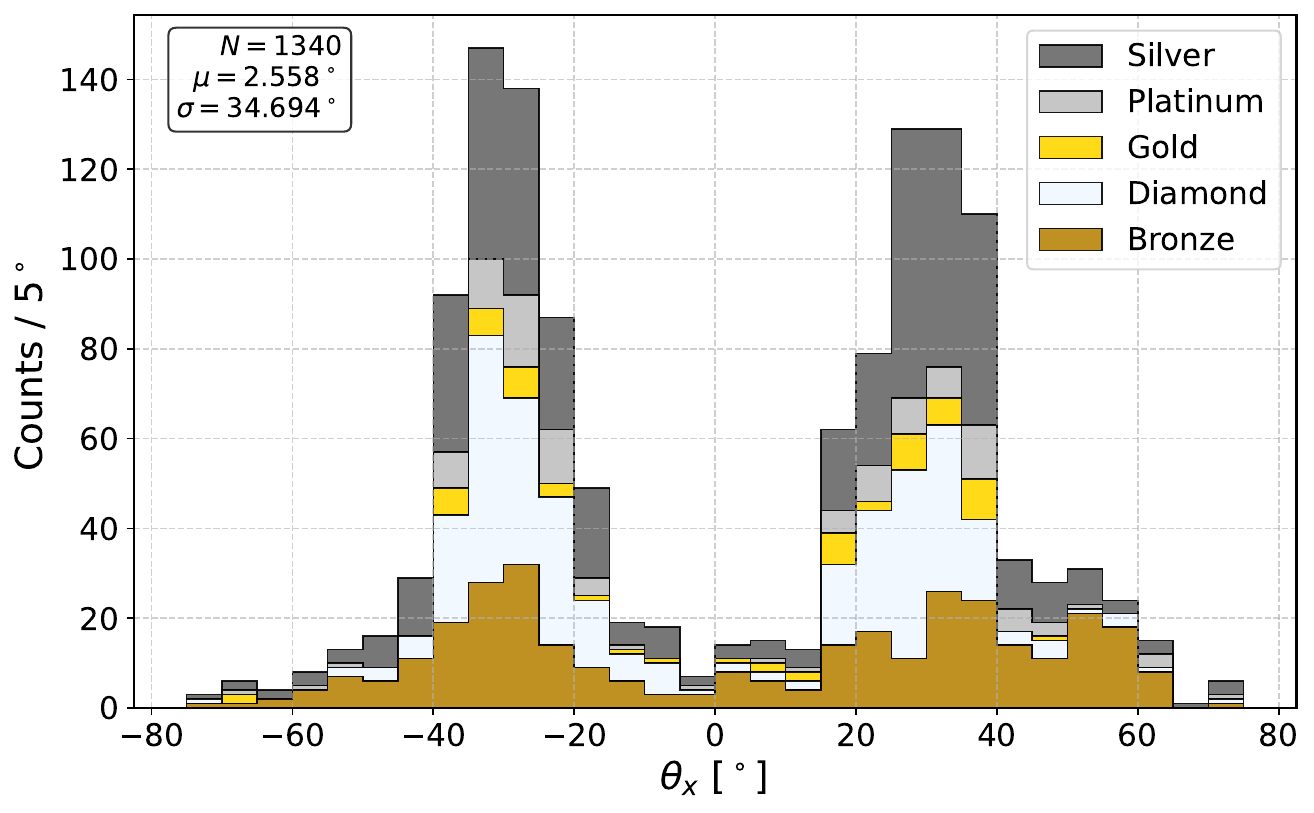}}     
    \caption{Angular distribution of reconstructed tracks for different shielding configurations: (\textbf{a}) D1 cosmic control dataset, (\textbf{b}) polyethylene facing the beam followed by lead, (\textbf{c}) lead facing the beam followed by polyethylene, and (\textbf{d}) configuration consisting solely of polyethylene. The distributions reflect the angular acceptance of the detector and the impact of shielding on the observed track populations.}
    \label{fig:D1_theta}
\end{figure}

The resulting angular distribution shown in Figure~\ref{fig:D1_theta} are generally consistent with expectations for cosmic muons: only a small fraction of reconstructed tracks exhibit near-horizontal orientations. The peak of the distribution occurs near small angles $\theta_x \in [-20^\circ, 20^\circ]$,  indicating that a larger fraction of events is concentrated around the forward direction. For the D1 beam data, the angular distribution closely resembles that observed in the D1 cosmic dataset. Overall, the angular distribution results do not provide conclusive evidence for the presence of beam-induced muons, which are expected to exhibit predominantly horizontal trajectories within the range $\theta_x \in [-5^\circ, 5^\circ]$. Instead, the tracks that successfully pass the reconstruction display an angular distribution similar to that of the D1 cosmic data, suggesting that the reconstructed tracks are dominated by cosmic muons rather than beam-related events.

\subsubsection{Timing Information in D1}

\begin{figure}[H]
    \centering
    \subfloat[D1 cosmic]{ \includegraphics[width=.48\linewidth]{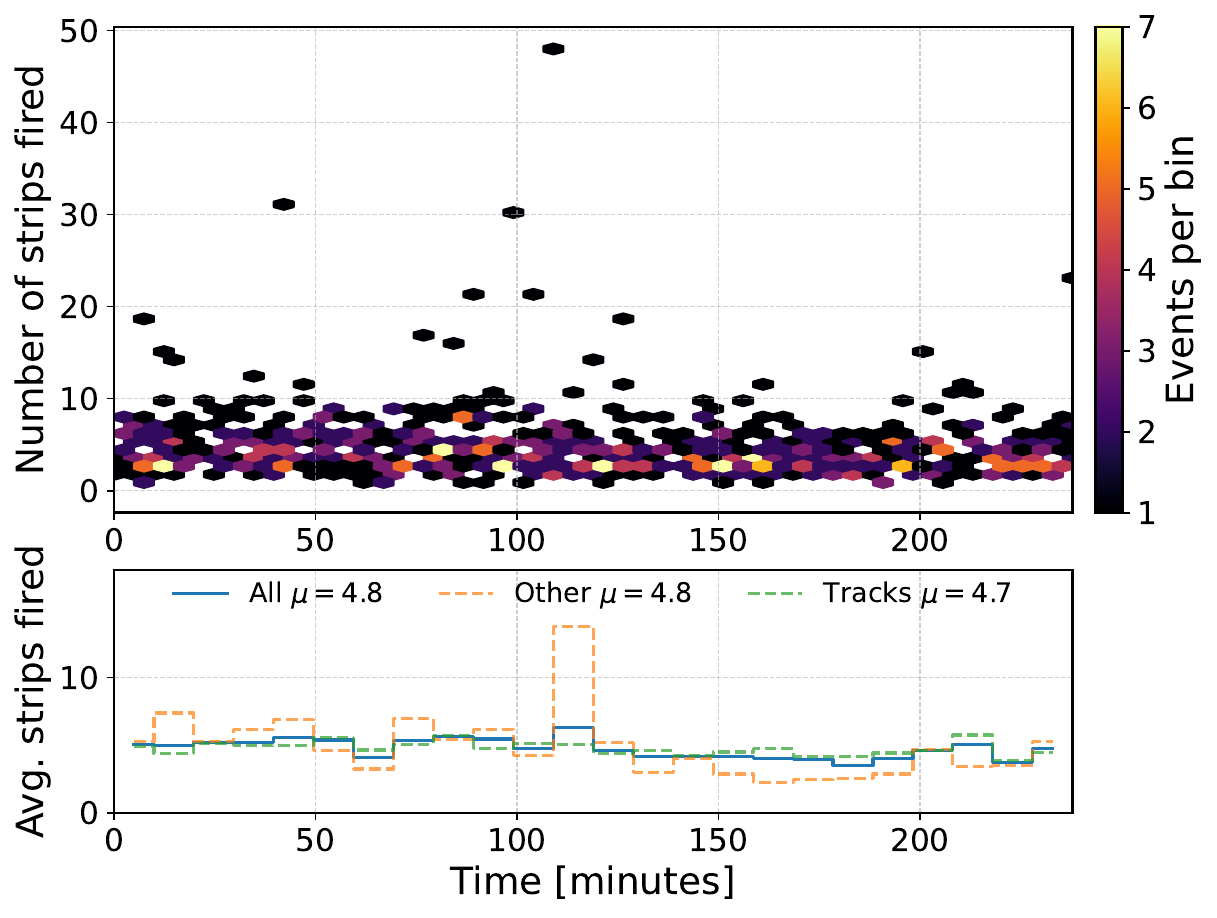}}
    \hfill
    \subfloat[Beam (PE–Pb)]{\includegraphics[width=.48\linewidth]{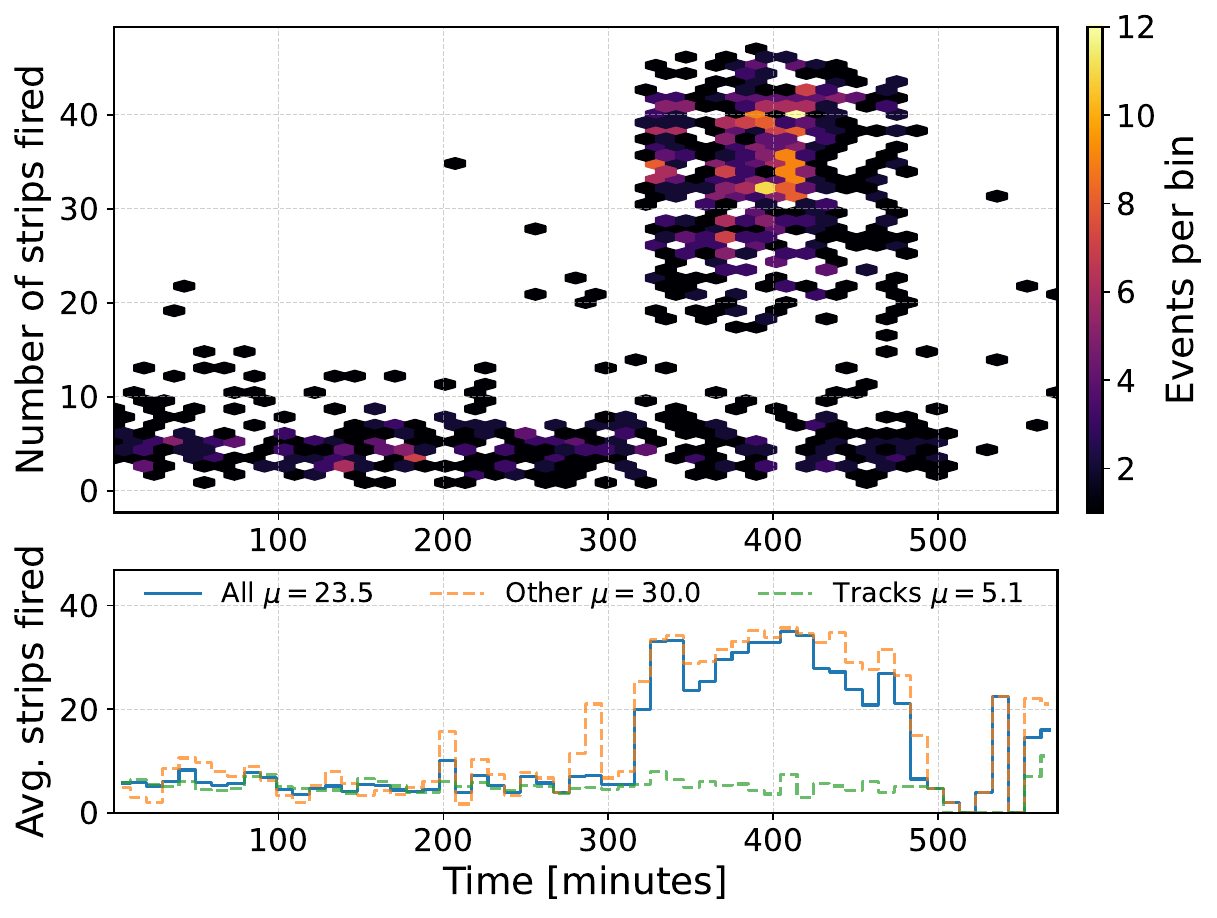}}\\
    \vspace{0.1cm}
    \subfloat[Beam (Pb–PE)]{\includegraphics[width=.48\linewidth]{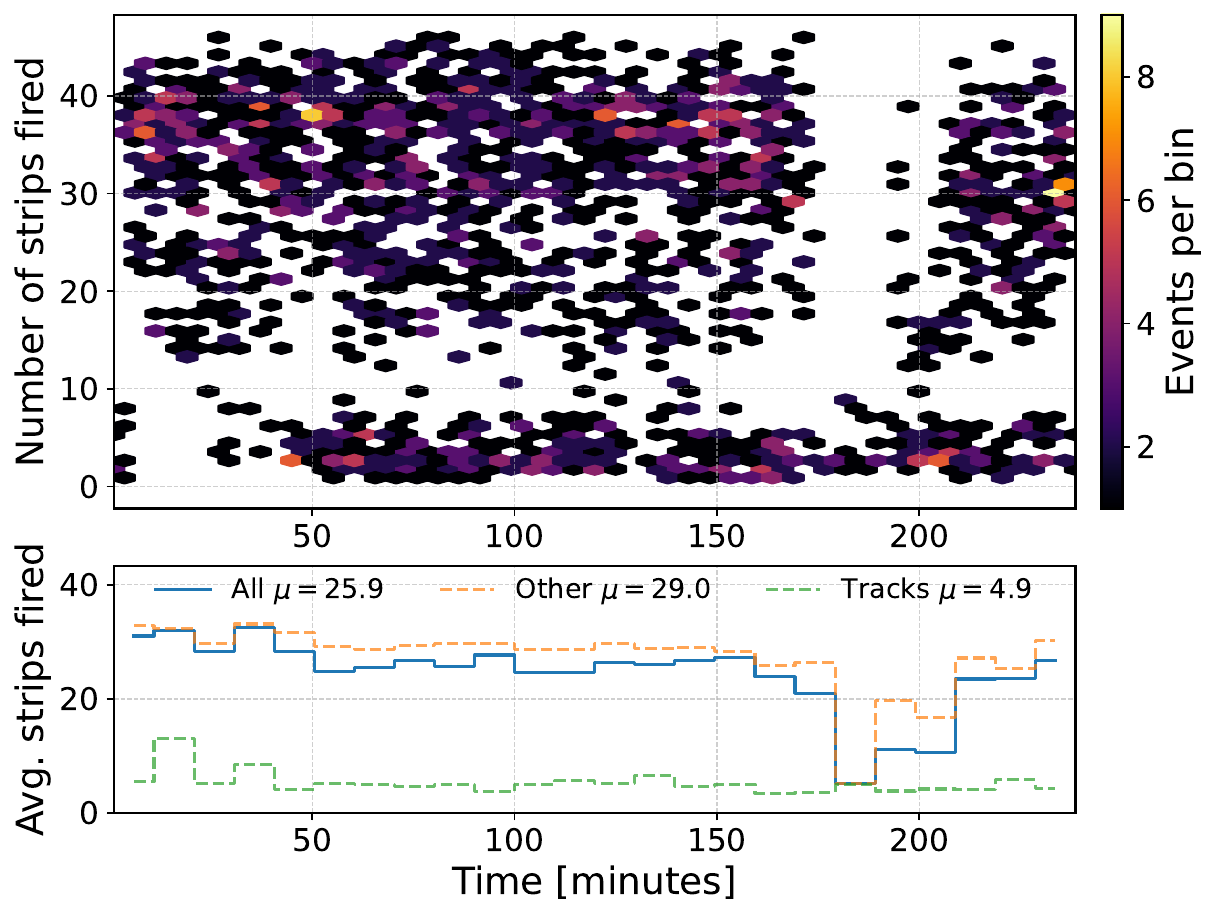}}
    \hfill
    \subfloat[Beam (PE only)]{\includegraphics[width=.48\linewidth]{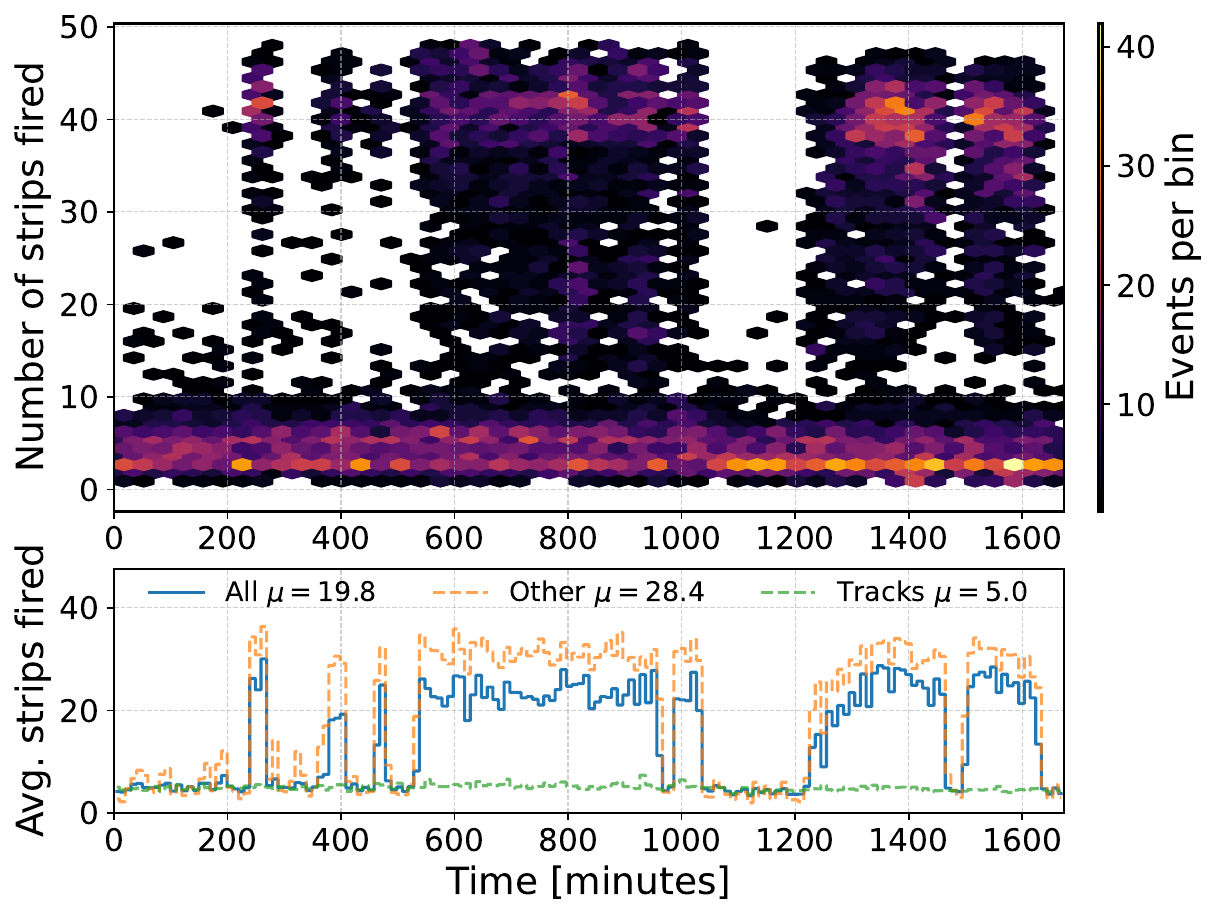}}
    \caption{Number of strips fired per event as a function of external scintillators trigger time for multiple shielding configurations: (\textbf{a}) D1 cosmic, (\textbf{b}) polyethylene facing the beam followed by lead (PE–Pb), (\textbf{c}) lead facing the beam followed by polyethylene (Pb–PE), and (\textbf{d}) configuration consisting solely of polyethylene.}
    \label{fig:D1_timing}
\end{figure}

\begin{figure}[H]
    \centering
    \subfloat[D1 cosmic]{\includegraphics[width=.48\linewidth]{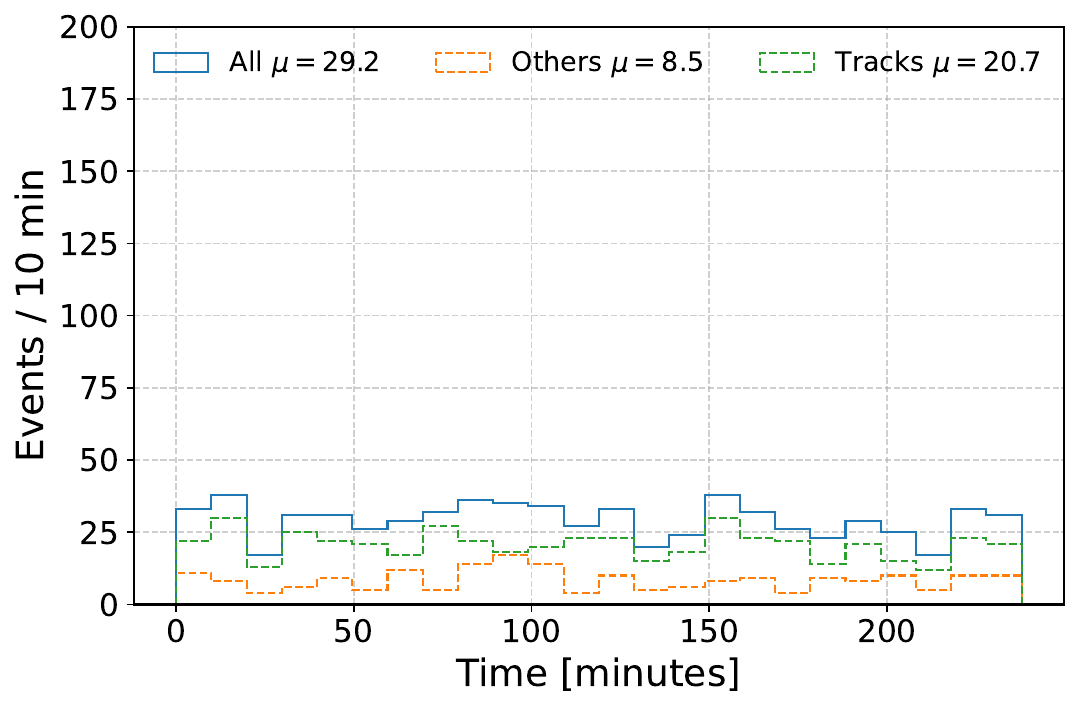}}     
    \hfill
    \subfloat[Beam (PE–Pb)]{\includegraphics[width=.48\linewidth]{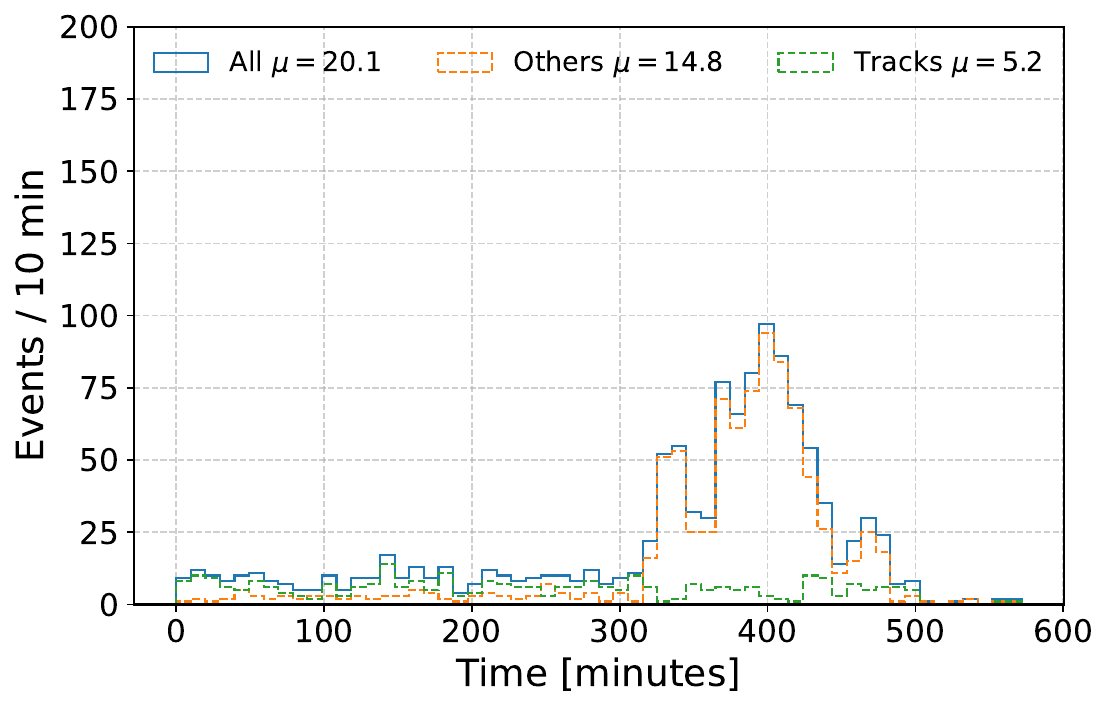}}\\
    \vspace{0.1cm}
    \subfloat[Beam (Pb–PE)]{\includegraphics[width=.48\linewidth]{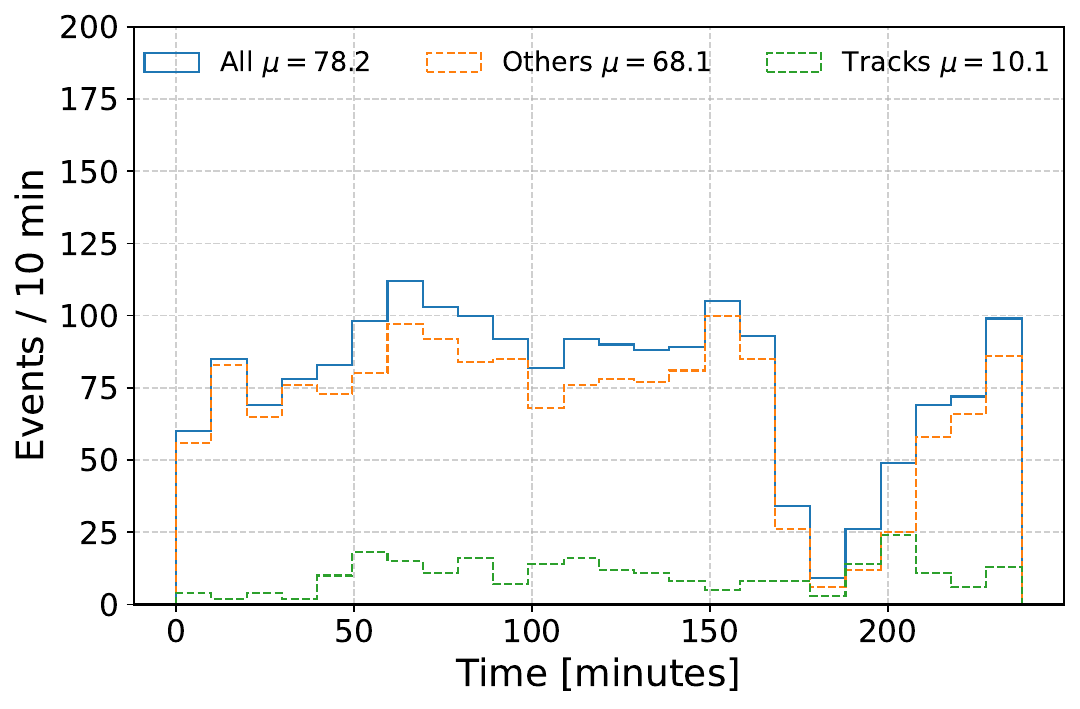}}   
    \hfill
    \subfloat[Beam (PE only)]{\includegraphics[width=.48\linewidth]{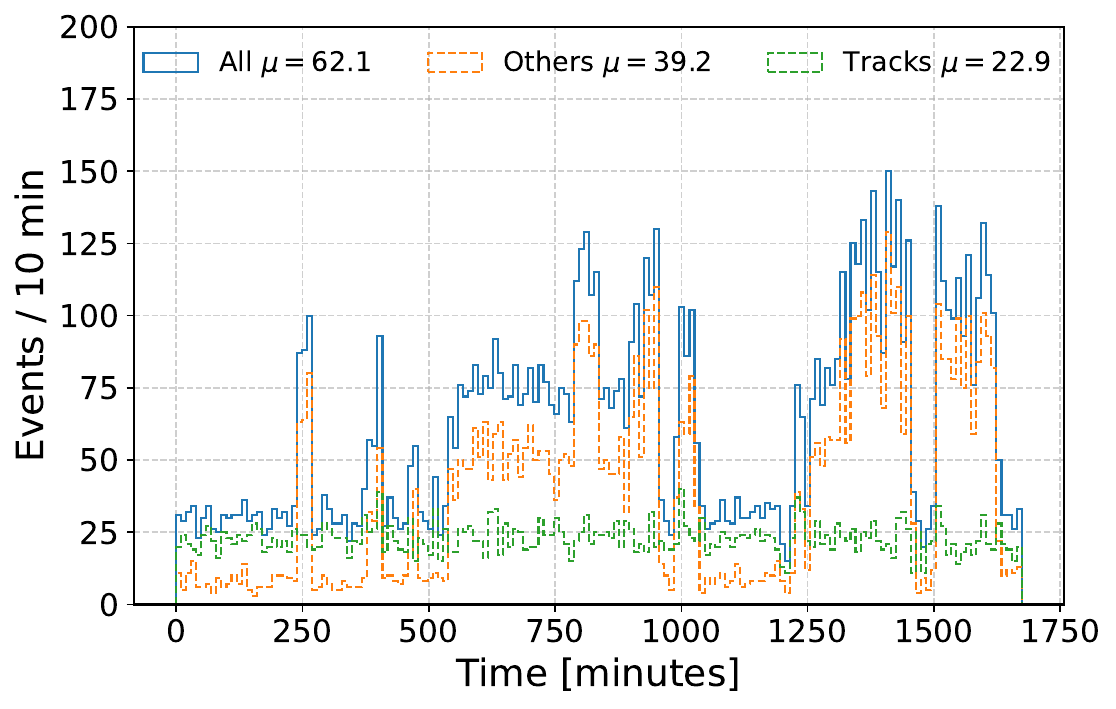}}
    \caption{Event rate as a function of external scintillators trigger time for multiple shielding configurations: (\textbf{a}) D1 cosmic, (\textbf{b}) polyethylene facing the beam followed by lead (PE–Pb), (\textbf{c}) lead facing the beam followed by polyethylene (Pb–PE), and (\textbf{d}) configuration consisting solely of polyethylene.}
    \label{fig:D1_event_rate}
\end{figure}

The external scintillator trigger time is recorded for all data-taking configurations, enabling a study of the temporal evolution of the detected events. Figure~\ref{fig:D1_timing} shows the number of fired strips as a function of trigger time for the four datasets, while Figure~\ref{fig:D1_event_rate} presents the corresponding event rates, distinguishing events that pass the tracking selection from those that do not as describe in section~\ref{sec:tracking}.
The D1 cosmic run exhibits a stable behavior over time, with a nearly constant track rate of approximately 20 events per 10 min and a steady strip multiplicity of about four strips per event. This stability is consistent with expectations for cosmic muons, whose flux at the detector location is effectively constant on the timescale of the measurement. The small fluctuations in the running average are normal statistical variations, while the occasional large spikes in the top panel likely come from unusual events, noise, or occasional broader charge sharing rather than the typical cosmic-track response. In contrast, the D1 beam dataset shows pronounced temporal fluctuations in both event rate and strip multiplicity. Several periods of enhanced activity are visible. During these intervals, both the event rate and the number of fired strips increase significantly, whereas in the intervening periods they return to levels comparable to those observed in the D1 cosmic run. This behavior strongly suggests intermittent beam-induced particle production superimposed on a steady cosmic background.

\subsubsection{Lessons Learnt}

In conclusion, this data-taking campaign gave the first operational experience for our team in a laser-beam laboratory, and made us understand several characteristics of the environment of Petawatt laser facilities such as ELI Beamlines. 
Most notably, we had to quickly learn how to adapt to the electromagnetic environment of the hall, improving the shielding until data were of sufficient quality. To that end, clean control datasets of cosmic muons (during no-beam periods) were crucial. 
Once the detectors gave sufficiently clean signals, the beam-on periods were essential to verify that our detectors are able to resist in the rough electromagnetic and radiation environment generated by the laser and by the beam's collision with the tungsten wall. 
The angular distribution of the reconstructed hits, including those in layer 1, confirms that the detectors were positioned downstream of the beam and were fully illuminated. The observed distribution of $\theta_x$ suggests a significant contribution from combinatorial background, which may obscure a potential signal from beam-induced muons. For this reason, it was decided for the subsequent data-taking campaign to prioritize three-point tracking capability by aligning all detectors with horizontally oriented strips, thereby enabling a more robust track reconstruction and allowing meaningful measurement of fit residuals.

\subsection{D2 and D3 Campaigns}

This section reports on the data acquired during the D2 and D3 data-taking campaign, whose details are presented in Table~\ref{table:summary}. 
During the D1 run, layer 1 was oriented orthogonally to layers 0 and 2. For D2 and D3, all layers were aligned in the same direction to allow for 3-point tracking. 
To ensure comparability across datasets, only data taken without shielding are considered. 
The effect of the different shielding on the detector response is further studied in Section~\ref{sec:shielding}.
The goal of the analysis is to determine whether muons produced by the beam can be detected. To address this question, we compare the strip occupancy, strip multiplicity, angular distributions of the D2 cosmic and D2 and D3 beam datasets (Section~\ref{sec:strip_occu}) to then analyse the track properties (Section~\ref{sec:tracking}). Finally, the timing structure of the raw detector data bring valuable insight on the nature of the recorded signals, and we discuss it in Section~\ref{sec:timing}. 

\begin{table}[H]
    \caption{Summary of the D2 Cosmic, D2 Beam, and D3 Beam data takings. The DAQ threshold settings were validated through a threshold scan performed upon arrival at ELI Beamlines, as described in Appendix \ref{sec:thresh_scan}. \label{table:summary}}
\begin{adjustwidth}{-\extralength}{0cm}
   \begin{tabularx}{\fulllength}{CcCCCCC} 
     \toprule
     \textbf{Data Taking} & \textbf{Shielding} & \textbf{Duration} & \textbf{Threshold [a.u.]} & \boldmath{$N_{\mathrm{events}}$} & \boldmath{$N_{\mathrm{tracks}}$} & \boldmath{$\frac{N_{\mathrm{tracks}}}{N_{\mathrm{events}}}$} \\
     \hline
     D2 Cosmic & None & 6 h & [35, 40] & 933 & 646 & 69.2\% \\ 
     D2 Beam & None & 3 h 45 min & [35, 40] & 878 & 477 & 54.3\% \\
     D2 Beam & Lead + polyethylene & 5 h 30 min & [35, 40] & - & - & — \\
     \hline
     D3 Beam & None & 10 h 30 min & 40 & 2704 & 1115 & 41.2\% \\
     D3 Beam & Lead & 15 h & 40 & 3567 & 1723 & 48.3\% \\
     D3 Beam & Lead + polyethylene & 7 h 30 min & 40 & 1976 & 883 & 40.3\% \\
     \bottomrule
    \end{tabularx}
\end{adjustwidth}
\end{table}

\subsubsection{Strip Occupancy and Multiplicity in D2 and D3}
\label{sec:strip_occu}

\begin{figure}[H]
    \centering
    \includegraphics[width=1.\linewidth]{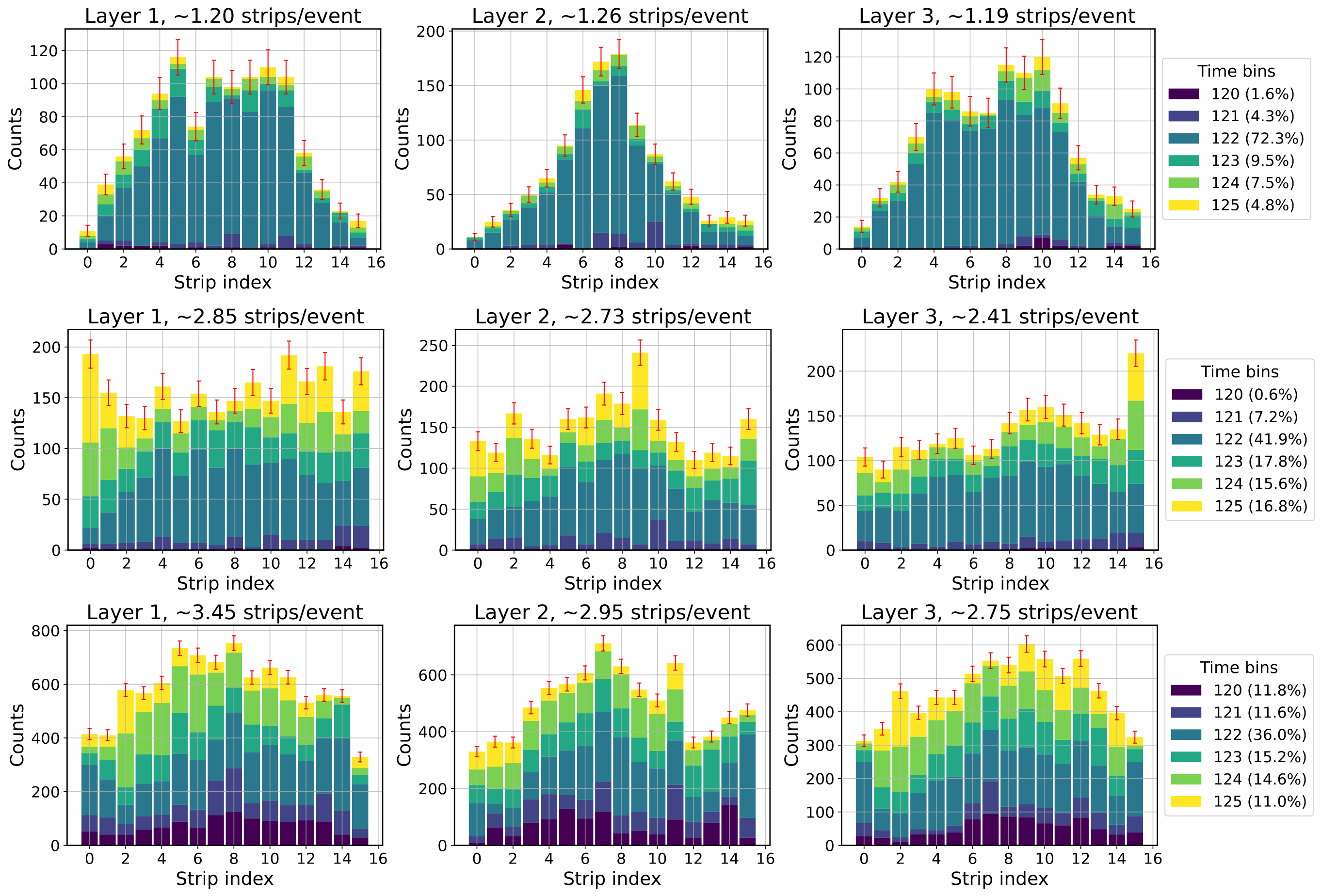}
    \caption{Strip occupancy of each of the three RPC layers, for the D2 cosmic (\textbf{top}), D2 beam (\textbf{middle}), and D3 data takings (\textbf{bottom}). Poissonian uncertainties are shown in red.}
    \label{fig:occupancies}
\end{figure}

Strip occupancy distributions corresponding to the three data takings are shown in Figure~\ref{fig:occupancies}. The D2 cosmic run is used as a baseline for evaluating the detector response to single muons. With the beam off, cosmic muons constitute the dominant component of the charged-particle flux. The cosmic strip-occupancy distribution indicates that the majority of fired strips appear in the 5 ns wide time bin 122 (corresponding to 610 ns after the scintillator trigger). In the central RPC layer, the occupancy exhibits the expected pattern for cosmic muons: an excess of central strips relative to the external strips, consistent with the predominantly non-horizontal zenith-angle distribution of the cosmic flux. The slight asymmetry observed in the first and last layers can be attributed to variations in strip response and environmental non-uniformities (the setup is located underground, and surrounding structures introduce anisotropies in the muon flux). While the D2 cosmic run yields approximately one fired strip per layer per event, the D2 beam dataset reaches up to 2.9 strips in the first layer (closer to the beam) and decreases to 2.4 strips in the third layer. 

Contrary to the cosmic case, the occupancy in the central layer is not consistent with a predominantly non-horizontal flux: external strips are fired as frequently as central strips. In addition, the timing distribution is broader: time bin 122, which accounted for 72\% of entries in the cosmic run, now represents only 42\%. These observations indicate that multiple charged particles enter the detector acceptance simultaneously during the trigger window. For the D3 cosmic run, the occupancy in the central layer is not consistent with a predominantly non-horizontal flux: external strips are fired as frequently as central strips. In addition, the timing distribution is broader: time bin 122, which accounted for 72\% of entries in the cosmic run, now represents only 39\%. These observations lead to the same conclusion as for the D2 beam data taking: multiple charged particles enter the detector acceptance simultaneously during the beam trigger window.\\

\begin{figure}[H]
    \centering
    \includegraphics[width=1.\linewidth]{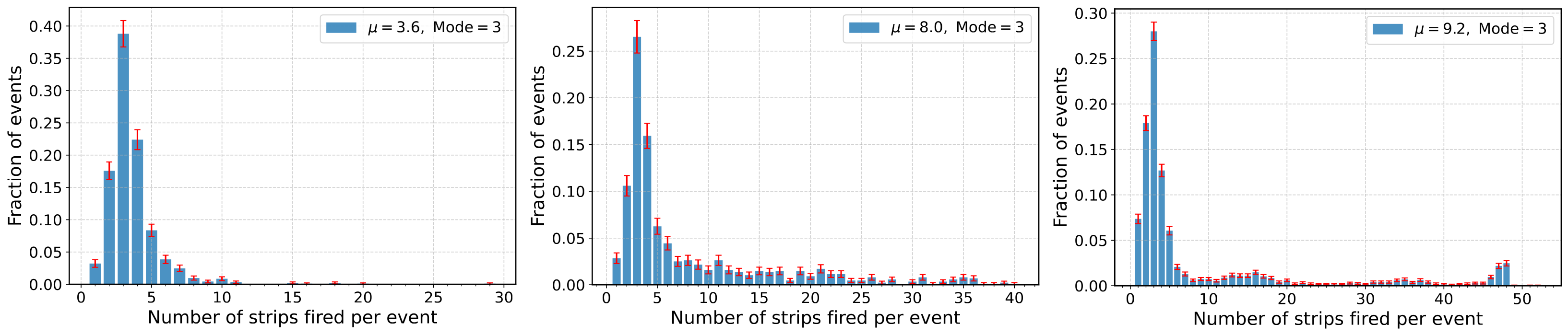}
    \caption{Number of strips fired per event for the D2 cosmic (\textbf{left}), D2 beam (\textbf{middle}) and D3 (\textbf{right}) data takings. The mode and mean $\mu$ of each distribution is presented in the legend. Poissonian uncertainties are shown in red.}
    \label{fig:strip_mult}
\end{figure}

The distributions of the number of strips fired per event shown Figure~\ref{fig:strip_mult}, corroborate the hypothesis of a higher number of charged particle entering the detector acceptance for the D2 beam and D3 beam data takings, for which the average number of strips fired per event respectively reach 8.0 and 9.2 against 3.6 for D2 cosmic. While the distribution means differs significantly, the mode of the distribution remains the same at 3.0 suggesting that cosmic muon contribution is dominant. It is interesting to note the difference between D2 beam and D3 beam, where D2 beam strip multiplicity distribution resembles the D2 cosmic one, but with an elongated tail, while D3 exhibits discontinuity and reaches up to 48 strips fired at a time. These observations indicate that, while cosmic muons remain a significant contribution, the beam-related data takings introduce additional charged-particle fluxes with substantially higher multiplicities. Ongoing simulation studies will help quantify the relative contributions of the different particle sources and further clarify the origin of these high-multiplicity events.

\subsubsection{Tracking}
\label{sec:tracking}

For the analysis of datasets D2 and D3, the event selection and track classification are applied as in D1, as detailed in Section~\ref{sec:event-selection-and-track-id}, with the fundamental difference that now all three detectors were aligned, meaning that we performed a 3-point track reconstruction in a 2D plane through a $\chi^2$ fit, allowing the calculation of hit residuals as a way to assess the quality of track reconstruction.
The $\chi^2$ value is utilized in the Bronze category: only the clusters granting the best fit are retained for further analysis, as shown in the Decision Tree of Figure~\ref{fig:decision_tree}. 

Event display for each of the six event classes are presented in Figure~\ref{fig:event_display_class}. Once fired strips have been selected, their IDs are converted into hit positions using the knowledge of the relative positions of all RPC chambers, as depicted in Figure~\ref{fig:setup}.

The proportion of the tracks corresponding to each class is presented in Table~\ref{table:class} along with the semi-coincidence efficiency $\epsilon_{\mathrm{semi}}$, full coincidence efficiency $\epsilon_{\mathrm{full}}$ and tracking efficiency $\epsilon_{\mathrm{track}}$. $\epsilon_{\mathrm{semi}}$ and $\epsilon_{\mathrm{full}}$ respectively correspond to the fraction of events that leave a hit in two and all chambers, and $\epsilon_{\mathrm{track}}$ correspond to the fraction of event that pass the tracking selection and with a linear track fit residuals below 1 cm. While $\epsilon_{\mathrm{semi}}$ and $\epsilon_{\mathrm{full}}$ have similar value for the three data takings, although a bit higher for D2 beam and D3 beam, the tracking efficiency decreases from 64.4 $\%$ for D2 cosmic to $49.8\%$ and $40.3\%$ for D2 and D3 beam. Such behavior is consistent with the observation from the previous section: because of the significantly higher number of strips fired per event, most of them end up being rejected during the tracking selection. 

\begin{table}[H]
    \caption{Proportion of tracks corresponding to each class, from highest (diamond) to lowest purity (bronze). Events classified as 'Other' are those that do not satisfy tracking requirements.}
    \label{table:class}
\begin{adjustwidth}{-\extralength}{0cm}
   \begin{tabularx}{\fulllength}{cCCCCCCCCC} 
     \toprule
     \textbf{Data Taking} & \textbf{Diamond} & \textbf{Platinum} & \textbf{Gold}  & \textbf{Silver} & \textbf{Bronze} & \textbf{Other} & \boldmath{$\epsilon_{\mathrm{semi}}$} & \boldmath{$\epsilon_{\mathrm{full}}$} & \boldmath{$\epsilon_{\mathrm{track}}$}\\ 
     \hline
     D2 Cosmic & 25.7\% & 7.6\% & 4.6\% & 11.9\% & 19.4\% & 30.8\% & 95.3\% & 70.4\% & 64.6\% \\ 
     D2 Beam   & 15.8\% & 4.3\% & 4.0\% & 8.8\%  & 21.4\% & 45.7\% & 95.7\% & 76.5\% & 49.8\% \\ 
     D3 Beam   & 14.0\% & 6.1\% & 2.0\% & 5.9\%  & 16.8\% & 55.3\% & 95.0\% & 73.4\% & 40.3\% \\ 
     \bottomrule
    \end{tabularx}
\end{adjustwidth}
\end{table}

\begin{figure}[H]
    \centering
    \includegraphics[width=\linewidth]{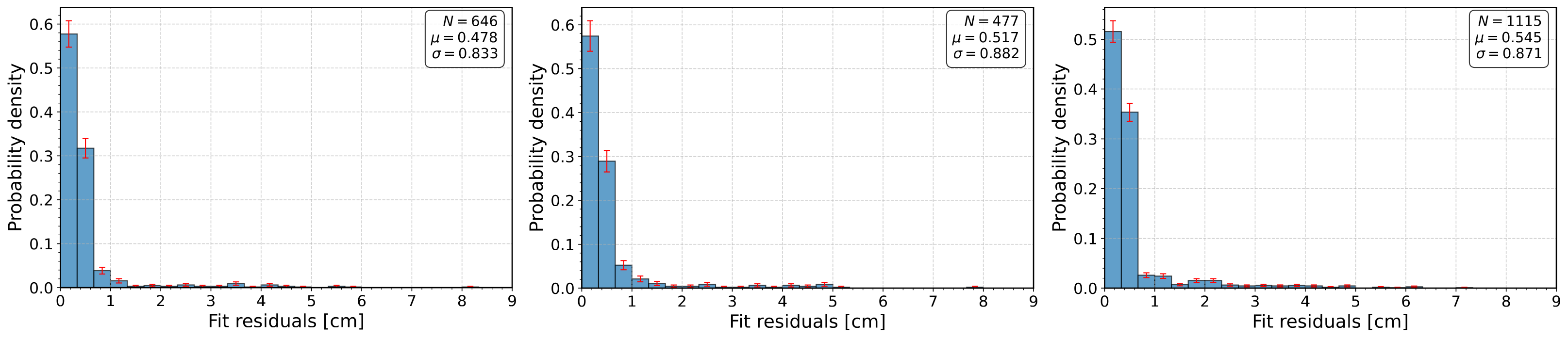}
    \caption{Linear fit residuals distribution for the D2 cosmic (\textbf{left}), D2 beam (\textbf{center}) and D3 beam (\textbf{right}) data takings. Poissonian uncertainties are shown in red.}
    \label{fig:res}
\end{figure}

The linear track fit residual distributions of the D2 and D3 beam data takings shown in Figure \ref{fig:res} have similar shape, mean $\mu\approx0.5$ cm and standard deviation $\sigma \approx 0.8$ as the D2 cosmic data taking. It confirms that the tracking procedure can effectively reconstruct muon tracks even in noisy environment. To ensure a good track quality, events with residuals above 1 cm are filtered out.

\begin{figure}[H]
    \centering
    \includegraphics[width=\linewidth]{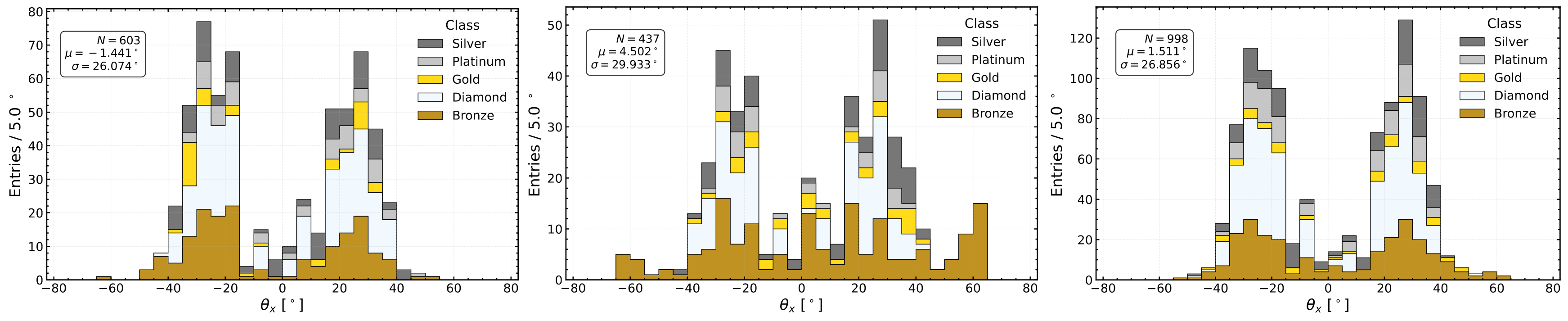}
    \caption{Distribution of angles with respect to the horizontal plane $\theta_x$ (horizontal muons have $\theta_x = 0$, and vertical muons have $\theta_x = 90^\circ$), for the D2 cosmic (\textbf{left}), D2 beam (\textbf{center}) and D3 beam (\textbf{right}) data takings.}
    \label{fig:zenith}
\end{figure}

\begin{figure}[H]
    \centering
    \includegraphics[width=0.49\linewidth]{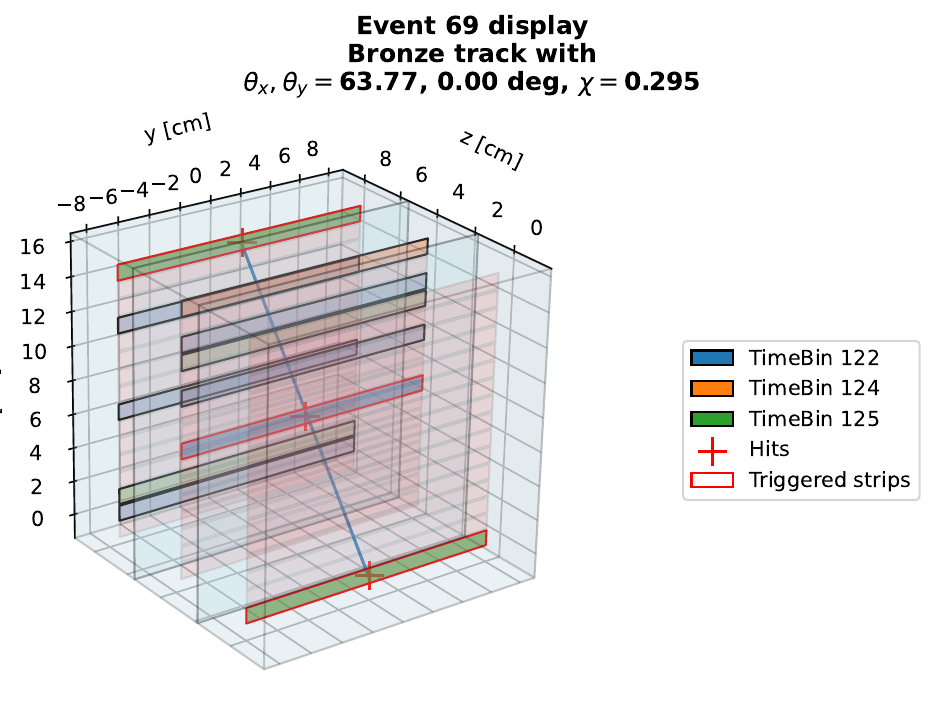}
    \includegraphics[width=0.49\linewidth]{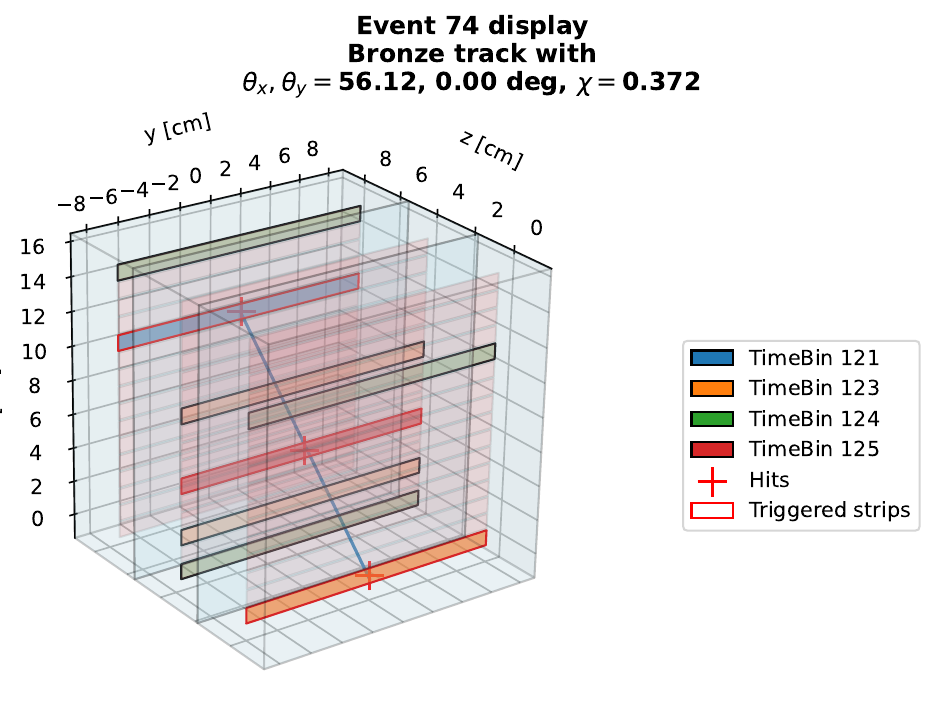}
    \caption{Example of low residual tracks obtained from combinatorial noise in the detector.}
    \label{fig:beam_low_res_bad_events}
\end{figure}

The resulting angular distribution are presented in Figure~\ref{fig:zenith} also agrees with expectations for cosmic muons: only a small subset of reconstructed tracks exhibits near-horizontal directions. However, low residuals are not sufficient to guarantee a physically meaningful track, particularly for Bronze-category tracks. As illustrated in Figure \ref{fig:beam_low_res_bad_events}, random combinations of fired strips can occasionally produce an apparently good linear fit. This mis-reconstruction inflates the number of Bronze events with large reconstructed angles relative to the horizontal in the D2 beam data taking. For D3 data taking, angular distribution resembles the D2 cosmic. Overall, the tracking results do not conclusively confirm the presence of beam-induced muons, which are expected to exhibit predominantly horizontal trajectories with $\theta_x \in [-5^\circ, 5^\circ]$. Tracks that pass the tracking exhibit the same angular distribution as for the D2 \textit{cosmic} data taking, which suggest that only cosmic muon tracks are properly reconstructed.

\subsubsection{Upper Limit on Beam-Induced Muon Flux}
\label{sec:upper-limit}

Statistical and systematic limitations prevent claiming that we are observing horizontal muons induced by the presence of the beam. Therefore, in this subsection we proceed with setting an upper limit on the horizontal muon flux, based on the data shown in Figure~\ref{fig:zenith}.

In Table~\ref{tab:signal_region_summary}, we count the number of tracks in the Signal Region (SR), defined as $\theta_x \in [-5^\circ, +5^\circ]$. Normalizing this count to the duration of each run, it can be noted that the SR rate is very similar for the two datasets where no beam-induced signal is expected (D2 Cosmic at $3.00$~tracks/h) or it is expected to be negligible (D3 Beam at $3.14$~tracks/h), with a mutual compatibility of $p = 0.88$, while it is more than double for D2 Beam ($7.73$~tracks/h).

\begin{table}[H]
\caption{Summary of reconstructed track counts and event rates in the signal region SR, defined as $\theta_x \in [-5^\circ, +5^\circ]$, across the D2 and D3 datasets.}
\label{tab:signal_region_summary}
\begin{tabularx}{\textwidth}{lcccccc}
\toprule
\textbf{Dataset} & \textbf{Beam State} & \textbf{Duration} & \textbf{Total Tracks} & \textbf{Tracks in SR} & \textbf{SR Rate} \\
 & & \textbf{[h]} &  &  & \textbf{[tracks/h]} \\
\hline
D2 Cosmic & Off & 6.00 & 646 & 18 & 3.00 \\
D2 Beam   & On (Multi-GeV) & 3.75 & 477 & 29 & 7.73 \\
D3 Beam   & On ($\le$1~GeV) & 10.50 & 1115 & 33 & 3.14 \\
\bottomrule
\end{tabularx}
\end{table}

To assess whether this increase in D2 Beam is statistically compatible with background fluctuations, we perform an exact two-sample Poisson test (Przyborowski--Wilenski test). Comparing D2 Beam ($29$~tracks in $3.75$~h) to D2 Cosmic ($18$~tracks in $6.00$~h) yields a one-sided $p$-value of $p = 8.5 \times 10^{-4}$ (equivalent to a $3.1\sigma$ excess). A similar test against the low-energy D3 Beam control dataset ($33$~tracks in $10.50$~h) yields $p = 3.5 \times 10^{-4}$ ($3.4\sigma$).

However, despite this statistically significant rate enhancement, this excess cannot be claimed as an observation of laser-induced muons. A large fraction of the tracks reconstructed in the horizontal direction during D2 Beam belong to the lowest-purity Bronze category (as seen in Figure~\ref{fig:zenith}), where high detector occupancy from beam-induced secondary showers creates combinatorial track artifacts. Therefore, we proceed with setting an upper limit on the horizontal muon flux.

To place this measurement on a solid quantitative footing, we extract a 95\% confidence level (CL) upper limit on the beam-induced muon yield using standard Poisson statistics with known background~\cite{FeldmanCousins1998}. Taking the expected background baseline rate as 3~tracks/h (as measured in both D2 Cosmic and D3 Beam) to estimate the expected background count during the $3.75$~h D2 Beam run ($b = 11.25$~tracks), and comparing it to the $n = 29$ observed tracks, the one-sided 95\% CL upper limit on the signal mean $S_{\text{up}}^{95\%}$ is determined from:
\begin{equation}
\sum_{k=0}^{n} \frac{(S_{\text{up}}^{95\%} + b)^k e^{-(S_{\text{up}}^{95\%} + b)}}{k!} = 0.05.
\end{equation}

Using the exact relationship with the chi-square distribution, $S_{\text{up}}^{95\%} + b = \frac{1}{2} F_{\chi^2}^{-1}(0.95; 2(n+1))$ with $2(n+1) = 60$ degrees of freedom, the total Poisson upper bound is $39.54$~tracks, yielding an upper limit on excess beam-induced signal tracks of $S_{\text{up}}^{95\%} = 39.54 - 11.25 = 28.3$~tracks (or $27.2$~tracks using the unified Feldman--Cousins ordering principle~\cite{FeldmanCousins1998}).

Accounting for the active detector cross section ($16 \times 16$~cm$^2$), a tracking reconstruction efficiency $\epsilon_{\text{track}} \approx 49.8\%$ (from Table~\ref{table:class}), and the total estimated shot count at the nominal $0.2$~Hz repetition rate ($N_{\text{shots}} \approx 2700$), this corresponds to an experimental upper bound on the horizontal muon fluence at $18$~m from the beam dump of:
\begin{equation}
\Phi_{\mu~\text{per shot}}^{\text{95\% CL}} \le 7.9 \times 10^{-5} \text{ cm}^{-2} \text{ per shot.}
\end{equation}

Because the RPC response does not distinguish between minimum ionizing electrons and muons, attributing all reconstructed tracks to muons provides a conservative upper bound.

Crucially, because the parasitic setup did not permit a simultaneous shot-to-shot measurement of the primary electron spectrum and charge, translating this limit to a per-electron basis carries intrinsic systematic uncertainty. Assuming a nominal bunch charge of $\sim$100~pC ($\sim$6.2 $\times 10^8$ primary electrons per shot) representative of typical high-energy operation at ELBA, this yields:
\begin{equation}
\Phi_{\mu~\text{per } e^-}^{\text{95\% CL}} \le 1.3 \times 10^{-13} \text{ cm}^{-2} \text{ per primary electron.}
\end{equation}

To compare this experimental limit with the FLUKA Monte Carlo simulations presented in Section~\ref{sec:beam-conditions}, one needs to account for the geometric attenuation between the simulation reference planes and the detector location. Assuming geometric dilution from an effective point-like interaction region in the beam dump (a behavior verified by the FLUKA fluence scaling between the $0.5$~m and $5$~m simulation planes), moving to the detector location at $18$~m introduces a geometric attenuation factor of $(5\text{ m} / 18\text{ m})^2 \approx 0.077$, corresponding to a $\approx 13$-fold reduction in fluence. Scaling the simulated high-energy muon fluence at $5$~m ($\sim 10^{-10}$--$10^{-9}$~cm$^{-2}/e^-$ for $E_\mu > 1$~GeV) by this factor yields an expected on-axis fluence at $18$~m of approximately $(0.8$--$7.7) \times 10^{-11}$~cm$^{-2}/e^-$.

Furthermore, during the parasitic run, the effective charge delivered to the dump was reduced by an additional factor of $2$--$20$ due to the spectrometer collimating slit sampling only $5$--$50$\% of the total beam (Section~\ref{sec:beam-conditions}), alongside shot-to-shot alignment jitter. More discussion about these systematic uncertainties can be found in Section~\ref{sec:systematics}. When combined with the high combinatorial background from secondary electromagnetic showers, these operational constraints explain why the true beam-induced muon signal remained well below the single-shot discovery threshold.

\subsubsection{Timing Information in D2 and D3}
\label{sec:timing}

\begin{figure}[H]
    \centering
    \includegraphics[width=\linewidth]{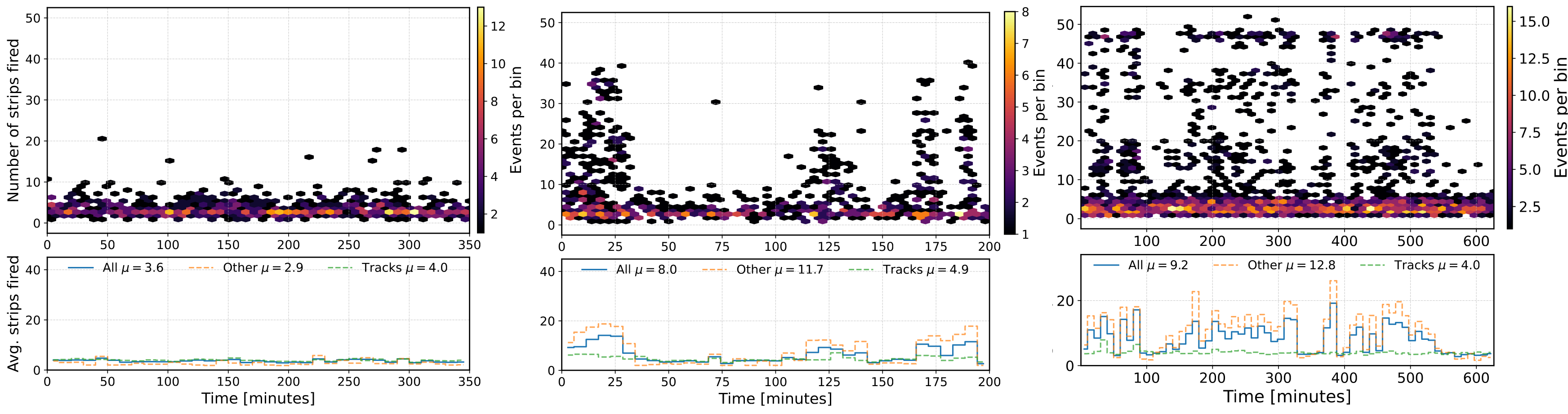}
    \caption{Number of strips fired per event as a function of external scintillators trigger time for the D2 cosmic (\textbf{left}), D2 beam (\textbf{center}) and D3 beam (\textbf{right}) datasets.}
    \label{fig:strips_time}
\end{figure}

\begin{figure}[H]
    \centering
    \includegraphics[width=\linewidth]{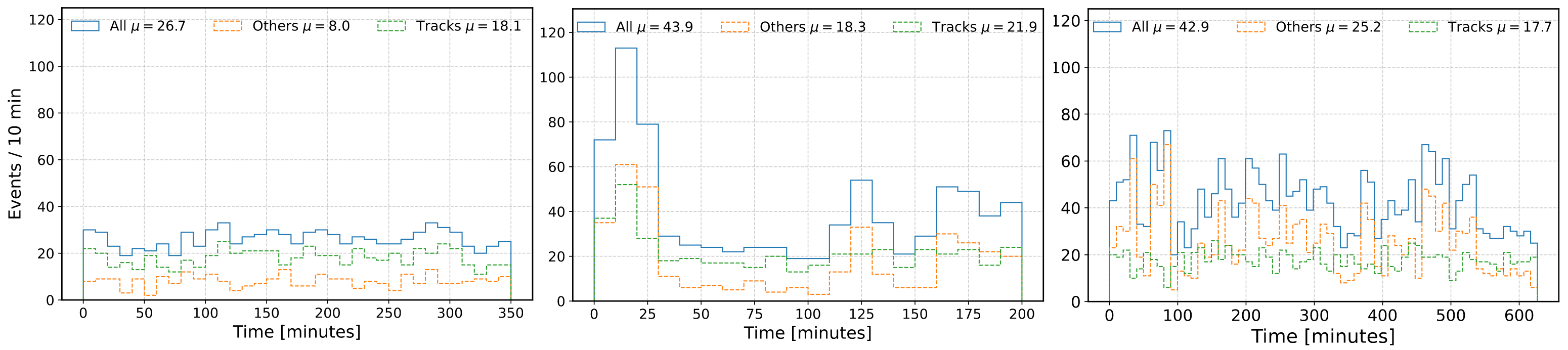}
    \caption{Event rate as a function of external scintillators trigger time for the D2 cosmic (\textbf{left}), D2 beam (\textbf{center}) and D3 beam (\textbf{right}) datasets.}
    \label{fig:event_rate}
\end{figure}

For each data-taking configuration, the time stamp of the external scintillator trigger is recorded, allowing us to study the temporal structure of the detected events. Figure~\ref{fig:strips_time} shows the number of fired strips as a function of trigger time for the three datasets, while Figure~\ref{fig:event_rate} presents the corresponding event rates, distinguishing events that pass the tracking selection from those that do not.

The D2 cosmic run exhibits a stable behavior over time, with a nearly constant track rate of approximately 18 events per 10 min and a steady strip multiplicity of about four strips per event. This stability is consistent with expectations for cosmic muons, whose flux at the detector location is effectively constant on the timescale of the measurement. In contrast, the D2 beam dataset shows pronounced temporal fluctuations in both event rate and strip multiplicity. Several periods of enhanced activity are visible, notably between 0–35 min, 100–140 min, and after 150 min. During these intervals, both the event rate and the number of fired strips increase significantly, whereas in the intervening periods they return to levels comparable to those observed in the D2 cosmic run. This behavior strongly suggests intermittent beam-induced particle production superimposed on a steady cosmic background. The D3 beam dataset exhibits a distinct timing profile. While the event rate also shows temporal discontinuities, the strip multiplicity reaches significantly higher values, up to a maximum of 48 strips, corresponding to the simultaneous firing of all strips in the detector. This indicates a more intense flux of charged secondary particles entering the detector acceptance, that could be consistent with a beam configuration that produces a higher particle multiplicity. Despite these differences, a notable feature common to all datasets is that events passing the tracking selection have a nearly constant strip multiplicity of about four, and a track rate close to 20 events per 10 min. 
This is consistent with the track sample being dominated by cosmic muons, which arrive continuously and at low multiplicity regardless of beam conditions. While this demonstrates that the tracking selection successfully rejects high-multiplicity beam-induced events, it is not sufficient to establish that beam-induced single-particle signatures are recovered.

\begin{figure}[H]
    \centering
    \includegraphics[width=1\linewidth]{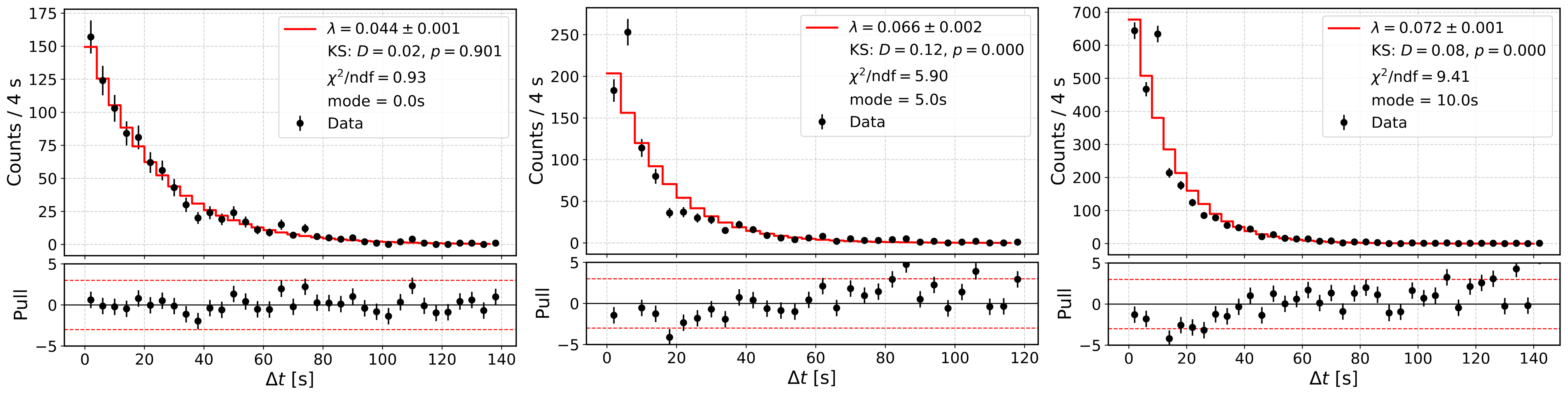}
    \caption{Fit of the time interval distribution $\Delta t$ for the D2 cosmic (\textbf{left}), D2 beam (\textbf{center}) and D3 beam (\textbf{right}) with an exponential distribution with unique parameter $\lambda$. The the unbinned $\Delta t$ mode, the fit's $\chi^2 / \mathrm{ndf}$, and the Kolmogorov–Smirnov statistic $D$ and its associated $p$-value are given in the legend. While the $\Delta t$ of the D2 comsic is fully compatible with the exponential model, D2 beam and D3 beam respectively exhibit deviations above 5$\sigma$ at 5 s and 10 s respectively.}
    \label{fig:dt}
\end{figure}

Given that the cosmic muon flux at sea level is approximately constant over time and that individual muon arrivals are statistically independent, the detection of cosmic muons can be modeled as a Poisson process. Under the additional assumption that the probability of detecting two muons within a time interval shorter than the detector dead time is negligible, the time interval $dt$ between successive events follows an exponential probability density function:

\begin{equation}
    f(dt; \lambda) = \lambda e^{-\lambda dt}
\end{equation}
with $\lambda$ the rate parameter. For each data taking, we compute the time interval and fit its distribution with $f(dt, \hat{\lambda})$ with $\hat{\lambda} = 1 / \langle dt\rangle$ where $\langle dt\rangle$ is the mean of the $dt$ distribution. Using this model as a reference, we fit the time interval distribution of the three data takings. The goodness of fit is assessed using the Kolmogorov–Smirnov (KS) test and the $\chi^2/\mathrm{ndf}$. We first consider the full event samples, irrespective of whether the tracking selection is satisfied. The fit results, shown in Figure~\ref{fig:dt}, indicate that the D2 cosmic data are well described by a Poisson process, with a fitted rate $\lambda_\mathrm{fit, cosmic}=0.044\pm0.001 \:\mathrm{s}^{-1}$ a KS distance $D = 0.02$, and the associated $p$-value of 0.9. This behavior is expected for an uncorrelated cosmic muon flux. In contrast, the D2 and D3 beam datasets deviate significantly from this behavior. In both cases, the time-interval distributions exhibit pronounced modes at 5.0 s and 10.0 s, respectively, deviating from the model fit by more than $5\sigma$. The KS test yields vanishing $p$-values for both data sets, and the large $\chi^2/\mathrm{ndf}$ values of 5.90 (D2 beam) and 9.41 (D3 beam) make the exponential model incompatible with the data. These deviations indicate a breakdown of the Poisson assumptions underlying the model. In particular, the observed structures in the time-interval distributions demonstrate that successive events are no longer statistically independent. Instead, events tend to cluster in time, reflecting temporal correlations characteristic to burst-like beam-induced particle production rather than random memory-less arrivals.

\begin{figure}[H]
    \centering
    \includegraphics[width=1\linewidth]{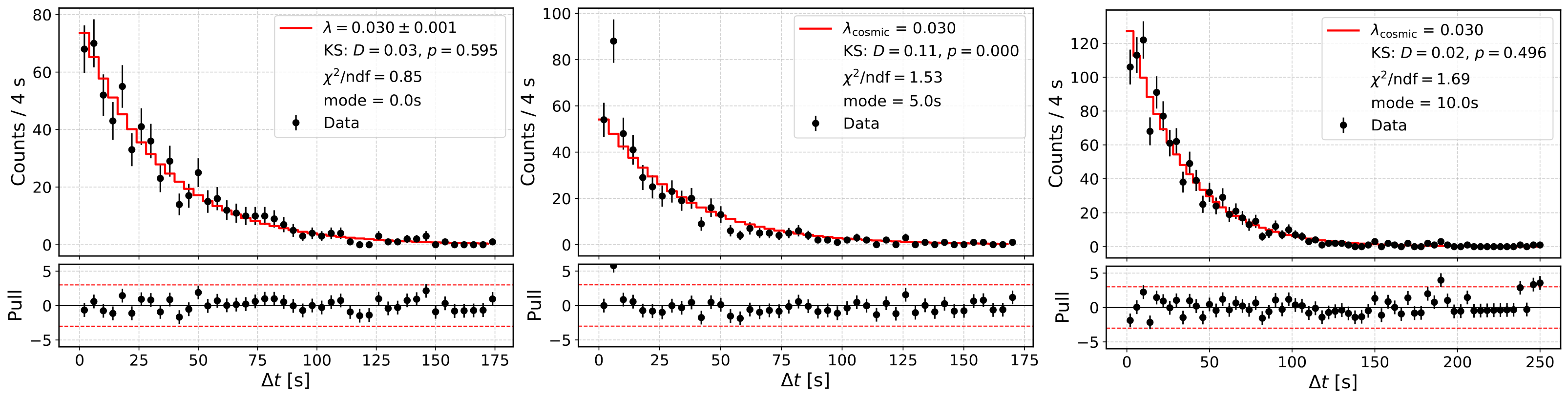}
    \caption{Fit of the time interval distribution $\Delta t$ for the D2 cosmic for events passing the tracking selection (\textbf{left}). The fitted exponential with $\lambda_{\mathrm{cosmic}}$ is reported on the time interval distribution of the D2 beam (\textbf{center}) and D3 beam datasets (\textbf{right}). The unbinned $\Delta t$ mode, the fit's $\chi^2 / \mathrm{ndf}$, and the Kolmogorov–Smirnov statistic $D$ and its associated $p$-value are given in the legend.}
    \label{fig:dt_tracks_only}
\end{figure}

To further isolate this effect, we repeat the analysis using only events that pass the tracking selection, which are compatible with single-particle signatures and for which contamination from secondaries and background is strongly suppressed. Fitting the D2 cosmic track sample with the exponential model yields a rate $\lambda_{\mathrm{tracks}}=0.030\pm0.001$, with a $\chi^2/\mathrm{ndf}=0.85$ and KS distance $D=0.03$ and $p$-value of 0.595, again fully consistent with the Poisson hypothesis, as shown in Figure~\ref{fig:dt_tracks_only}. Using this reference model, we compare the track-only time-interval distributions from the D2 and D3 beam datasets. In both cases, residual excesses are observed at characteristic time intervals of 5.0 s (D2 beam) and 10.0 s (D3 beam), with significances of approximately 5$\sigma$ and 3$\sigma$, respectively. Although the overall $\chi^2/\mathrm{ndf}$ values are closer to unity, indicating reasonable global agreement with the exponential model, the presence of localized peaks away from $dt=0$ reveals a violation of the Poisson hypothesis. Specifically, the assumption of independent event arrivals is broken, even for track-selected events, pointing to residual temporal structuring induced by the beam conditions.

\begin{figure}[H]
    \centering
    \includegraphics[width=\linewidth]{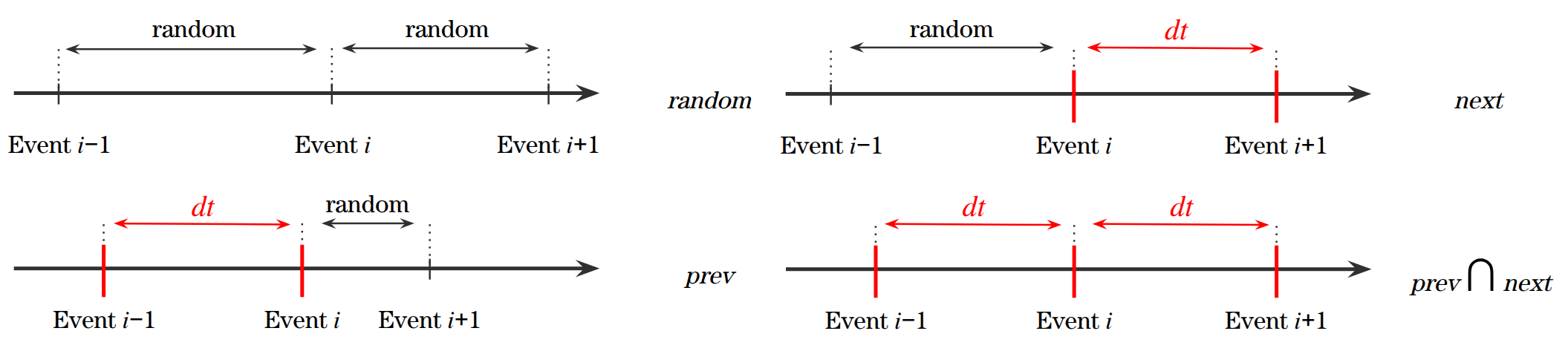}
    \caption{Classes of events based on the inter-arrival time with the previous or following event.}
    \label{fig:interarrival_sketch}
\end{figure}

The presence of localized excesses at well-defined time intervals suggests that these characteristic delays can be used to further classify events according to their temporal correlation with neighboring events. In particular, the D2 and D3 beam datasets exhibit sharply defined inter-arrival times centered at 5~s and 10~s, respectively. These intervals are consistent with the nominal repetition rates of the ELBA and GAMMATRON lasers during these campaigns, directly linking the temporal structure to laser operation. Using these characteristic intervals, each event can be categorized based on the time difference with the immediately preceding and following events. Four mutually exclusive temporal categories are defined, as illustrated in Figure~\ref{fig:interarrival_sketch}:

\begin{itemize}
    \item Random: the time intervals to both the previous and next events fall outside the characteristic interval, indicating no temporal correlation.
    \item Previous: the time interval to the previous event lies within the characteristic interval, while the interval to the next event does not.
    \item Next: the time interval to the next event lies within the characteristic interval, while the interval to the previous event does not.
    \item Previous $\cap$ Next: the time intervals to both the previous and next events lie within the characteristic interval, indicating that the event is part of a temporally correlated sequence.
\end{itemize}

Additionally, the union category Previous $\cup$ Next includes all events for which at least one of the two adjacent inter-arrival times falls within the characteristic interval, thereby providing a measure of the overall fraction of temporally correlated events. Although the finite bin width used in the time-interval histograms limits the visible resolution, the characteristic inter-arrival times are in fact extremely precise. By selecting narrow windows of $dt_{D2} = [4.9999,\,5.0001]\,$s for the D2 beam dataset and $dt_{D3} = [9.9999,\,10.0001]\,$s for the D3 beam dataset, corresponding to a tolerance of $\pm 100~\upmu$s, a significant fraction of events are found to exhibit temporal correlations. The ratio of events in the Previous $\cup$ Next category relative to the Random category is measured to be 29.5\% for the D2 beam dataset and 19.2\% for the D3 beam dataset. These results demonstrate that a substantial fraction of beam-associated events occur at highly regular and reproducible time intervals, reflecting an underlying periodic or structured beam-related process. This temporal classification therefore provides a powerful tool to distinguish temporally correlated beam-induced activity from the uncorrelated cosmic background, which follows Poisson statistics and does not exhibit such preferred time intervals.

For each time class, the total number of events and the number of events satisfying the tracking selection are reported in Table~\ref{tab:time_class}. As expected, the Random category is largely dominated by cosmic muons, which typically produce low-multiplicity signatures involving only a few fired strips. These events are therefore highly likely to satisfy the tracking criteria, resulting in a large fraction of track-selected events within this class. A striking asymmetry is observed between the Prev and Next categories. In both the D2 and D3 beam datasets, events in the Prev category are significantly less likely to satisfy the tracking selection than those in the Next category. For example, in the D2 beam dataset, only 29.2\% of Prev events pass the tracking selection, compared to 59.7\% of Next events. A similar trend is observed in the D3 beam dataset, where the tracking efficiency is 6.8\% for Prev events and 27.7\% for Next events. Naively, one would expect these two categories to exhibit similar detector signatures, as both are beam-induced. The observed difference therefore points to an underlying asymmetry in the detector response or in the temporal structure of the beam itself.

\begin{table}[H] 
    \caption{Number of events corresponding to each time class, and number of events passing the tracking requirements for each time class. \label{tab:time_class}}
   \begin{tabularx}{\textwidth}{CCCCCC}
     \toprule
     \textbf{Data Taking} & \textbf{Time Class} & \boldmath{$N_{\mathrm{class}}$} & \boldmath{$N_{\mathrm{tracks}\,\cap\,\mathrm{class}}$} & \boldmath{$\frac{N_{\mathrm{tracks}\,\cap\,\mathrm{class}}}{N_{\mathrm{tracks}}}$} & \boldmath{$\frac{N_{\mathrm{tracks}\,\cap\,\mathrm{class}}}{N_{\mathrm{class}}}$} \\ 
     \hline
     \multirow{5}{*}{D2 beam} 
      & Random & 678 & 395 & 82.8\% & 58.3\% \\
      & Prev $\cup$ Next & 200 & 82 & 17.2\% & 41.0\% \\ 
      & Prev & 72 & 21 & 4.4\% & 29.2\% \\ 
      & Next & 72 & 43 & 9.0\% & 59.7\% \\ 
      & Prev $\cap$ Next & 56 & 18 & 3.77\% & 32.1\% \\ 
       \hline
     \multirow{5}{*}{D3 beam} 
      & Random & 2269 & 1054 & 94.5\% & 46.52\% \\
      & Prev $\cup$ Next & 435 & 61 & 5.5\% & 14.0\% \\
      & Prev & 148 & 10 & 0.9\% & 6.8\% \\ 
      & Next & 148 & 41 & 3.7\% & 27.7\% \\ 
      & Prev $\cap$ Next & 139 & 10 & 0.9\% & 7.2\% \\ 
     \bottomrule
    \end{tabularx}
\end{table}

\begin{figure}[H]
    \centering
    \includegraphics[width=\linewidth]{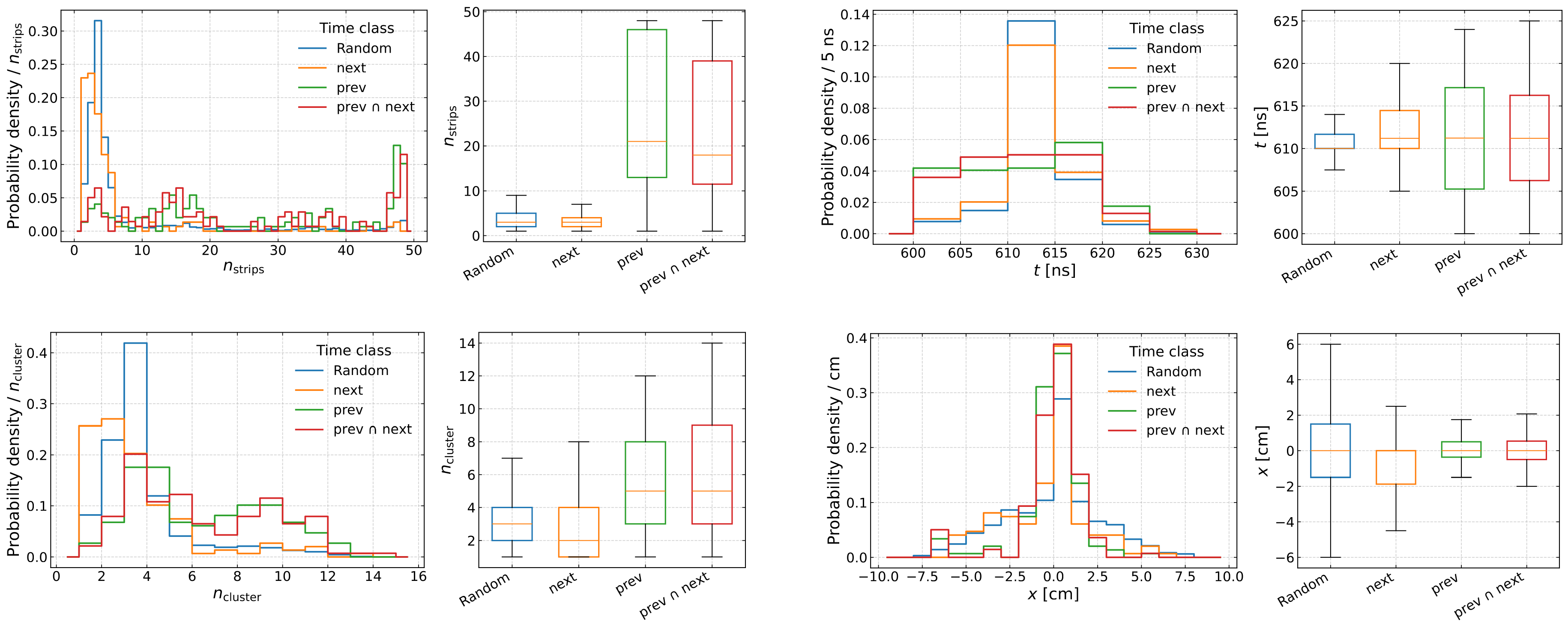}
    \caption{Distribution of raw detector variables from the D3 beam data for 4 different classes of events, based on the characteristic inter-arrival time $dt_{D3} = [9.9999,\,10.0001]$: total number of strips fired (\textbf{top left}), total number of clusters (\textbf{bottom left}), the average timing are which strips are fired with respect to the external scintillator trigger (\textbf{top right}), and the average $x$ position of fired strips (\textbf{bottom right}). Apart from the $x$ position, the Random and Next, and the Prev and Prev $\cap$ Next categories appear to have the same distribution.}
    \label{fig:time_class_d3}
\end{figure}

To further investigate this effect, we examine the distributions of several raw detector observables for each time class, including the number of fired strips, the number of clusters, the average strip timing relative to the external scintillator trigger, and the spatial distribution of fired strips. The results for the D2 and D3 beam datasets are shown in Figures~\ref{fig:time_class_d2} and~\ref{fig:time_class_d3}, respectively. These distributions reveal clear and systematic differences between the Prev and Next categories. The Next category closely resembles the Random category, with both exhibiting low-multiplicity signatures characterized by approximately four fired strips and clusters per event, consistent with single-particle tracks. In contrast, the Prev and Prev $\cap$ Next categories are dominated by high-multiplicity events, with significantly larger numbers of fired strips and clusters. These events also exhibit distinct timing characteristics, reflected in the distribution of the average strip time relative to the trigger, indicating a different temporal development of the detector response.

\begin{figure}[H]
    \centering
    \includegraphics[width=\linewidth]{figs/timing/5s_time_class_comparison.pdf}
    \caption{Distribution of raw detector variables from the D2 beam data for 4 different classes of events, based on the characteristic inter-arrival time $dt_{D2} = [4.9999,\,5.0001]~\rm{s}$: total number of strips fired (\textbf{top left}), total number of clusters (\textbf{bottom left}), the average timing are which strips are fired with respect to the external scintillator trigger (\textbf{top right}), and the average $x$ position of fired strips (\textbf{bottom right}). Apart from the $x$ position, the Random and Next, and the Prev and Prev $\cap$ Next categories appear to have the same distribution.}
    \label{fig:time_class_d2}
\end{figure}

The consistency of this behavior across both independent beam data takings strongly suggests that it originates from a physical or detector-related mechanism rather than a statistical fluctuation. Two main hypotheses can be considered. The first is a detector-related effect: the passage of a beam-induced particle burst may leave a transient remnant signal in the detector, such as residual charge, delayed electronics response, or increased noise occupancy. As a result, events occurring shortly after a beam burst would exhibit artificially elevated strip multiplicities and degraded track reconstruction efficiency. The second hypothesis is that the beam itself exhibits temporal correlations, with the particle composition or intensity depending on the recent beam history. In this scenario, events immediately following a beam burst would correspond to a different particle population than those initiating a burst, leading to intrinsically different detector signatures.

To assess the validity of the first hypothesis, we examine events occurring immediately after pairs of temporally correlated events and divide them into two distinct populations: (i) events that also follow the characteristic beam inter-arrival time and are therefore classified as beam-related, and (ii) events that occur after the same sequence but with a random time interval, consistent with cosmic-ray origin. This classification allows us to disentangle detector-induced effects from intrinsic beam properties. If the detector-related hypothesis is correct, any event occurring shortly after a beam-induced burst should exhibit elevated strip multiplicity and altered timing characteristics, regardless of whether the event itself originates from the beam or from a cosmic muon. In contrast, if the effect arises from intrinsic beam correlations, only beam events following a beam burst should exhibit increased multiplicity, while cosmic events occurring afterward should retain the low-multiplicity signature characteristic of single-particle tracks. The distributions of strip multiplicity, number of clusters, and average strip timing for these two populations are found to be statistically consistent. In particular, both populations exhibit similarly elevated multiplicities and comparable timing profiles. This observation indicates that the observed effect does not depend on the nature of the subsequent particle but rather on the recent history of detector activity. These results therefore support the detector-related hypothesis: the passage of a beam-induced particle burst leaves a transient remnant signal in the detector, which temporarily modifies its response and leads to increased apparent strip multiplicity in subsequent events. 

This multi-second persistence of elevated strip multiplicity can be understood by considering two complementary physical mechanisms, both of which are specific to the beam-induced environment and absent during cosmic-ray-only data taking.

The first mechanism concerns the glass electrodes of our RPCs. Float glass has a bulk resistivity of $\rho \approx 10^{12}~\Omega\cdot\text{cm}$, yielding a dielectric relaxation time $\tau = \rho \varepsilon_0 \varepsilon_r \approx 0.7$--$1.0$~s (for $\varepsilon_r \approx 8$--$10$). This governs the rate at which surface charges deposited by an intense ionizing burst are neutralized through the resistive bulk, restoring the nominal electric field across the $1$~mm gas gap. A sub-nanosecond shower of secondary particles from a laser shot deposits dense charge across the active area, temporarily depressing the local field; full recovery requires several relaxation time constants ($3\tau$--$5\tau \approx 3$--$5$~s). The front-end ASIC electronics, with pulse durations and baseline recovery times below $100$~ns~\cite{Abbrescia2000,CMSmuonTDR}, operate on timescales many orders of magnitude shorter, ruling out electronics saturation as the cause of the multi-second effect.

The second mechanism, likely dominant in the present experimental conditions, involves delayed signals from beam-dump-produced neutrons. In a beam-dump experiment, abundant secondary neutrons are produced through photo- and electro-nuclear spallation reactions, as confirmed by our FLUKA simulations (Section~\ref{sec:beam-conditions}). Fast neutrons propagating into the surrounding material are moderated on timescales of seconds, depending on the moderator composition and neutron energy; upon capture, they excite nuclear states whose de-excitation products (gamma rays and recoil particles) are emitted with characteristic delays that can extend to several seconds. Additionally, fast neutrons interacting directly within the detector materials produce recoil protons or alpha particles through elastic scattering or capture reactions, and the resulting excited nuclear states may decay on similarly extended timescales. This delayed secondary emission produces detector hits that are physically correlated with the original beam shot but arrive well after the prompt particle burst. The sensitivity of glass-based RPCs to such recoil products makes them susceptible to this class of delayed signals. Importantly, this mechanism is entirely absent during cosmic-ray-only running, since cosmic muons do not produce a significant neutron flux at sea level, which explains why the multi-second memory effect is observed exclusively in beam-on datasets. 
This interpretation is further supported by the shielding study presented in Section~\ref{sec:shielding}: polyethylene, a well-known neutron moderator, substantially reduces the fraction of strips fired in the earliest time bin (from $12\%$ without shielding to $1\%$ with combined Pb$+$PE shielding), consistent with suppression of the prompt neutron-induced component.

Disentangling the relative contributions of these two mechanisms---glass relaxation and neutron-induced delayed emission---will require dedicated measurements under controlled beam conditions, including systematic variation of shielding composition and thickness guided by Monte Carlo simulations. 
Future measurements performed under controlled beam conditions, with fixed acquisition times and a known number of beam pulses, will allow a direct comparison of event rates and detector signatures across time classes, enabling a definitive identification of the underlying mechanism.

\begin{figure}[H]
    \centering
    \includegraphics[width=\linewidth]{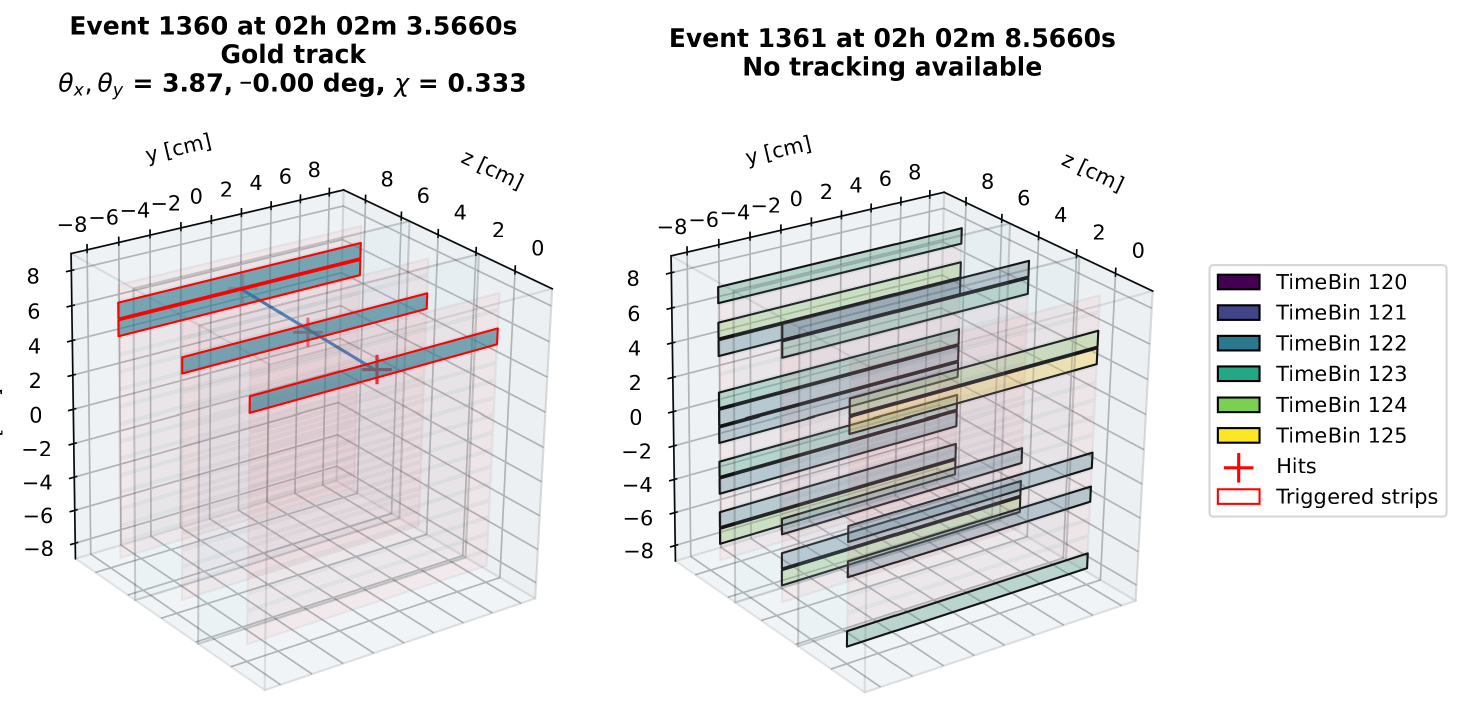}
    \caption{Event display of event 1360, a beam induced muon candidate from the D2 beam data-taking. Event 1361 coming 5.00001 s after is also shown.}
    \label{fig:5 s_candidate}
\end{figure}

\begin{figure}[H]
    \centering
    \includegraphics[width=\linewidth]{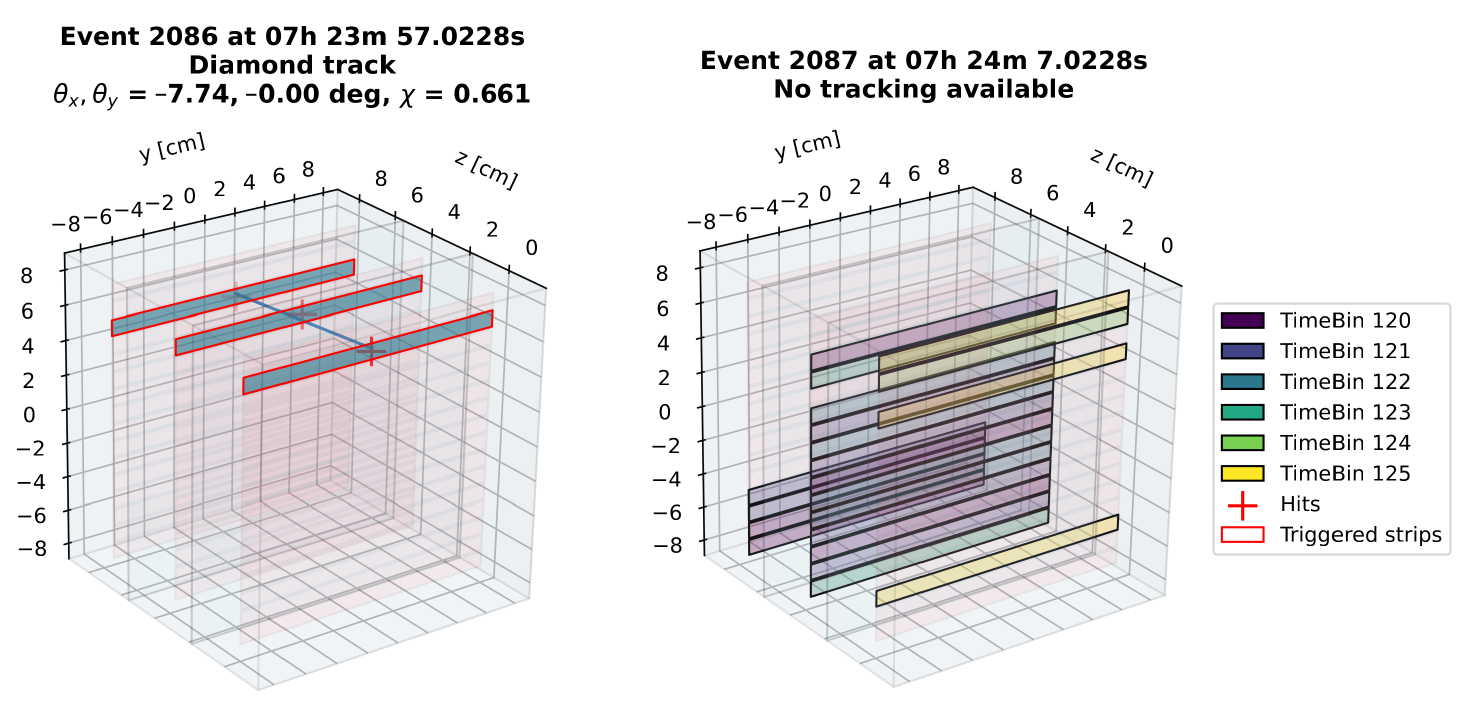}
    \caption{Event display of event 2086, a beam induced muon candidate from the D3 beam data-taking. Event 2087 coming 10.00001 s after is also shown.}
    \label{fig:10 s_candidate}
\end{figure}

In summary, beam-associated events exhibit clear temporal correlations, characterized by highly precise inter-arrival times that reflect the periodic structure of the beam pulses. Within this population, events belonging to the \textit{Prev} category are typically associated with high strip multiplicities, which are not compatible with the signature expected from a single particle traversing the detector. In contrast, events in the \textit{Next} category exhibit significantly lower strip multiplicities, consistent with the passage of a single particle through the detector acceptance. Although the available statistics are not yet sufficient to perform a robust quantitative analysis, it is instructive to examine individual event displays from the \textit{Next} category to investigate their topology. In particular, we search for track-like patterns compatible with horizontal trajectories, which would be consistent with beam-induced muons entering the detector. Representative examples of such events are shown in Figures~\ref{fig:5 s_candidate} and~\ref{fig:10 s_candidate} for the D2 beam and D3 beam datasets, respectively. The observed trajectories are horizontal and occur at the characteristic beam inter-arrival times, making them strong candidates for beam-induced muon tracks. However, the identification of these tracks as muons cannot be established solely from individual event displays. A statistically robust classification will require a more systematic analysis combined with dedicated shielding studies to better characterize the particle composition of the beam-induced flux.

\subsubsection{Laser Tag}
\label{Psec:laser_tag}

\begin{figure}[H]
    \centering
    \includegraphics[width=0.49\linewidth]{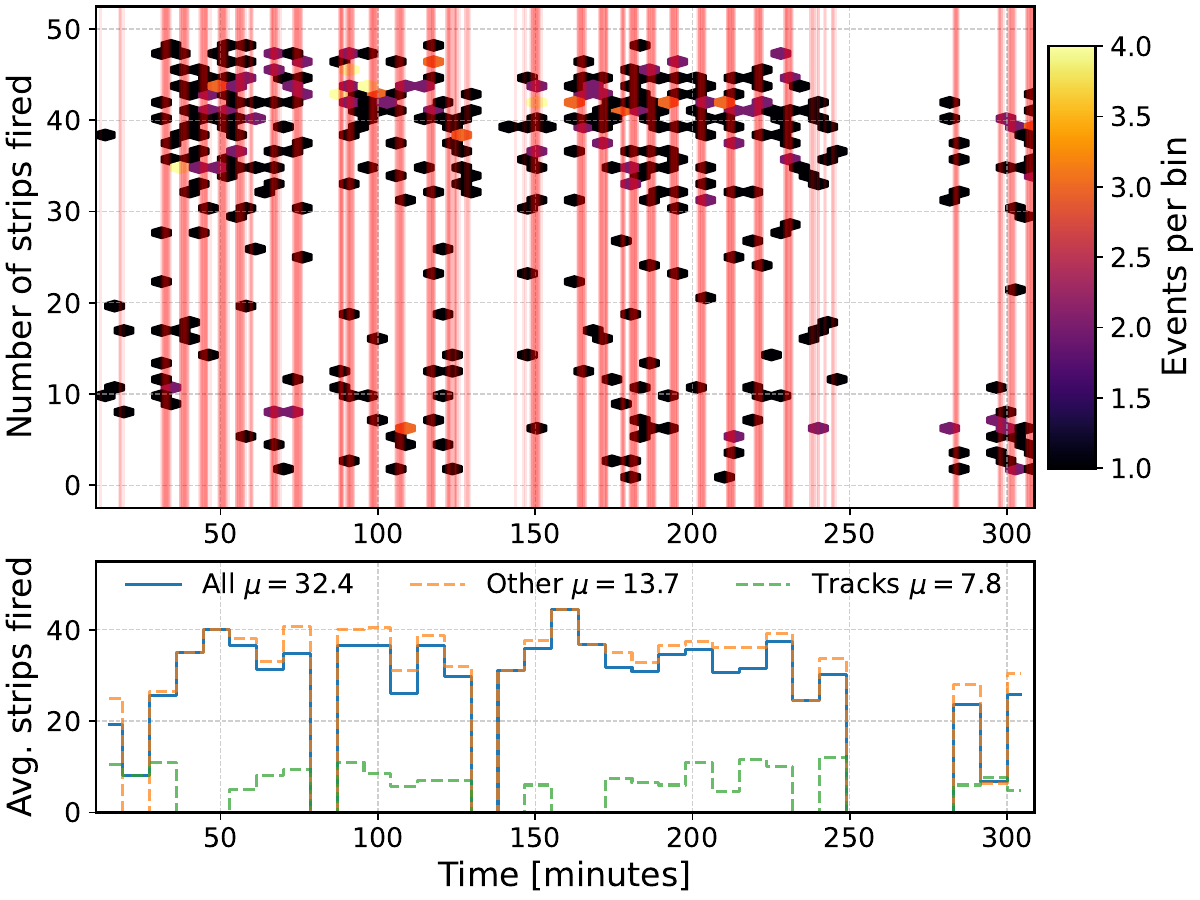}
    \includegraphics[width=0.49\linewidth]{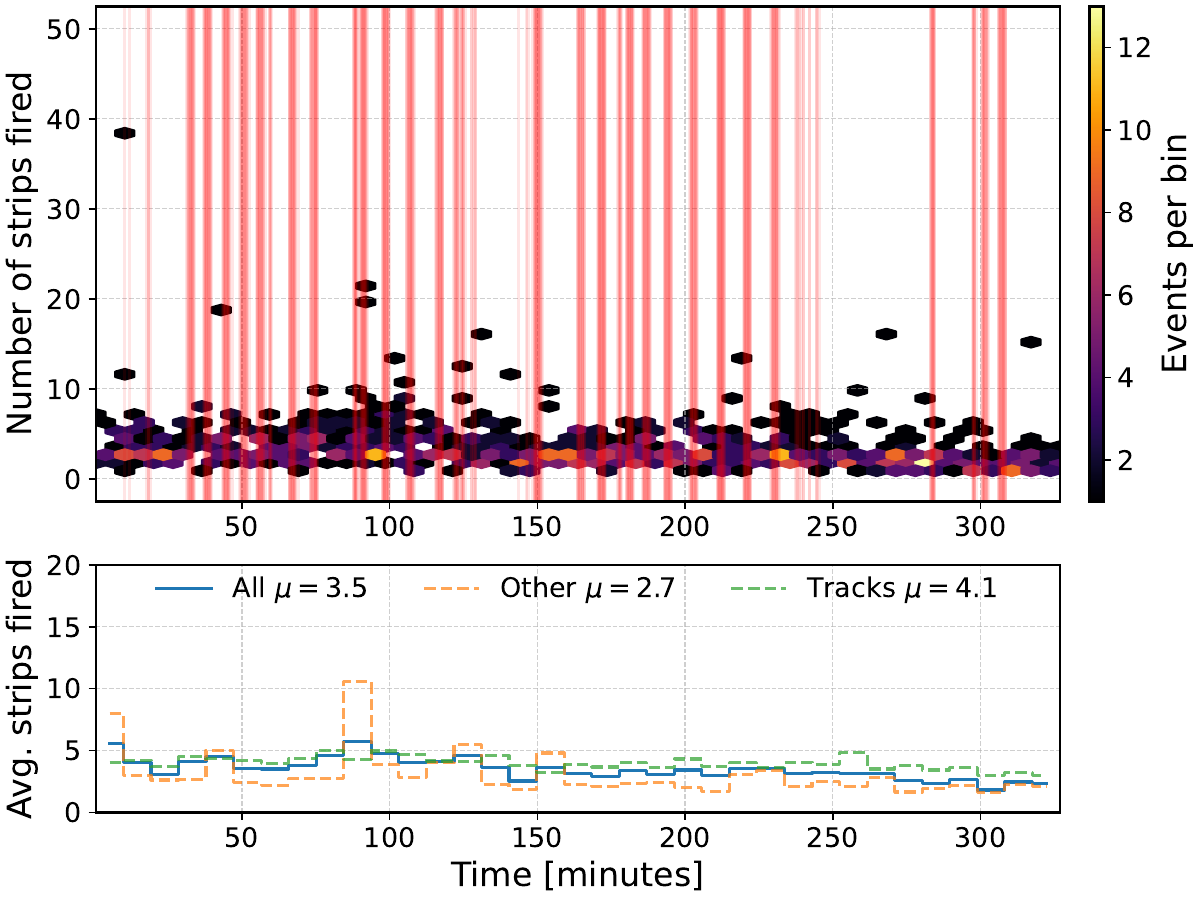}
    \caption{Number of strips fired per event as a function of external scintillators trigger time for laser-tagged (\textbf{left}) and non-tagged events (\textbf{right}). Times at which the laser is triggered are represented by the transparent red lines.}
    \label{fig:laser1}
\end{figure}

During the D2 data-taking period, an additional 5.5-h run was carried out with the beam enabled and the laser trigger signal recorded. The laser tag directly encodes the beam state: when the laser is on, a laser-induced muon beam is delivered to the experimental area, whereas a missing laser tag corresponds to periods with no beam. This allows the dataset to be cleanly partitioned into two populations: beam-on (laser-tagged) and beam-off (non-tagged) events. The laser time stamps are superimposed as transparent red lines on the strip multiplicity versus time distribution shown in Figure~\ref{fig:laser1}. 

A strong correlation is observed between the beam state and the detector response. When the laser is off, the average number of fired strips for tracked events is 4.1, consistent with the D2 cosmic data-taking conditions, and events with large strip multiplicity are rare. This behavior indicates that the recorded activity is dominated by isolated cosmic muons. In contrast, when the laser is on, the average strip multiplicity increases dramatically,  and average number of fired strips reaches 32. This sharp increase provides direct evidence that the laser tag successfully identifies periods of intense beam-induced activity, during which a large number of secondary charged particles are produced and simultaneously detected.

Taken together, these results demonstrate that the laser tag provides a powerful and reliable handle for separating beam-induced particle bursts from the steady cosmic muon background. They also confirm that the observed increases in strip multiplicity and the emergence of temporal correlations are intrinsic signatures of beam activity, rather than artifacts of detector performance or data acquisition effects.

\subsubsection{Detector Shielding and Background Characterization}
\label{sec:shielding}

\begin{figure}[H]
    \centering
    \includegraphics[width=\linewidth]{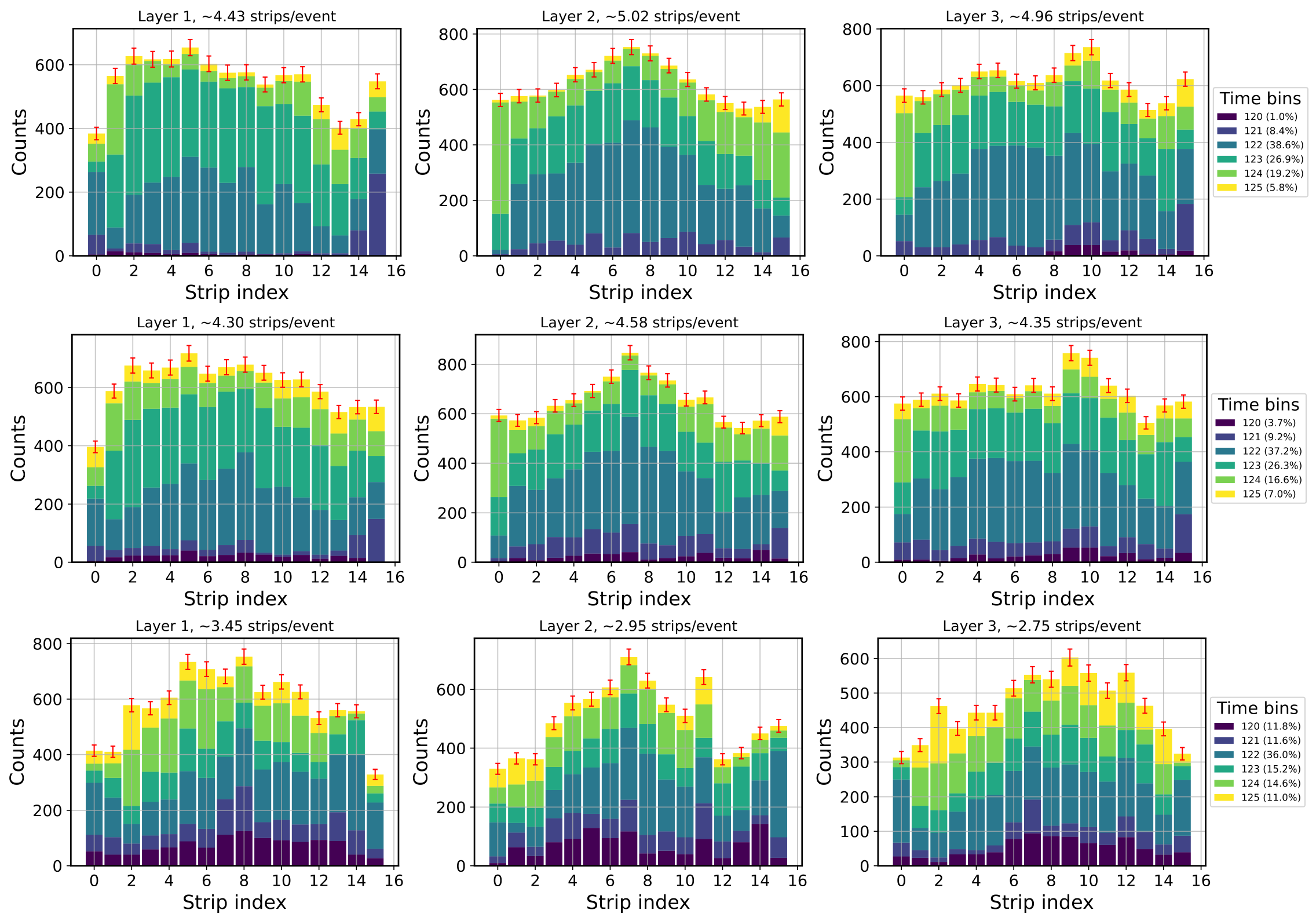}
    \caption{Strip occupancy of each of the three RPC layers for the D3 data taking with lead and polyethylene shielding (\textbf{top}), lead shielding (\textbf{middle}) and without shielding (\textbf{bottom}). Poissonian uncertainties are shown in red.}
    \label{fig:mult_shielding}
\end{figure}

During the D3 data-taking period, several shielding configurations were investigated: polyethylene (PE) combined with lead, lead only, and no shielding. The impact of the shielding configuration on the per-event strip multiplicity is shown in Figure~\ref{fig:mult_shielding}. For data taken without shielding, the average number of fired strips per event in layers 1, 2, and 3 is 3.45, 2.95, and 2.75, respectively. This average increases to 4.30, 4.58, and 4.35 when lead shielding is used, and further to 4.43, 5.02, and 4.96 when both lead and PE shielding are installed. This increase in strip multiplicity suggests the production of charged secondary particles within the shielding materials. The shielding configuration also significantly affects the temporal distribution of fired strips. The D2 cosmic run, which serves as a reference for the intrinsic detector response, shows a negligible fraction of strips fired in time bin 120, below 1\%. In contrast, for the D3 data-taking without shielding, this fraction rises sharply to approximately 12\%. The use of shielding substantially mitigates this effect: the contribution at time bin 120 is reduced to 3.7\% with lead shielding and to 1.0\% with the combined lead and PE shielding.

\begin{figure}[H]
    \centering
    \includegraphics[width=\linewidth]{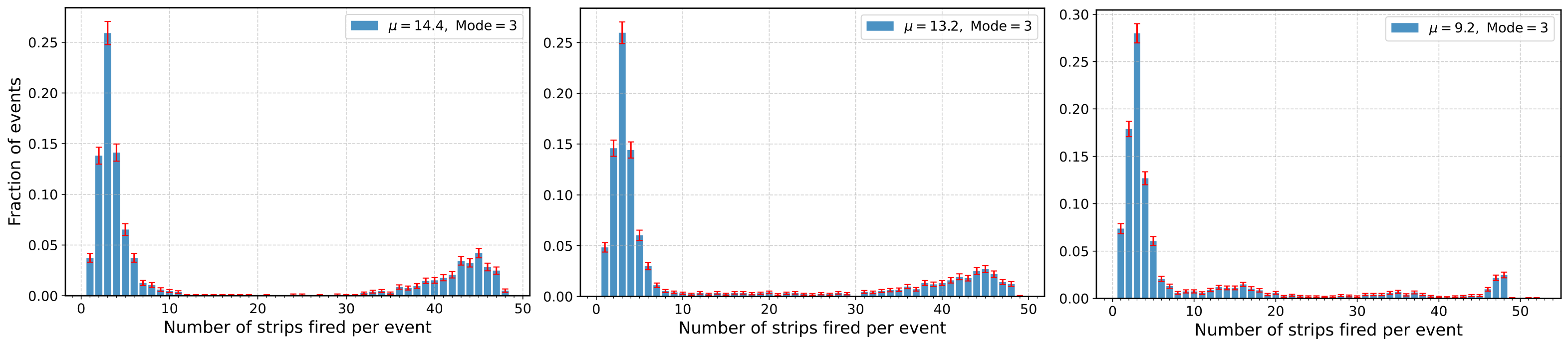}
    \caption{Number of strips fired per event for D3 data taking with lead and polyethylene shielding (\textbf{left}), lead shielding (\textbf{middle}) and without shielding (\textbf{right}). The mode and mean $\mu$ of each distribution is presented in the legend. Poissonian uncertainties are shown in red.}
    \label{fig:strips_shield}
\end{figure}

Further evidence for the production of charged secondaries in the shielding materials is provided by the distribution of the number of fired strips per event, shown in Figure~\ref{fig:strips_shield}. For the unshielded configuration, the high-multiplicity tail of the distribution begins around 10 strips per event. When shielding is present, this tail is shifted toward significantly higher multiplicities, extending beyond 30 strips per event for both lead-only and lead-plus-PE configurations. In addition, the presence of shielding reduces the fraction of low-multiplicity events: the proportion of events firing fewer than three strips decreases from 24\% without shielding to below 20\% when either lead or lead plus PE shielding is used. This reduction in low-multiplicity events translates into improved semi-coincidence, full-coincidence, and tracking efficiencies for data sets acquired with shielding, as shown in Table~\ref{table:class_shield}. However, the observed increase in charged secondary production highlights the need to balance the shielding of low-energy particles against the generation of secondaries in dense materials. The optimal shielding configuration should therefore be guided by detailed simulations, in preparation for the next campaigns. 

\begin{table}[H]
    \caption{Proportion of tracks corresponding to each class, from highest (diamond) to lowest purity (bronze) for the three shielding configurations. Events classified as other are the ones that do not satisfy tracking requirements.}
    \label{table:class_shield}
\begin{adjustwidth}{-\extralength}{0cm}
\centering
\begin{tabularx}{\fulllength}{cCCCCCCCCC}
     \toprule
     \textbf{Shielding} & \textbf{Diamond} & \textbf{Platinum} & \textbf{Gold}  & \textbf{Silver} & \textbf{Bronze} & \textbf{Other} & \boldmath{$\epsilon_{\mathrm{semi}}$} & \boldmath{$\epsilon_{\mathrm{full}}$} & \boldmath{$\epsilon_{\mathrm{track}}$}\\ 
     \hline
     Lead + PE & 14.0\% & 6.1\%  & 2.0\%  & 5.9\%  & 16.8\% & 55.3\% & 95.0\%  & 73.4\% & 40.3\%\\ 
     Lead & 15.8\% & 5.2\%  & 2.4\%  & 6.5\%  & 13.9\% & 56.2\% & 93.5\%  & 72.3\% & 40.6\%\\ 
     None & 15.6\% & 5.7\%  & 2.2\%  & 6.2\%  & 11.6\% & 58.8\% & 88.3\%  & 61.6\% & 36.9\%\\ 
     \bottomrule
    \end{tabularx}
\end{adjustwidth}
\end{table}

To further investigate the impact of shielding configurations on detector signatures, we examine the distribution of events in the PCA space introduced in Appendix~\ref{sec:pca}. Figure~\ref{fig:pca_shielding_comp} shows the event populations in the $\mathrm{PC}_1$–$\mathrm{PC}_2$ plane for the different shielding configurations. Focusing on the high-activity regime ($\mathrm{PC}_1 > 4$), beyond the cosmic-dominated region, distinct structures emerge depending on the shielding material.

\begin{figure}[H]
    \centering
    \includegraphics[width=\linewidth]{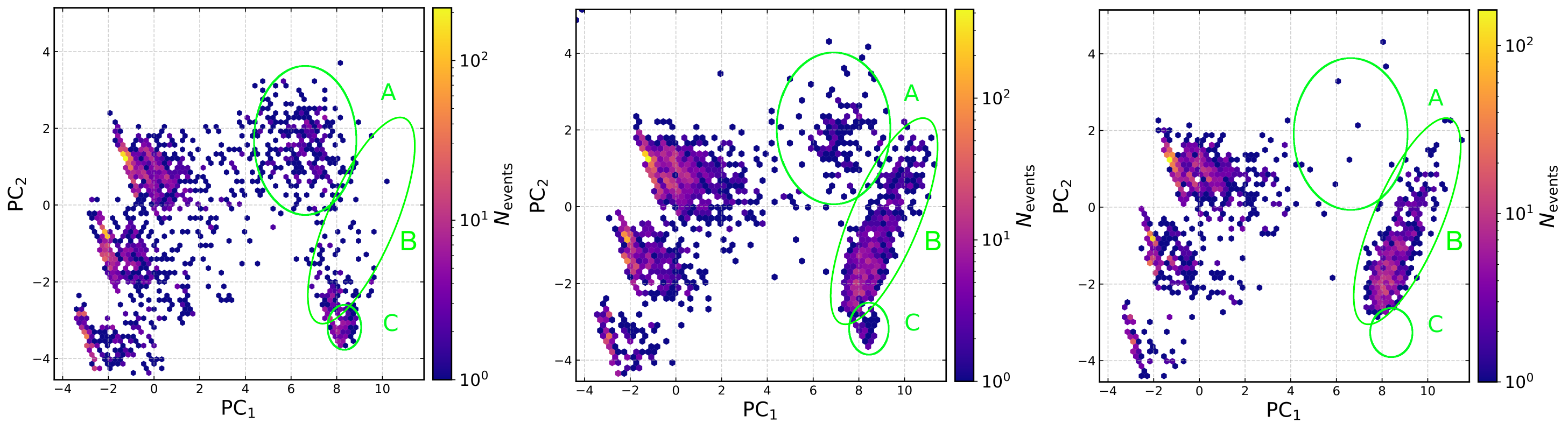}
    \caption{Representation of D3 beam data corresponding to different shielding configurations in the PCA plane: None (\textbf{left}), lead (\textbf{center}) and lead and poly-ethylene (\textbf{right}). While the region with PC$_1$ < 4 appears to be similar for each shielding configuration, the PC$_1$ > 4 exhibits different clusters labeled A, B and C each corresponding to a different cluster and strip multiplicity regime.}
    \label{fig:pca_shielding_comp}
\end{figure}

For the lead-only configuration, three well-defined clusters can be identified in this region:

\begin{itemize}
    \item \textbf{Region A}: events with more than eight clusters and fewer than 35 fired strips
    \item \textbf{Region B}: events with more than 35 fired strips and fewer than 12 clusters
    \item \textbf{Region C}: events with more than 45 fired strips.
\end{itemize}

\begin{figure}[H]
    \centering
    \includegraphics[width=\linewidth]{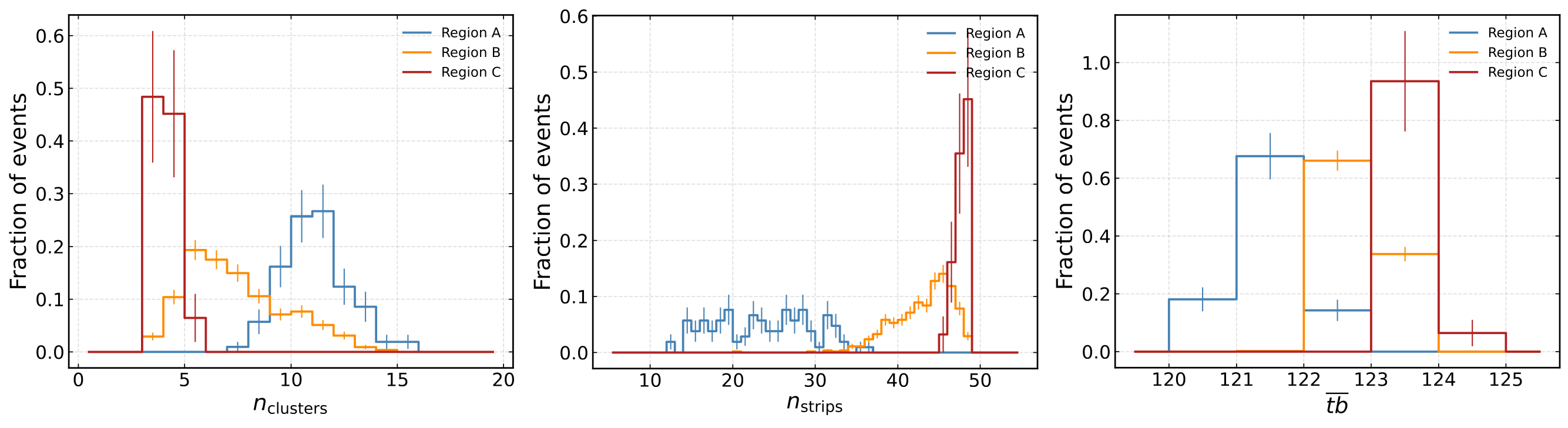}
    \caption{Normalized distributions of key event-level observables for the three regions identified in the PCA plane for the lead only shielding configuration. (\textbf{Left}): total number of clusters per event. (\textbf{Center}): total number of fired strips per event. (\textbf{Right}): average time bin of the fired strips within the event. The three regions (A, B, and C) exhibit clearly distinct topological and temporal signatures.} 
    \label{fig:shielding_comp_lead_only}
\end{figure}

The associated number of clusters and fired strips as well as the average event-wise time bin distributions are presented in Figure~\ref{fig:shielding_comp_lead_only}. Region~A is characterized by moderate cluster multiplicity and relatively low strip counts, Region~B by large strip multiplicity with fewer clusters, and Region~C by very high strip multiplicity. 
The differences in both multiplicity and timing distributions support the interpretation that these regions correspond to event classes with different underlying particle origins or interaction mechanisms. Region B is also present in the lead + polyethylene (PE) configuration, while it is largely absent in the unshielded dataset. This suggests that the corresponding signature is associated with secondary particles produced in the lead shielding. In contrast, Region A is observed more prominently in the unshielded data than in the lead-only dataset, indicating that the presence of lead suppresses this event class. Region C appears in both the lead-only and unshielded configurations but disappears when both lead and PE shielding are installed.

While comprehensive particle-level simulations of the shielding setup are left for future work, the topological features of these three regions can be tentatively interpreted through standard particle interaction mechanisms:  
\begin{itemize}
\item    Region A: The combination of moderate cluster counts and low strip multiplicity is consistent with localized, point-like energy depositions. These may arise from fast photoneutrons undergoing nuclear recoil or hard bremsstrahlung photons undergoing Compton scattering within the detector materials, producing isolated hits with minimal charge sharing.
\item    Region B: The increase in strip multiplicity alongside moderate cluster counts is characteristic of secondary electromagnetic showers initiated in the lead absorber. High-energy gamma rays converting via pair production or Compton scattering generate secondary electrons and positrons that emerge from the back of the lead slab over a broader area, firing contiguous groups of strips simultaneously. This provides a plausible rationale for the emergence of Region B whenever lead shielding is installed. 
\item    Region C: The very high multiplicity regime likely reflects dense, unattenuated particle cascades reaching the active volume. The disappearance of Region C under combined lead and polyethylene shielding suggests that the combined absorbers successfully moderate and absorb this soft shower component. Future dedicated simulations will be essential to test and refine these tentative identifications.
\end{itemize}

\subsection{Systematic Uncertainties and Experimental Limitations}
\label{sec:systematics}

In the section, we outline some sources of systematic uncertainty and experimental limitations which need to be addressed in future studies, distinguishing between those that are dominant in this study and those that are swamped by the statistical uncertainty due to the limited dataset.

\begin{itemize}
    \item \textbf{Electron beam and laser jitter:} As described in Section~\ref{sec:beam-conditions}, shot-to-shot fluctuations in electron energy ($0.5$--$5$~GeV) and charge ($10$--$100$~pC) represent the largest systematic uncertainty. Because secondary production rates scale non-linearly with primary electron energy near threshold, these variations preclude an exact shot-by-shot normalization of particle yields. Future dedicated campaigns with active beam diagnostics will allow full control and systematic propagation of the beam-related experimental uncertainties.
    \item \textbf{Collimator conversion:} The spectrometer slit intercepted $50--95\%$ of the total beam charge upstream of the main beam dump (Section~\ref{sec:beam-conditions}). This component acted as an unmonitored secondary source, introducing spatial and spectral broadening into the forward particle flux.
    \item \textbf{Detector positioning and geometry:} An uncertainty of 20~cm on the distance between the beam dump and the RPC stack (18~m) translates into an acceptance uncertainty of approximately $2.2\%$. Internal relative alignment of the three RPC layers was maintained within $\pm 1$~mm, and tracking artifacts from residual misalignments were constrained by the $1.0$~cm track-fit residual cut. These effects are negligible in comparison with the statistical uncertainties due to the limited statistics of tracks in the present study, but will become important and will need to be addressed in future runs at very high muon intensity.
\end{itemize}

\section{Summary}
\label{sec:summary}

This work constitutes a proof-of-principle demonstration of the deployment of a portable RPC telescope in the environment of a laser-plasma accelerator. The measurements were performed during parasitic operation of the ELBA facility, where the accelerator configuration and beam parameters were continuously optimized for the primary experiment and therefore remained outside our control. Within these constraints, the detector operated reliably throughout three separate campaigns demonstrating robustness and ease of deployment. 

The analysis yielded valuable insights across several diverse datasets: a beam-off cosmic baseline, three datasets with varying shielding configurations to characterize secondary background composition collected at Gammatron, and the multi-GeV dataset collected at ELBA. While the high-energy beam run exhibited a statistically significant rate enhancement in the near-horizontal signal region ($p = 8.5 \times 10^{-4}$ relative to cosmics and Gammatron datasets), the dominance of low-purity tracks polluted by combinatorial artifacts from beam-induced showers prevents determining its signal content, and no claim of muon observation is made. 
Based on this sample, we established a 95\% CL upper limit on the horizontal muon fluence at $18$~m of $7.9 \times 10^{-5}$~cm$^{-2}$ per shot ($1.3 \times 10^{-13}$~cm$^{-2}$ per primary electron assuming nominal $100$~pC bunches), which is consistent with FLUKA Monte Carlo expectations scaled for geometric attenuation and parasitic beam sampling.

These preliminary results validate the RPCs' operational stability and tracking robustness in intense radiation and electromagnetic environments, while defining the sensitivity threshold for this experimental approach. The lessons learned strongly motivate future dedicated campaigns with full beam parameter control. To overcome the background limitations identified in this study, the upcoming run will focus on two key experimental enhancements:
\begin{enumerate}
    \item \textbf{Direct Shot-to-Shot Laser Tagging:} Establishing a robust hardware laser trigger directly registered by the DAQ system will enable tight event gating locked to the laser arrival time. Because cosmic muons arrive randomly and uniformly across the continuous acquisition window, restricting the analysis to the immediate beam-pulse window will eliminate virtually the entire cosmic-ray background.
    \item \textbf{Simulation-Guided Shielding Design:} Detailed Monte Carlo particle transport simulations will be conducted to optimize the composition, geometry, and placement of passive absorbers (combining high-$Z$ materials such as lead with hydrogenous moderators like polyethylene). This will allow us to maximize the suppression of soft electromagnetic cascades and photoneutrons while mitigating the generation of secondary knock-on charged particles within the shielding itself.
\end{enumerate}

Together, these improvements will provide the background rejection and statistical sensitivity required to isolate and characterize the laser-driven muon signal definitively.

\vspace{6pt} 

\authorcontributions{Conceptualization, A.C., A.G., E.C.G., and R.V.;
methodology, A.C., A.G., A.J.S., E.C.G., E.R., H.M.M., J.E.S., J.S., M.L., S.I. and R.V.; 
software, M.L., P.D. and S.I.; 
validation, D.A., M.L. and S.I.; 
formal analysis, D.A., E.R., J.E.S., M.L. and S.I.; 
investigation, A.J.S., D.A., E.R., F.V., I.Z., J.E.S., J.S., M.L., M.N., S.I. and S.L.; 
resources, E.C.G., F.V., J.S., M.N., P.D. and {U.C.}; 
data curation, M.L. and S.I.; 
writing---original draft preparation, A.G., E.R., J.E.S., R.V. and S.I.; 
writing---review and editing, A.C., A.G., E.R., G.M.G., H.M.M., J.E.S., R.V. and S.I.; 
visualization, M.L. and S.I.; 
supervision, A.C., A.G., E.C.G., R.V. and M.T.; 
project administration, A.G., E.C.G., M.T.; 
funding acquisition, A.G. and H.M.M. 
All authors have read and agreed to the published version of the manuscript.}

\funding{The CP3 team (A.G., E.C.G., M.L., P.D. and S.I) acknowledges funding by the Fonds de la Recherche Scientifique - FNRS under Grant No. J.0043.22.
 The Maryland team acknowledges the support of the US Dept. of Energy (DE-SC0015516, DE-SC0024406, and DE-SC0024398).}

\dataavailability{Dataset available on request from the authors.}

\acknowledgments{We thank Mehar Ali Shah, Helio Nogima, Mapse Barroso Ferreira Filho, Salvatore Buontempo, and Kevin Mota Amarilo for their friendly collaboration during the joint data-taking campaigns reported in this document and the fruitful discussions on the interpretation of our data. 
The interpretation of our results also benefited from simulation studies of the ELBA experimental environment by Hamid Basiri, which will be reported in future publications. We also thank the electronics and mechanical workshops of the CRC at UCLouvain, and in particular Quentin de Smedt for his help in preparing our detector setup for these beam tests, and Nicolas Szilasi for the technical drawings of our RPC detectors.}

\conflictsofinterest{The authors declare no conflicts of interest.}

\abbreviations{Abbreviations}{
The following abbreviations are used in this manuscript:\\

\noindent 
\begin{tabular}{@{}ll}
RPC & Resistive Plate Chamber\\
LPA & Laser-Plasma Accelerator\\
LWFA & Laser Wake-Field Accelerator\\
ELBA   & ELectron Beam Accelerator for fundamental science
\end{tabular}
}

\appendix

\section{Threshold Scan}
\label{sec:thresh_scan}

\begin{figure}[H]
    \centering
    \includegraphics[width=0.7\linewidth]{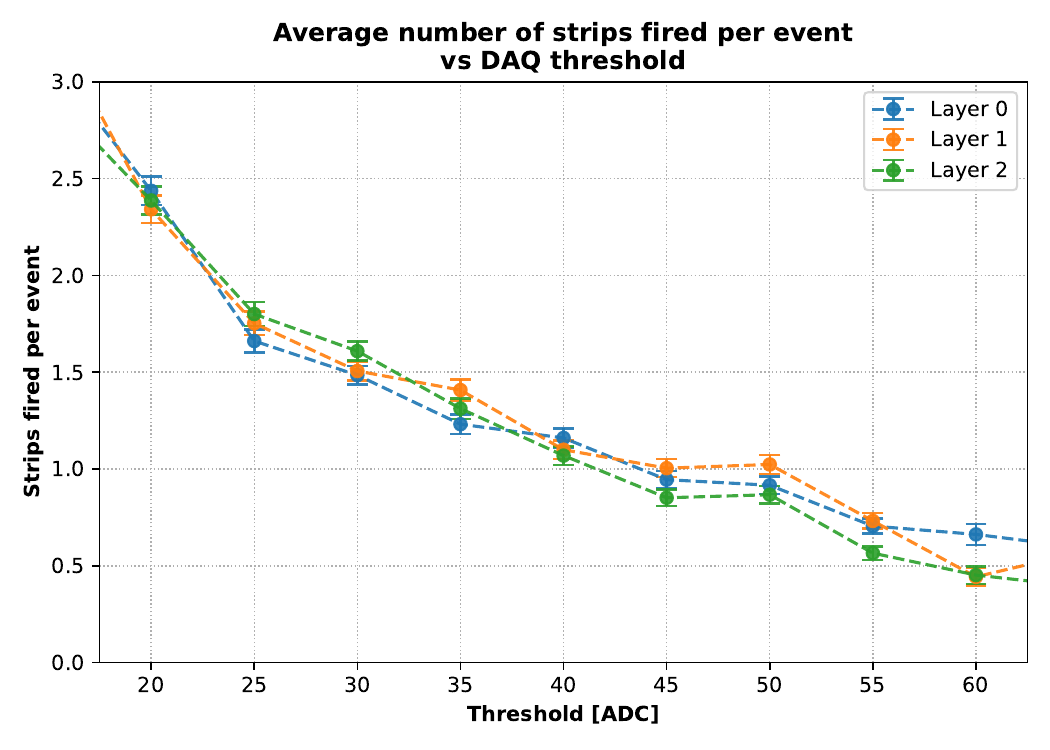}
    \caption{Average number of strips fired per event as a function of the DAQ threshold value, for each of the three RPC chambers. Below a threshold value of 40, the average number of strips fired per event goes below 1.}
    \label{fig:thresh_scan}
\end{figure}

The DAQ threshold settings were validated through a threshold scan. As shown in Figure \ref{fig:thresh_scan}, the mean number of fired strips per event and per chamber remains above 1, consistent with the expectation for minimally ionizing cosmic muons and supporting the chosen thresholds.

\section{Validation with Cosmic Data at CP3}
\label{sec:cosmics_validation}

\begin{figure}[H]
    \centering
    \subfloat[ELI-like setup]{\includegraphics[width=.5\linewidth]{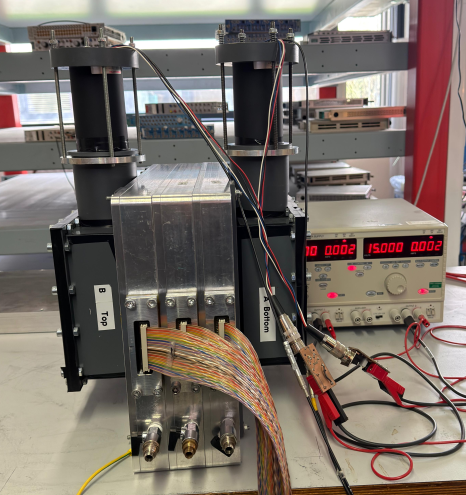}}
    \hfill
    \subfloat[Vertical-flux setup]{\includegraphics[width=.4\linewidth]{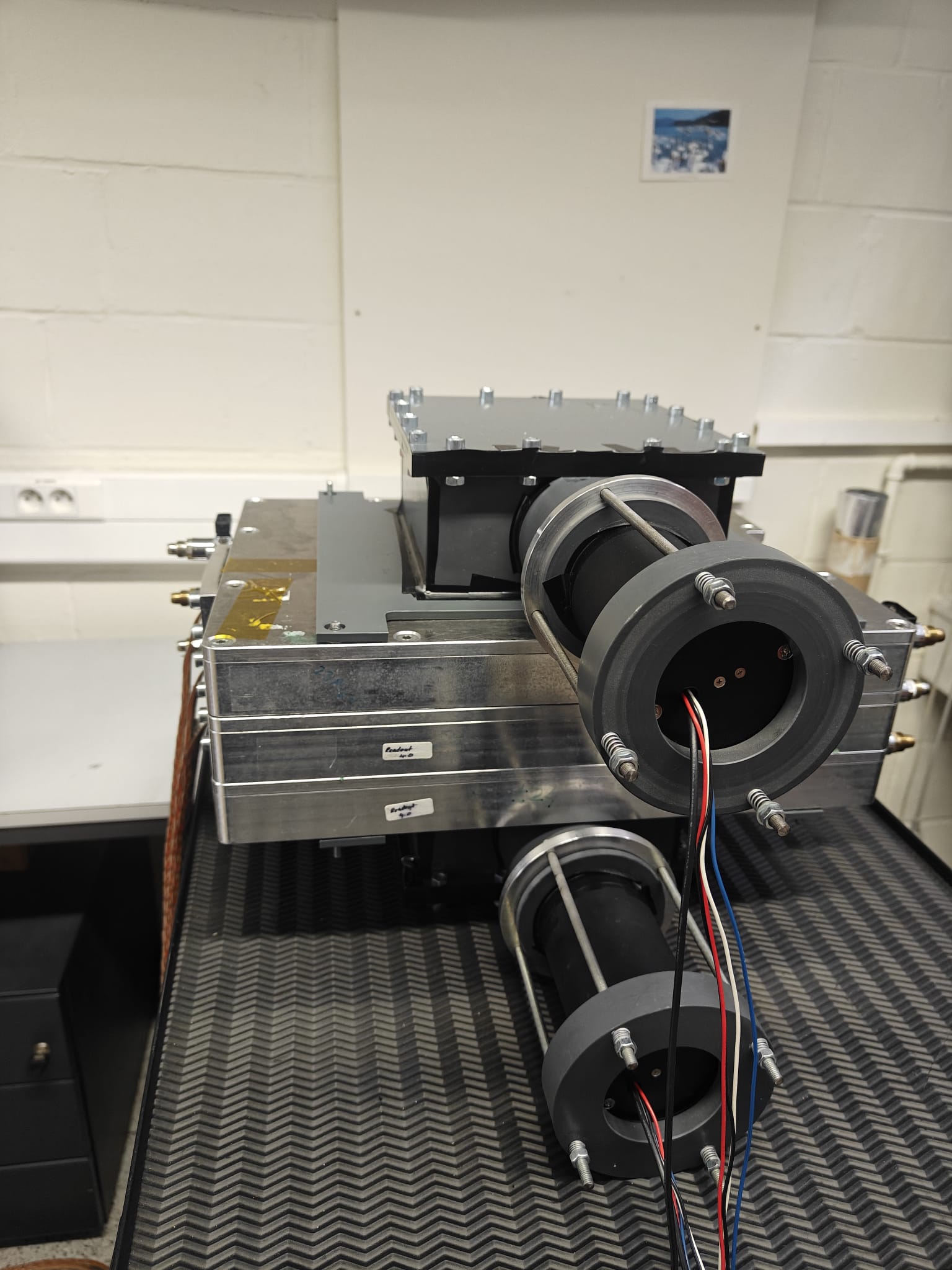}}     
    \caption{Experimental configurations used for the cosmic-ray measurements: (\textbf{a}) ELI-like setup and (\textbf{b}) vertical-flux setup used for cosmic-ray data acquisition in the laboratory at CP3.}
    \label{fig:detectorsetup_CP3}
\end{figure}

Cosmic-ray data have been used as a crucial validation step before and during data-taking in all campaigns at ELI Beamlines.
Additionally, two simple measurements were performed in the laboratory at CP3 after the end of the last campaign, in order to validate the data acquisition at ELI Beamlines and test the analysis framework in clean and controlled conditions. 
The first measurement was carried out using exactly the same detector arrangement as in the D2 campaign at ELI Beamlines, while in the second one, to collect more statistics, the detectors were rotated by 90 degrees such to face the sky. The two setups are shown in Figure~\ref{fig:detectorsetup_CP3} and we refer to them, respectively, as ``ELI-like'' and ``vertical-flux'' setups.

\subsection{ELI-Like Setup at CP3}

\begin{figure}[H]
    \centering
    \includegraphics[width=\textwidth]{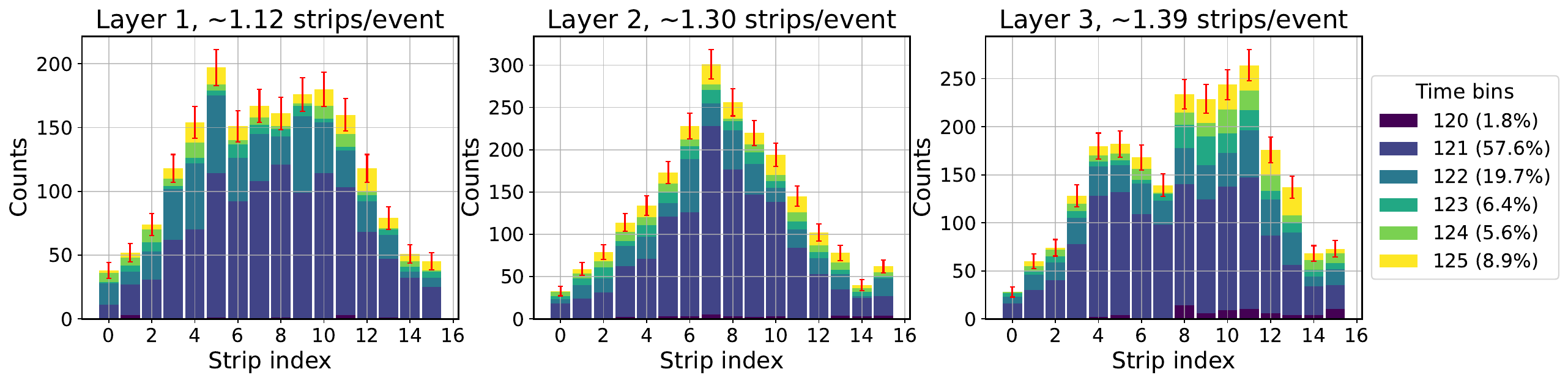}
    \caption{Strip-occupancy distributions for the cosmic-ray dataset recorded with the ELI-like at CP3.}
    \label{fig:elisetup_lab}
\end{figure}

Figure~\ref{fig:elisetup_lab} presents the strip-occupancy distributions obtained from the cosmic-ray dataset recorded in the ELI-like setup at CP3 and compares them with the reference cosmic-ray occupancy distributions shown in the top panel of Figure~\ref{fig:occupancies}. The strong agreement between the two measurements demonstrates a consistent detector response and confirms the reproducibility of the detector performance under both data-taking configurations.

\begin{figure}[H]
    \centering
    \subfloat[]{\includegraphics[width=.48\linewidth]{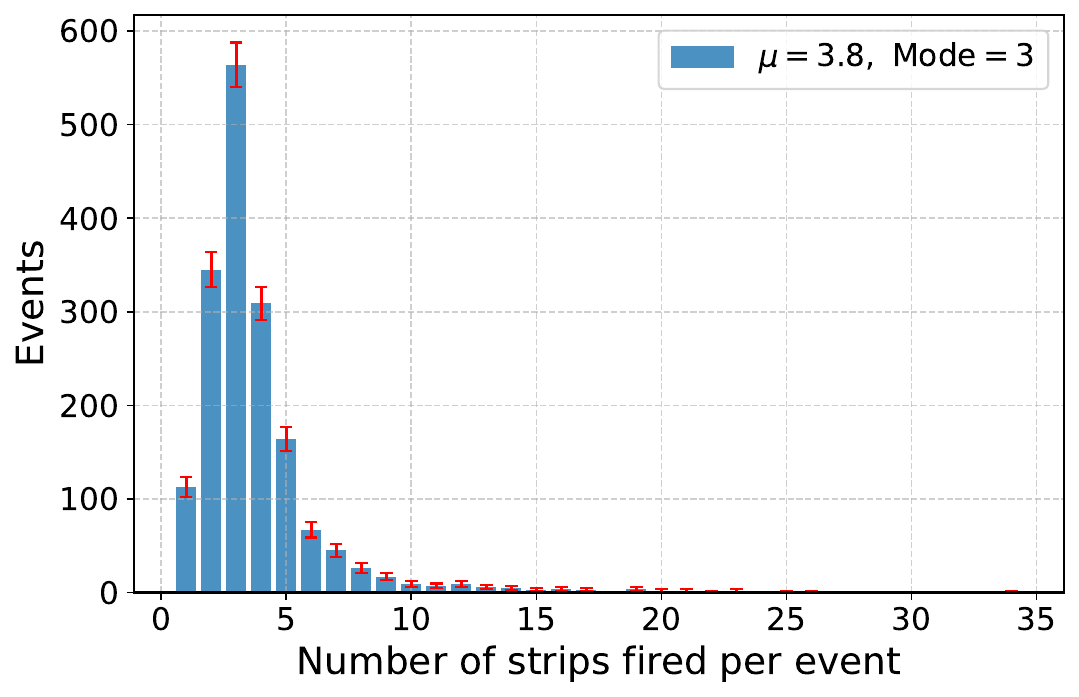}}    
    \hfill
    \subfloat[]{\includegraphics[width=.48\linewidth]{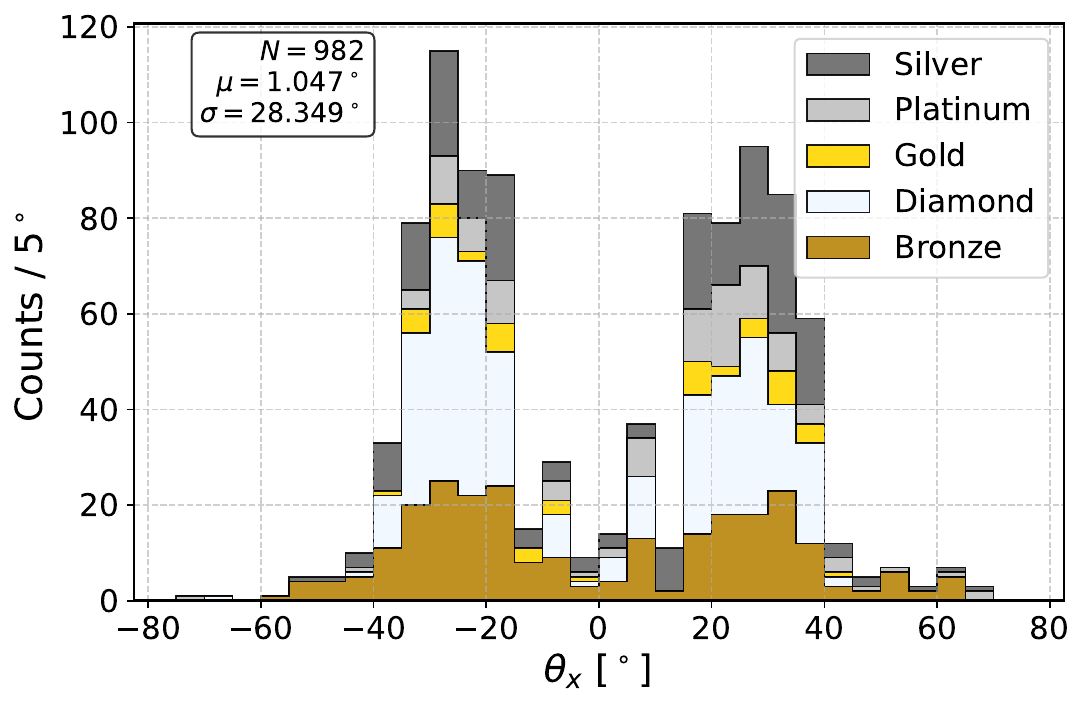}}\\
    \vspace{0.1cm}
    \subfloat[]{ \includegraphics[width=.48\linewidth]{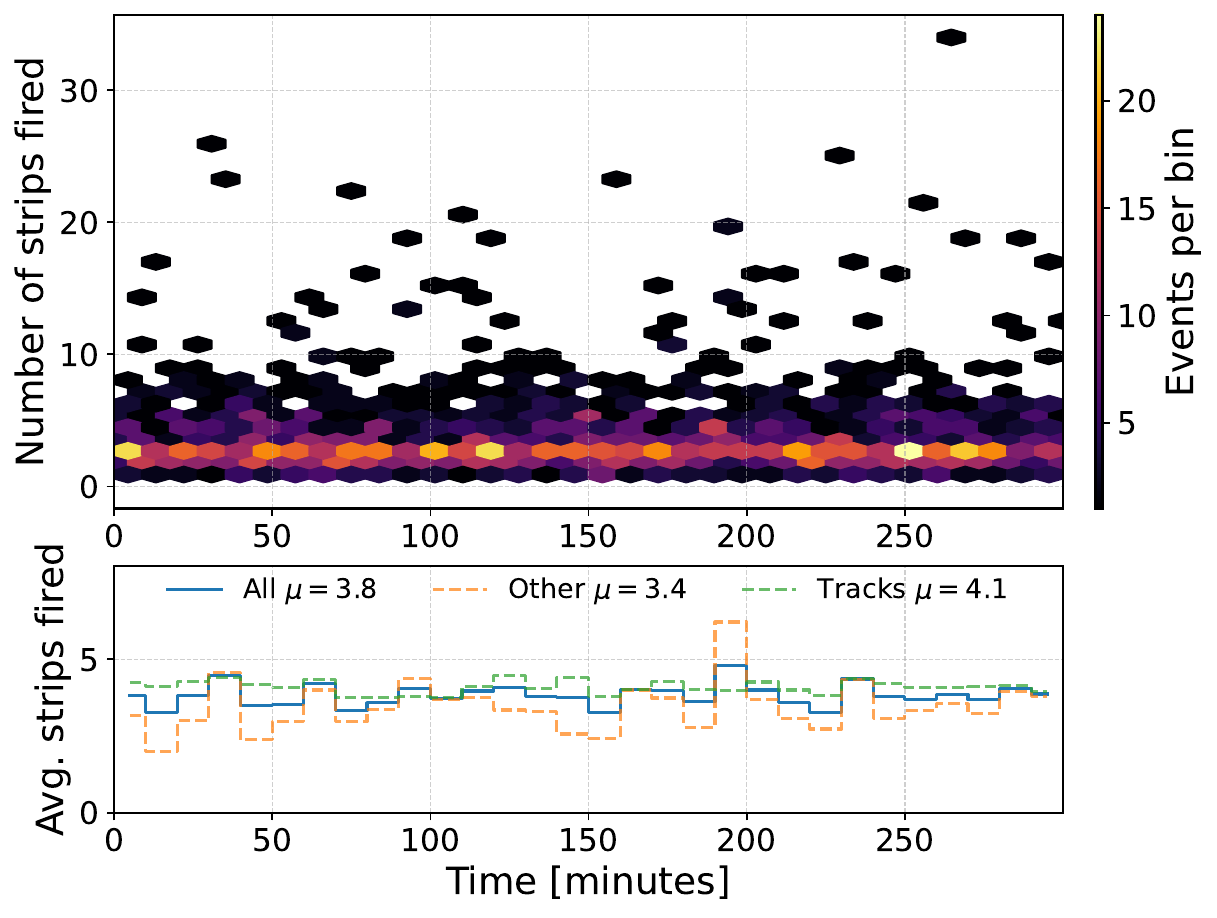}}
    \hfill
    \subfloat[]{\includegraphics[width=.48\linewidth]{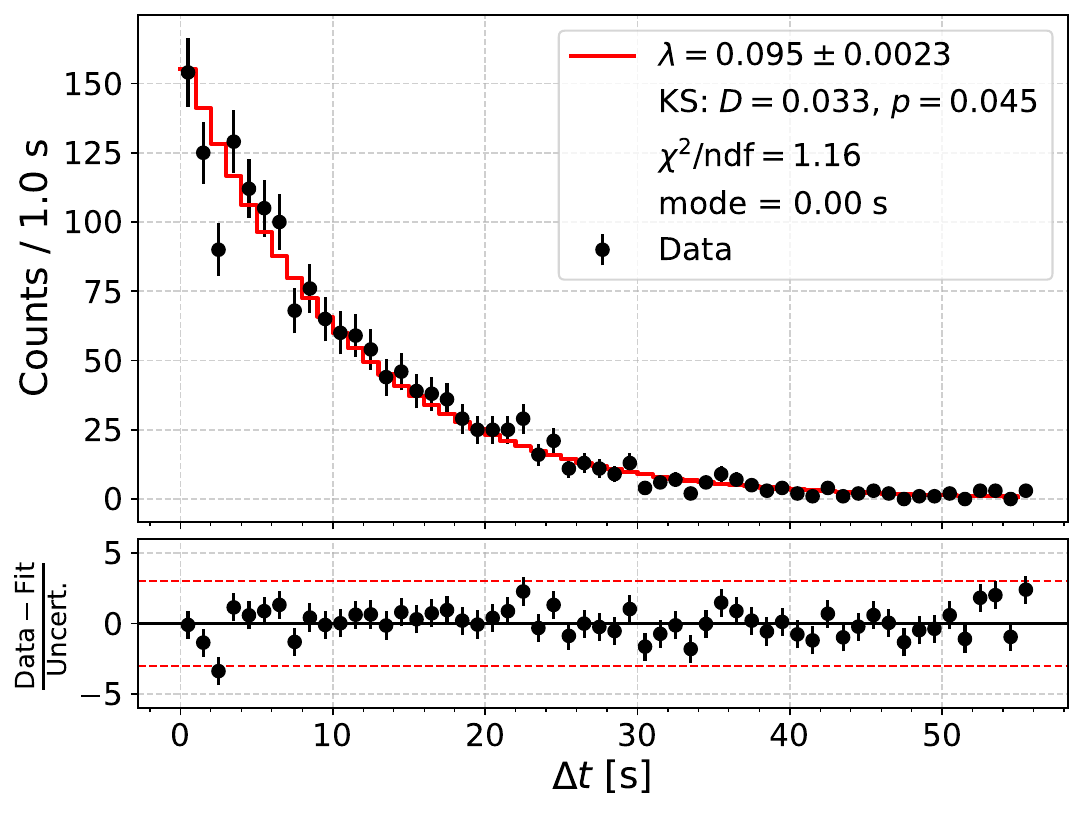}}
    \caption{Validation plots showing the strip multiplicity, zenith-angle distribution, time dependence, and inter-arrival time behaviour for the cosmic-ray dataset recorded with the ELI beam-test setup at CP3. (\textbf{a}) Strip multiplicity for the D1 cosmic-ray dataset. (\textbf{b}) Zenith-angle distribution for laboratory cosmic rays. (\textbf{c}) Strip multiplicity as a function of time. (\textbf{d}) Inter-arrival time distribution.}
    \label{fig:validation_plots_eli}
\end{figure}

Figure~\ref{fig:validation_plots_eli} presents the validation plots for the laboratory cosmic-ray dataset recorded in the ELI-like setup, including the strip multiplicity, zenith-angle distribution, time dependence, and inter-arrival time behaviour. All observed distributions are consistent with the corresponding measurements obtained during the ELI Beamlines beam-test campaign, demonstrating stable detector performance.

The strip-multiplicity distribution shown in Figure~\ref{fig:validation_plots_eli}a exhibits a pronounced peak at low multiplicity, with a mode of approximately three strips, indicating that most events are associated with localized detector activity. The strip-occupancy distribution further shows that the majority of fired strips are concentrated in time bin 121, corresponding to approximately 605\,ns after the scintillator trigger for a 5\,ns time-bin width.

The zenith-angle distributions shown in Figure~\ref{fig:validation_plots_eli}b are consistent with the expected behaviour for the D1, D2, and D3 cosmic-ray datasets. The distributions peak near small angles, $\theta_x \in [-20^\circ, 20^\circ]$, indicating that a significant fraction of events is concentrated around the forward direction. Furthermore, the angular distributions observed for the CP3 cosmic-ray data closely resemble those measured for the D1, D2, and D3 detector configurations.

The time-dependent distributions remain stable throughout the full data-taking period, indicating a consistent detector response and stable operating conditions during the measurement campaign. The measured inter-arrival time follow the exponential behavior expected from a Poisson process, indicating that cosmic-ray detections occur randomly and independently, with no significant correlated background contribution.

\subsection{Vertical-Flux Setup at CP3}

\begin{figure}[H]
    \centering
    \subfloat[ ]{\includegraphics[width=1\textwidth]{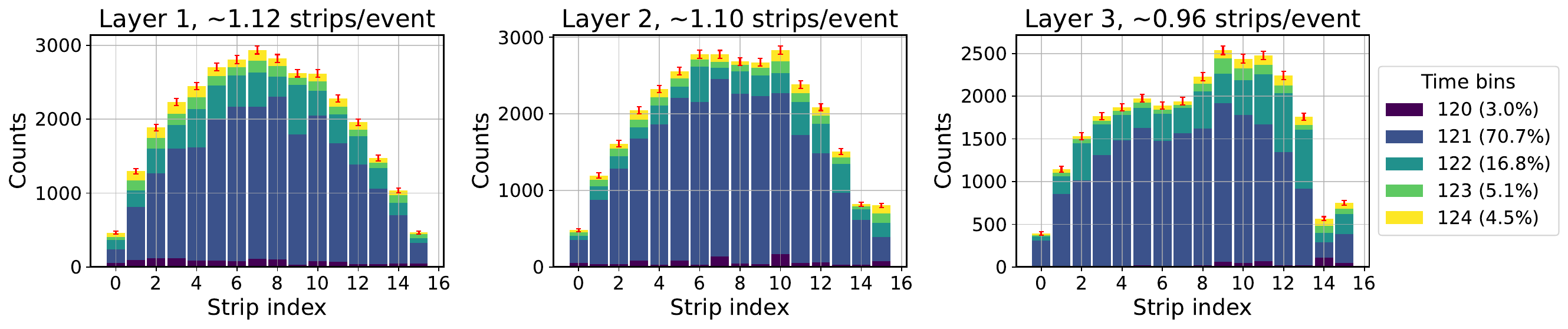}}\\
    \vspace{0.1cm}
    \subfloat[ ]{\includegraphics[width=.48\linewidth]{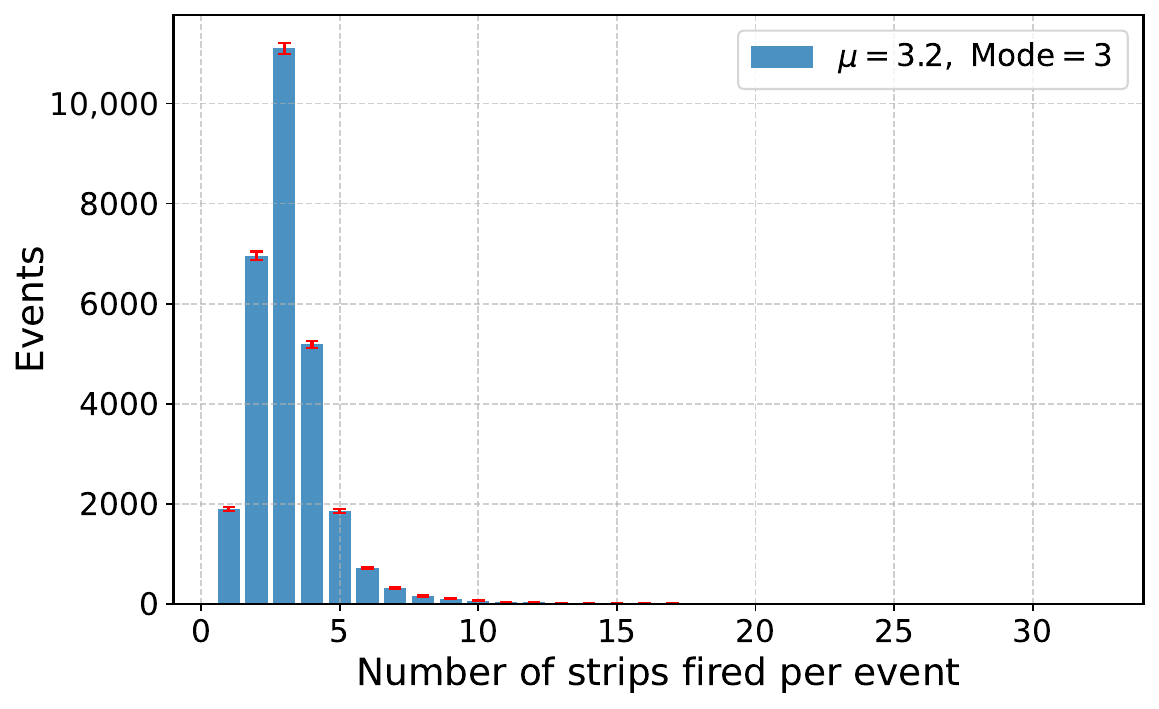}}     
    \hfill
    \subfloat[ ]{\includegraphics[width=.48\linewidth]{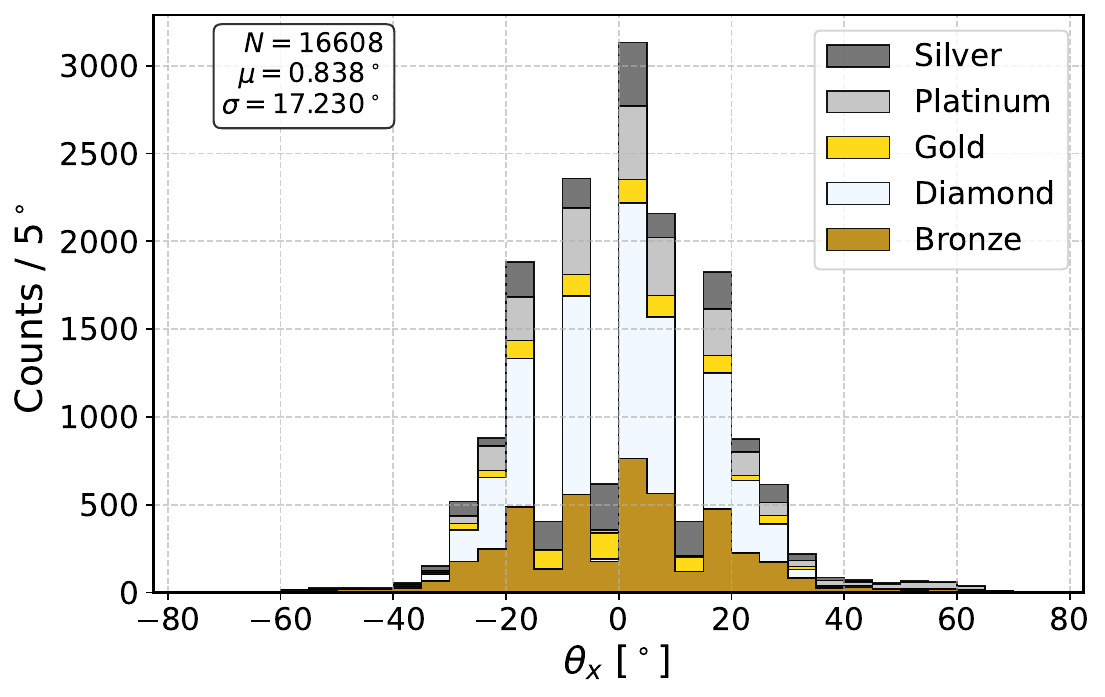}}\\
    \vspace{0.1cm}
    \subfloat[ ]{\includegraphics[width=.48\linewidth]{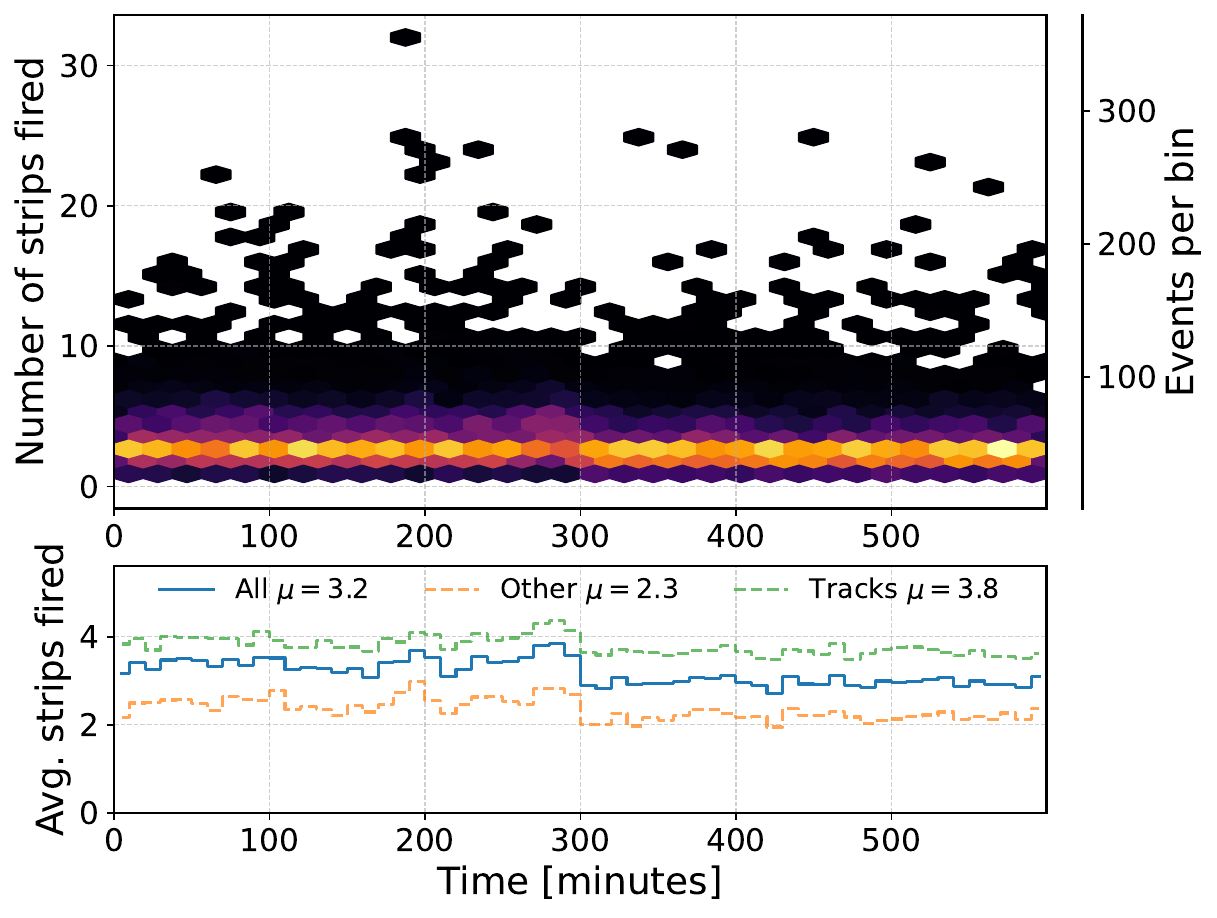}}  
    \hfill
    \subfloat[ ]{\includegraphics[width=.48\linewidth]{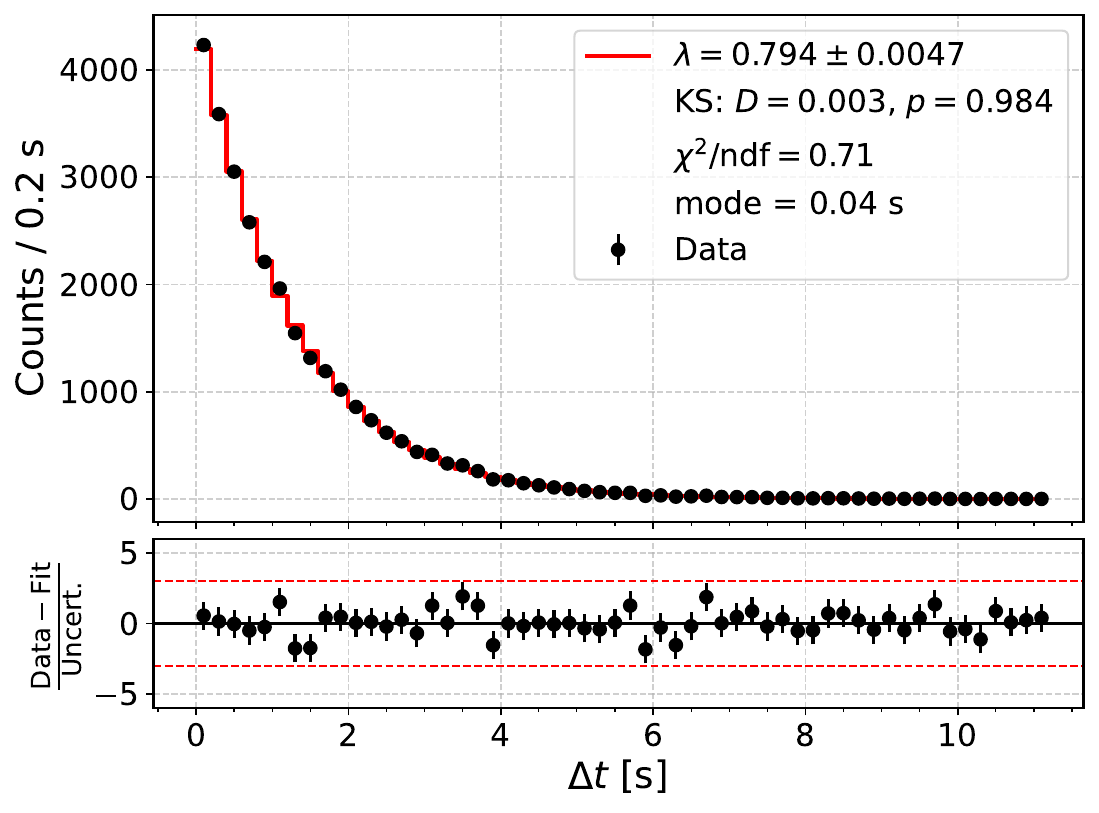}}
    \caption{Validation plots showing the strip multiplicity, zenith-angle distribution, time dependence, and inter-arrival time behaviour for the laboratory cosmic-ray dataset. (\textbf{a}) Strip occupancy distributions for the cosmic-ray dataset recorded with the vertical-flux setup at CP3. (\textbf{b}) Strip multiplicity for the D1 cosmic-ray dataset. (\textbf{c}) Zenith-angle distribution for laboratory cosmic rays. (\textbf{d}) Strip multiplicity as a function of time. (\textbf{e}) Inter-arrival time distribution.}
    \label{fig:validation_plots}
\end{figure}

Figure~\ref{fig:validation_plots}a presents the strip-occupancy distributions obtained from the cosmic-ray dataset recorded in the vertical-flux configuration (i.e., with the detectors disposed horizontally) at CP3. The observed occupancy pattern is consistent with the expected detector response for atmospheric cosmic-ray measurements and demonstrates stable detector behaviour during the data-taking period. The strip-occupancy distribution further shows that the majority of fired strips are concentrated in time bin 121, corresponding to approximately 605~ns after the scintillator trigger for a 5~ns time-bin width.

Figure~\ref{fig:validation_plots} shows the validation plots for the laboratory cosmic-ray dataset recorded in the vertical-flux setup, including the strip multiplicity, zenith-angle distribution, time dependence, and inter-arrival time behavior. The strip-multiplicity distribution shown in Figure~\ref{fig:validation_plots}b exhibits a pronounced peak at low multiplicity, with a mode of approximately three strips, indicating that most events are associated with localized detector activity.

The angular (zenith-angle) distribution further validates the detector response, showing a strong concentration of events around $\theta_x \approx 0^\circ$. This behaviour is expected for the vertical-flux configuration, where atmospheric cosmic rays predominantly arrive close to the vertical direction, resulting in a higher fraction of near-normal incidence tracks. This agreement confirms the geometrical and physical consistency of the reconstruction framework.

The zenith-angle distributions shown in Figure~\ref{fig:validation_plots}c exhibit a significantly larger population around $\theta_x \approx 0^\circ$ compared to the ELI-like setup. This difference reflects the expected change in angular acceptance between configurations and further supports the validity of the reconstruction pipeline.

The time-dependent distributions remain stable throughout the full data-taking period, indicating a consistent detector response and stable operating conditions during the measurement campaign. The inter-arrival time distribution is consistent with a Poisson process, exhibiting the expected exponential behaviour, which further confirms the random nature of cosmic-ray arrivals and the absence of correlated background contributions.

\section{Principal Component Analysis for Data Categorization}
\label{sec:pca}

Datasets used in this analysis are high-dimensional, both in terms of the number of events and the number of detector observables. Each event is characterized by multiple detector response features, including hit multiplicities, spatial distributions, and timing information. Because the analysis is fully data-driven, with no prior assumptions about the plasma beam properties or the energy spectrum of secondary particles, the structure of the data must be inferred directly from the measured observables. This motivates the use of dimensionality reduction techniques, which allows for the extraction of the dominant patterns in the data while retaining the most relevant physical information, thus facilitating physical interpretation. In this study, we focus on raw detector observables, computed before the application of any tracking or event-selection algorithms, in order to minimize model-dependent biases. All events that trigger the external scintillator system are considered. The event-wise features are grouped below according to their physical meaning.

\begin{itemize}
    \item Global event-level observables
    \begin{itemize}
        \item $n_{\mathrm{layers}}$: number of detector layers registering at least one fired strip.
        \item $n_{\mathrm{strips}}$: total number of fired strips across all layers.
        \item $\Delta t_{i-1}$: time interval between the current event and the previous event.
        \item $\Delta t_{i+1}$: time interval between the current event and the subsequent event.
    \end{itemize}
    \item Layer-level multiplicity observables
    \begin{itemize}
        \item $n_{\mathrm{strips},i}$: number of fired strips in layer $i$.
        \item $n_{\mathrm{cluster},i}$: number of clusters in layer $i$, where a cluster is defined as a set of neighboring fired strips.
        \item $\mathrm{hit}_i$: binary variable indicating whether layer $i$ records at least one hit.
    \end{itemize}
    \item Timing observables
    \begin{itemize}
        \item $\overline{tb_i}$: mean time-bin value of fired strips in layer $i$. Time bins are 5 $ns$ wide, and give the time interval between the scintillator trigger and the firing of a strip.
        \item $\mathrm{std}(tb_i)$: standard deviation of the time-bin distribution in layer $i$.
    \end{itemize}
    \item Spatial observables
    \begin{itemize}
        \item $\overline{x_i}$: mean $x$-position of fired strips in layer $i$.
        \item $\mathrm{std}(x_i)$: standard deviation of the $x$-position distribution in layer $i$.
    \end{itemize}
\end{itemize}

\begin{figure}[H]
    \centering
    \includegraphics[width=\linewidth]{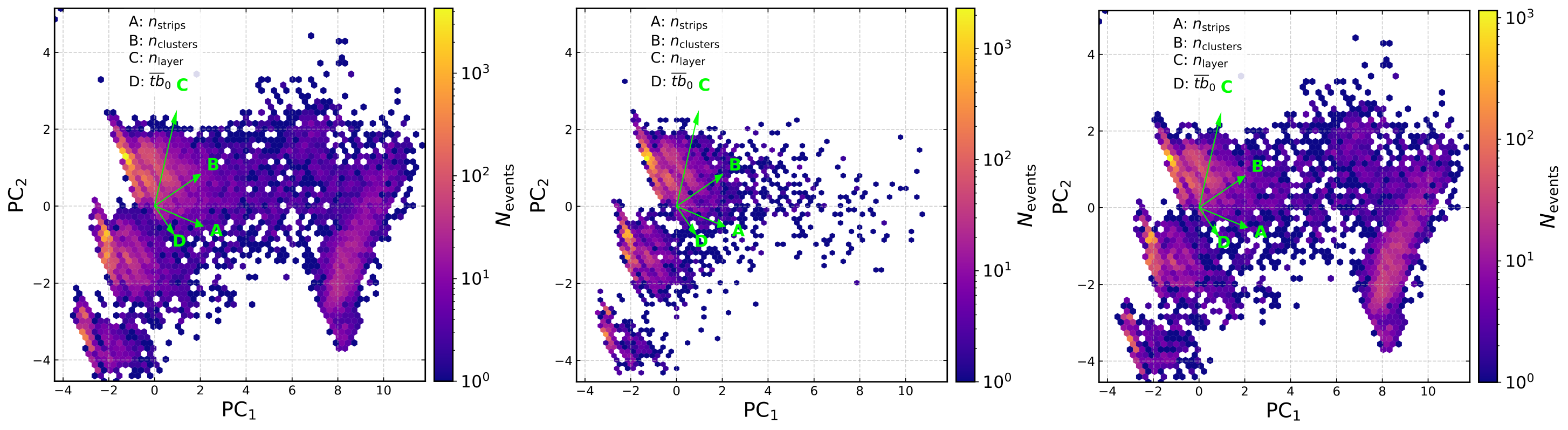}
    \caption{Distribution of events in the $\mathrm{PC}_1$–$\mathrm{PC}_2$ plane for combined D2 and D3 datasets (\textbf{left}), cosmic data (\textbf{center}), and beam data (\textbf{right}). Hexagonal binning is used to highlight event density. Arrows indicate the directions of selected original feature vectors projected onto the PCA space. The beam dataset exhibits an additional population of high-activity events, characterized by large hit and cluster multiplicities, which is absent in the cosmic sample.}
    \label{fig:pca_beam_vs_data}
\end{figure}

To reduce the dimensionality of this feature space, we apply Principal Component Analysis (PCA)~ \cite{pca_review}, a widely used linear technique that transforms the original correlated variables into a new set of orthogonal components ordered by decreasing variance. The principal components are linear combinations of the original features and correspond to the eigenvectors of the covariance matrix of the data. Prior to applying PCA, each feature $x$ is standardized by centering and rescaling:
\begin{equation}
x \;\rightarrow\; \frac{x - \mu_x}{\sigma_x},
\end{equation}
where $\mu_x$ and $\sigma_x$ denote the mean and standard deviation of the feature, respectively. Using only the first two principal components, approximately 47.6\% of the total variance of the dataset is preserved. Although a substantial fraction of the information resides in higher-dimensional components, this level of variance retention is sufficient for exploratory visualization and qualitative comparison between data samples. Figure~\ref{fig:pca_beam_vs_data} shows the distribution of events from the D2 and D3 data-taking periods in the $\mathrm{PC}_1$–$\mathrm{PC}_2$ plane, separately for beam and cosmic datasets. A distinct population of events appears in the beam data that is absent in the cosmic sample. This population is characterized by high detector activity, reflected in large hit multiplicities and cluster counts. Conversely, a second population with low multiplicity is present in both beam and cosmic data, exhibiting features consistent with cosmic muon signatures. At this stage, it is not possible to unambiguously determine whether these low-multiplicity events in the beam dataset originate from beam-related muons, cosmic muons, or secondary particles produced in beam interactions.

\begin{figure}[H]
    \centering
    \includegraphics[width=\linewidth]{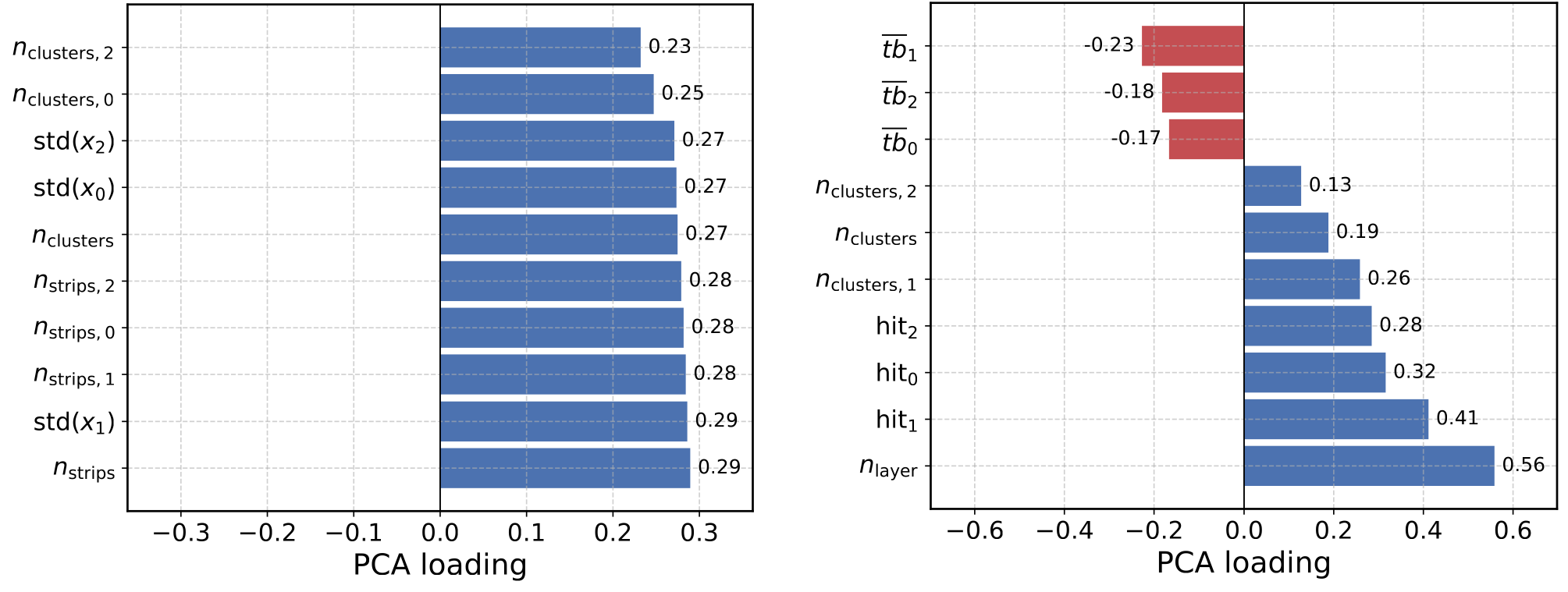}
    \caption{Loadings of the first two principal components, $\mathrm{PC}_1$ and $\mathrm{PC}_2$, for the ten most influential features. Global multiplicity observables (total number of fired strips, number of active layers, and number of clusters), together with layer-level strip multiplicities, dominate the variance captured by the leading components.}
    \label{fig:pca_loadings}
\end{figure}

\begin{figure}[H]
    \centering
    \includegraphics[width=\linewidth]{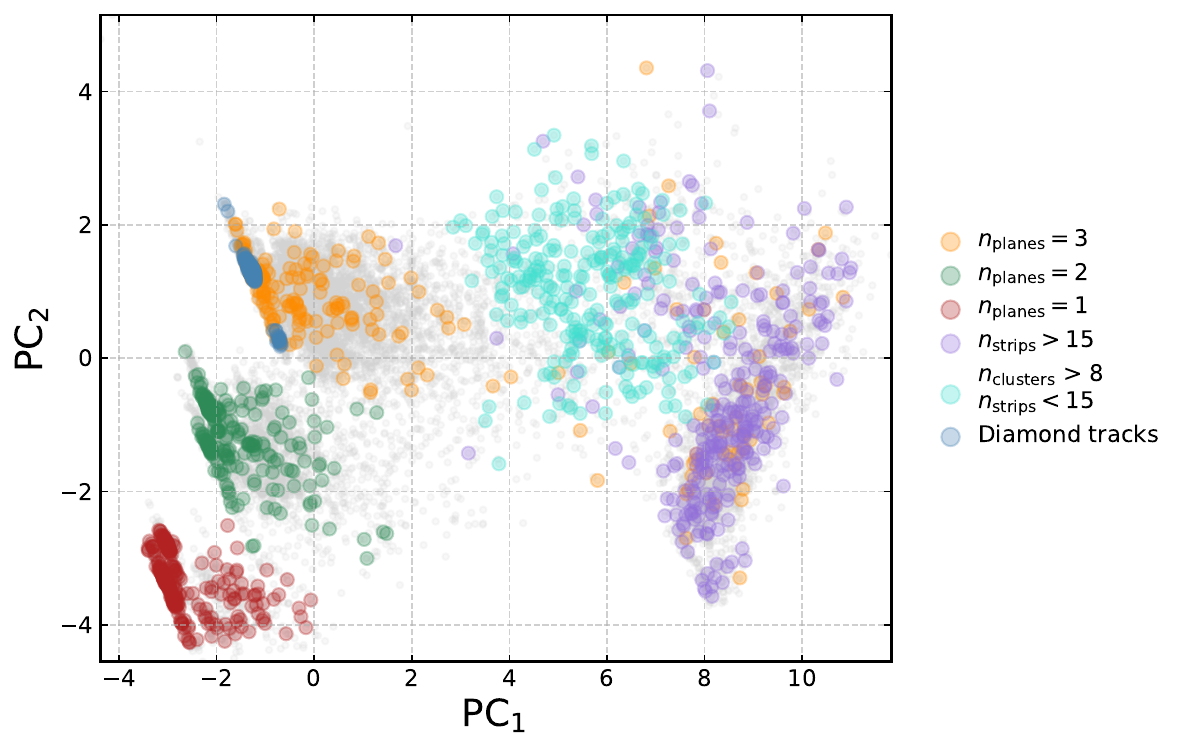}
    \caption{Distribution of events in the $\mathrm{PC}_1$–$\mathrm{PC}_2$ plane, highlighting regions corresponding to different detector topologies and multiplicity regimes. Distinct clusters associated with different event topologies are clearly visible, illustrating the ability of PCA to separate event populations based on raw detector observables. Only 400 events are shown for each category, and remaining events are displayed in grey.}
    \label{fig:pca_mapping}
\end{figure}

To interpret these structures in terms of detector observables, it is necessary to examine the contribution of the original features to the principal components. The loading of a feature on a principal component quantifies its weight in the corresponding linear combination and therefore indicates which detector observables dominate the variance captured along each principal direction. The loadings of the first two principal components, $\mathrm{PC}_1$ and $\mathrm{PC}_2$, are shown in Figure~\ref{fig:pca_loadings}, where only the ten features with the largest absolute contributions are displayed. Beyond global trends, PCA also reveals how specific event topologies populate the reduced phase space. Figure~\ref{fig:pca_mapping} shows the distribution of events in the $\mathrm{PC}_1$–$\mathrm{PC}_2$ plane, highlighting regions corresponding to different detector signatures, such as the number of active planes, strip and cluster multiplicities, and high-quality “diamond” tracks (see Section~\ref{sec:tracking}). Events with similar topologies cluster together, illustrating the ability of PCA to capture meaningful structure in an unsupervised manner. In particular, events hitting one, two, and three planes form well-separated clusters with minimal overlap. The region at large $\mathrm{PC}_1$ values ($\mathrm{PC}_1 > 4$) is dominated by high-multiplicity events, both in terms of fired strips and clusters. Within this region, events with many fired strips and those with many clusters but fewer strips occupy distinct sub-regions, indicating different underlying detector responses. High-quality diamond tracks are highly localized in the PCA plane, reflecting their characteristic and stable detector signature. 

Although this reduced representation does not provide a complete event classification on its own, it offers a powerful and intuitive framework for visualizing correlations among detector observables and for identifying distinct event populations. It therefore serves as a valuable intermediate step toward more refined event categorization and physical interpretation.

\begin{adjustwidth}{-\extralength}{0cm}

\reftitle{References}

\end{adjustwidth}

\begin{thebibliography}{999}

\bibitem[Zhang et~al.(2025)Zhang, Deng, et~al.]{zhang2025proof}
Zhang, F.; Deng, L.; Ge, Y.; Wen, J.; Cui, B.; Feng, K.; Wang, H.; Wu, C.; Pan, Z.; Liu, H.; et al. Proof-of-principle demonstration of muon production with an ultrashort high-intensity laser.  \emph{Nat. Phys.} \textbf{2025}, \textit{21}, 1050–1056. \url{https://doi.org/10.1038/s41567-025-02872-2}.

\bibitem[Terzani et~al.(2025)Terzani, Kisyov, et~al.]{terzani2025measurement}
Terzani, D.; Kisyov, S.; Greenberg, S.; Le Pottier, L.; Mironova, M.; Picksley, A.; Stackhouse, J.; Tsai, H.-E.; Li, R.; Rockafellow, E.; et al.
Measurement of directional muon beams generated at the Berkeley Lab Laser Accelerator. \emph{Phys. Rev. Accel. Beams} \textbf{2025}, \emph{28}, 103401. \url{https://doi.org/10.1103/kxjr-h7zs}.

\bibitem[Calvin et~al.(2026)Calvin, Gerstmayr, et~al.]{Calvin2026}
Calvin, L.; Gerstmayr, E.; Arran, C.; Tudor, L.; Foster, T.; Fleck, K.; Bergmann, B.; Doria, D.; Kettle, B.; Maguire, H.; et al.
Experimental evidence of production of directional muons from a laser-wakefield accelerator. \emph{Plasma Phys. Control. Fusion} \textbf{2026}, \emph{68}, 035015. \url{https://doi.org/10.1088/1361-6587/ae4d05}.

\bibitem[Tanaka et~al.(2023)Tanaka, Bozza, et~al.]{tanaka2023muography}
Tanaka, H.K.M.; Bozza, C.; Bross, A.; Cantoni, E.; Catalano, O.; Cerretto, G.; Giammanco, A.; Gluyas, J.; Gnesi, I.; Holma, M.;  et al. Muography. \emph{Nat. Rev. Methods Primers} \textbf{2023}, \emph{3}, 88.

\bibitem[{International Atomic Energy Agency}(2022)]{IAEA2022}
International Atomic Energy Agency. Muon imaging: Present Status and Emerging Applications. IAEA TECDOC 2012, IAEA, Vienna, 2022.

\bibitem[Procureur(2018)]{procureur2018muon}
Procureur, S. Muon radiography: Overview of the technique and recent applications in geophysics. \emph{Nucl. Instruments Methods Phys. Res. Sect. A Accel. Spectrometers Detect. Assoc. Equip.} \textbf{2018}, \emph{878}, 169--179. \url{https://doi.org/10.1016/j.nima.2017.08.004}.

\bibitem[Kaiser(2018)]{Kaiser2018muography}
Kaiser, R. Muography: Overview and future directions. \emph{Philos. Trans. R. Soc. A Math. Phys. Eng. Sci.} \textbf{2018}, \emph{377}, 20180049. \url{https://doi.org/10.1098/rsta.2018.0049}.

\bibitem[Bonomi et~al.(2020)Bonomi, Checchia, D'Errico, Pagano, and
  Saracino]{bonomi2020applications}
Bonomi, G.; Checchia, P.; D’Errico, M.; Pagano, D.; Saracino, G. Applications of cosmic-ray muons. \emph{Prog. Part. Nucl. Phys.} \textbf{2020}, \emph{112}, 103768. \url{https://doi.org/10.1016/j.ppnp.2020.103768}.

\bibitem[Bae et~al.(2026)Bae, Borozdin, Hecht, Miyadera, Morris, Stuke, and
  Tanaka]{bae2026muography}
Bae, J.; Borozdin, K.; Hecht, A.; Miyadera, H.; Morris, C.; Stuke, M.; Tanaka, H. Muography: Discoveries, innovations, and applications. \emph{J. Appl. Phys.} \textbf{2026}, \emph{140}, 020401. \url{https://doi.org/10.1063/5.0346450}.

\bibitem[Giammanco et~al.(2025)Giammanco, {Al Moussawi},
  et~al.]{muonsCH-Giammanco2024}
Giammanco, A.; Al Moussawi, M.; Boone, M.; De Kock, T.; De Roy, J.; Huysmans, S.; Kumar, V.; Lagrangev, M.; Tytgat, M. Toward using cosmic rays to image cultural heritage objects. \emph{iScience} \textbf{2025}, \emph{28}, 2589. \url{https://doi.org/10.1016/j.isci.2025.112094}.

\bibitem[Blundell et~al.(2021)Blundell, De~Renzi, Lancaster, and
  Pratt]{MakingMuons2021}
Blundell, S.J.; De Renzi, R.; Lancaster, T.; Pratt, F.L. Making muons. In \emph{Muon Spectroscopy: An Introduction}; Oxford University Press: Oxford, UK, 2021. \url{https://doi.org/10.1093/oso/9780198858959.003.0013}.

\bibitem[Ninomiya et~al.(2015)]{ninomiya2015nondestructive}
Ninomiya, K.; Kubo, M.K.; Nagatomo, T.; Higemoto, W.; Ito, T.U.; Kawamura, N.; Strasser, P.; Shimomura, K.; Miyake, Y.; Suzuki, T.;  et al.  Nondestructive elemental depth-profiling analysis by muonic X-ray measurement. \emph{Anal. Chem.} \textbf{2015}, \emph{87}, 4597--4600.

\bibitem[Reidy et~al.(1978)Reidy, Hutson, Daniel, and Springer]{reidy1978use}
Reidy, J.J.; Hutson, R.L.; Daniel, H.; Springer, K. Use of muonic X-rays for nondestructive analysis of bulk samples for low Z constituents. \emph{Anal. Chem.} \textbf{1978}, \emph{50}, 40--44.

\bibitem[Cataldo et~al.(2022)Cataldo, Clemenza, Ishida, and
  Hillier]{cataldo2022novel}
Cataldo, M.; Clemenza, M.; Ishida, K.; Hillier, A.D. A novel non-destructive technique for cultural heritage: Depth profiling and elemental analysis underneath the surface with negative muons. \emph{Appl. Sci.} \textbf{2022}, \emph{12}, 4237.

\bibitem[Biswas et~al.(2023)]{biswas2023non}
Biswas, S.; Megatli-Niebel, I.; Raselli, L.; Simke, R.; Cocolios, T.E.; Deokar, N.; Elender, M.; Gerchow, L.; Hess, H.; Khasanov, R.;  et al. The non-destructive investigation of a late antique knob bow fibula (Bügelknopffibel) from Kaiseraugst/CH using Muon Induced X-ray Emission (MIXE). \emph{Herit. Sci.} \textbf{2023}, \emph{11}, 43.

\bibitem[Hofmann et~al.(2023)]{Hofmann2023}
Hofmann, B.A.; Schreyer, S.B.; Biswas, S.; Gerchow, L.; Wiebe, D.; Schumann, M.; Lindemann, S.; García, D.R.; Lanari, P.; Gfeller, F.;  et al. An arrowhead made of meteoritic iron from the late Bronze Age settlement of Mörigen, Switzerland and its possible source. \emph{J. Archaeol. Sci.} \textbf{2023}, \emph{157}, 105827. \url{https://doi.org/10.1016/j.jas.2023.105827}.

\bibitem[Santonico and Cardarelli(1981)]{santonico1981development}
Santonico, R.; Cardarelli, R. Development of resistive plate counters. \emph{Nucl. Instrum. Methods Phys. Res.} \textbf{1981}, \emph{187}, 377--380.

\bibitem[Santonico(2004)]{SANTONICO20041}
Santonico, R. RPC understanding and future perspectives. \emph{Nucl. Instruments Methods Phys. Res. Sect. A Accel. Spectrometers Detect. Assoc. Equip.} \textbf{2004}, \emph{533}, 1--6. \url{https://doi.org/10.1016/j.nima.2004.06.160}.

\bibitem[Fonte(2002)]{fonte2002review}
Fonte, P. A review of resistive plate chambers. \emph{IEEE Trans. Nucl. Sci.} \textbf{2002}, \emph{49}, 881--887. \url{https://doi.org/10.1109/TNS.2002.1039583}.

\bibitem[{CMS Collaboration}(1997)]{CMSmuonTDR}
CMS Collaboration. \emph{The CMS Muon Project}; Technical Design Report CERN-LHCC-97-032; CERN: Geneva, Switzerland, 1997.

\bibitem[{ATLAS Collaboration}(1997)]{ATLASmuonTDR}
ATLAS Collaboration. \emph{ATLAS Muon Spectrometer}; Technical Design Report CERN-LHCC-97-022; CERN: Geneva, Switzerland, 1997.

\bibitem[Aielli et~al.(2006)Aielli, Assiro, et~al.]{aielli2006layout}
Aielli, G.; Assiro, R.; Bacci, C.; Bartoli, B.; Bernardini, P.; Bi, X.; Biondo, B.; Bleve, C.; Bricola, S.; Budano, F.;  et al. Layout and performance of RPCs used in the Argo-YBJ experiment. \emph{Nucl. Instruments Methods Phys. Res. Sect. A Accel. Spectrometers Detect. Assoc. Equip.} \textbf{2006}, \emph{562}, 92--96. \url{https://doi.org/10.1016/j.nima.2006.02.136}.

\bibitem[Abreu et~al.(2018)Abreu, Andringa, et~al.]{abreu2018marta}
Abreu, P.; Andringa, S.; Assis, P.; Blanco, A.; Martins, V.B.; Brogueira, P.; Carolino, N.; Cazon, L.; Cerda, M.; Cernicchiaro, G.;  et al.  MARTA: A high-energy cosmic-ray detector concept for high-accuracy muon measurement. \emph{Eur. Phys. J. C} \textbf{2018}, \emph{78}, 333. \url{https://doi.org/10.1140/epjc/s10052-018-5820-2}.

\bibitem[C{\^a}rloganu et~al.(2013)C{\^a}rloganu, Niess, B{\'e}n{\'e}, Busato,
  Dupieux, Fehr, Gay, Miallier, Vulpescu, Boivin, Combaret, Labazuy, Laktineh,
  L{\'e}nat, Mirabito, and Portal]{carloganu2013towards}
Cârloganu, C.; Niess, V.; Béné, S.; Busato, E.; Dupieux, P.; Fehr, F.; Gay, P.; Miallier, D.; Vulpescu, B.; Boivin, P.; et al. Towards a muon radiography of the Puy de Dôme. \emph{Geosci. Instrum. Methods Data Syst.} \textbf{2013}, \emph{2}, 55--60. \url{https://doi.org/10.5194/gi-2-55-2013}.

\bibitem[Giammanco et~al.(2022)Giammanco, Andringa,
  et~al.]{RPC-MuographyBook2022}
Giammanco, A.; Andringa, S.; Gil, E.C.; Tytgat, M. Resistive Plate Chambers in Muography. In \emph{Muography}; American Geophysical Union (AGU): Washington, DC, USA, 2022; Volume 18, pp. 253--265. \url{https://doi.org/10.1002/9781119722748.ch18}.

\bibitem[Mourou et~al.(2011)Mourou, Korn, et~al.]{eliwhitebook}
Mourou, G.; Korn, G. (Eds.) \emph{ELI---Extreme Light Infrastructure WHITEBOOK Science and Technology with Ultra-Intense Lasers}; THOSS Media GmbH: Berlin, Germany, 2011.

\bibitem[Haefner et~al.(2017)Haefner, Bayramian, et~al.]{l3hapls1}
 Haefner, C.L.; Bayramian, A.; Betts, S.; Bopp, R.; Buck, S.; Cupal, J.; Drouin, M.; Erlandson, A.; Horáček, J.; Horner, J.;  et al. High average power, diode pumped petawatt laser systems: A new generation of lasers enabling precision science and commercial applications. In \emph{Proceedings of the Research Using Extreme Light: Entering New Frontiers with Petawatt-Class Lasers III}; Korn, G., Silva, L.O., Eds.; International Society for Optics and Photonics, SPIE: Bellingham, WA, USA, 2017; Volume 10241. \url{https://doi.org/10.1117/12.2281050}.

\bibitem[Nov{\'a}k et~al.(2025)Nov{\'a}k, Szotkowski, et~al.]{l3hapls2}
Novák, J.; Szotkowski, P.; Link, E.S.; McLoughlin, J.J.; Kiani, L.S.; Galvin, T.C.; Velas, K.M.; Černý, J.; Sedmidubská, J.; Cuhra, M.;  et al. Towards the petawatt capability of the L3 HAPLS laser system. In \emph{Proceedings of the Research Using Extreme Light Infrastructures: New Frontiers with Petawatt-Level Lasers VI}; International Society for Optics and Photonics, SPIE: Bellingham, WA, USA, 2025; Volume PC13535. \url{https://doi.org/10.1117/12.3056595}.

\bibitem[Feder et~al.(2020)Feder, Miao, et~al.]{Feder2021}
Feder, L.; Miao, B.; Shrock, J.E.; Goffin, A.; Milchberg, H.M. Self-waveguiding of relativistic laser pulses in neutral gas channels. \emph{Phys. Rev. Res.} \textbf{2020}, \emph{2}, 043173. \url{https://doi.org/10.1103/PhysRevResearch.2.043173}.

\bibitem[Miao et~al.(2022)Miao, Shrock, et~al.]{Miao2022}
Miao, B.; Shrock, J.E.; Feder, L.; Hollinger, R.C.; Morrison, J.; Nedbailo, R.; Picksley, A.; Song, H.; Wang, S.; Rocca, J.J.;  et al. Multi-GeV Electron Bunches from an All-Optical Laser Wakefield Accelerator. \emph{Phys. Rev. X} \textbf{2022}, \emph{12}, 031038. \url{https://doi.org/10.1103/PhysRevX.12.031038}.

\bibitem[Sisma et~al.(2026)Sisma, Nevrlka, et~al.]{Sisma2025}
Šišma, J.; Nevrkla, M.; Vitha, F.; Lorenz, S.; Zymak, I.; Špádová, A.; Kollárová, A.; Jech, M.; Jančárek, A.; Peceli, D.;  et al. High-repetition-rate, all-reflective optical guiding and electron acceleration in helium using an off-axis axicon. \emph{High Power Laser Sci. Eng.} \textbf{2026}, \emph{14}, e52. \url{https://doi.org/10.1017/hpl.2026.10154}.

\bibitem[Tajima and Dawson(1979)]{Tajima1979}
Tajima, T.; Dawson, J. Laser Electron Accelerator. \emph{Phys. Rev. Lett.} \textbf{1979}, \emph{43}, 267--270. \url{https://doi.org/10.1103/PhysRevLett.43.267}.

\bibitem[Esarey et~al.(2009)Esarey, Schroeder, and Leemans]{Esarey_review}
Esarey, E.; Schroeder, C.; Leemans, W. Physics of laser-driven plasma-based electron accelerators. \emph{Rev. Mod. Phys.} \textbf{2009}, \emph{81}, 1229--1285. \url{https://doi.org/10.1103/RevModPhys.81.1229}.

\bibitem[Durfee and Milchberg(1993)]{Durfee1993}
Durfee, C.; Milchberg, H. Light pipe for high intensity laser pulses. \emph{Phys. Rev. Lett.} \textbf{1993}, \emph{71}, 2409--2412. \url{https://doi.org/10.1103/PhysRevLett.71.2409}.

\bibitem[Rockafellow et~al.(2025)Rockafellow, Miao, et~al.]{Rockafellow2025}
Rockafellow, E.; Miao, B.; Shrock, J.E.; Sloss, A.; Le, M.S.; Hancock, S.W.; Zahedpour, S.; Hollinger, R.C.; Wang, S.; King, J.;  et al. Development of a high charge 10 Gev laser electron accelerator. \emph{Phys. Plasmas} \textbf{2025}, \emph{32}. \url{https://doi.org/10.1063/5.0265640}.

\bibitem[Picksley et~al.(2024)Picksley, Stackhouse, et~al.]{Picksley2024}
Picksley, A.; Stackhouse, J.; Benedetti, C.; Nakamura, K.; Tsai, H.E.; Li, R.; Miao, B.; Shrock, J.E.; Rockafellow, E.; Milchberg, H.M.;  et al. Matched Guiding and Controlled Injection in Dark-Current-Free, 10-GeV-Class, Channel-Guided Laser-Plasma Accelerators. \emph{Phys. Rev. Lett.} \textbf{2024}, \emph{133}, 255001. \url{https://doi.org/10.1103/PhysRevLett.133.255001}.

\bibitem[Miao et~al.(2025)Miao, Shrock, et~al.]{Miao2025}
Miao, B.; Shrock, J.E.; Rockafellow, E.; Sloss, A.J.; Milchberg, H.M.  Meter-scale supersonic gas jets for multi-gev laser-plasma accelerators. \emph{Rev. Sci. Instrum.} \textbf{2025}, \emph{96}. \url{https://doi.org/10.1063/5.0248959}.

\bibitem[Pak et~al.(2010)Pak, Marsh, et~al.]{Pak2010}
Pak, A.; Marsh, K.A.; Martins, S.F.; Lu, W.; Mori, W.B.; Joshi, C. Injection and Trapping of Tunnel-Ionized Electrons into Laser-Produced Wakes. \emph{Phys. Rev. Lett.} \textbf{2010}, \emph{104}, 025003. \url{https://doi.org/10.1103/PhysRevLett.104.025003}.

\bibitem[Miao et~al.(2020)Miao, Feder, et~al.]{miao2020}
Miao, B.; Feder, L.; Shrock, J.E.; Goffin, A.; Milchberg, H.M. Optical Guiding in Meter-Scale Plasma Waveguides. \emph{Phys. Rev. Lett.} \textbf{2020}, \emph{125}, 074801. \url{https://doi.org/10.1103/PhysRevLett.125.074801}.

\bibitem[Ahdida et~al.(2022)Ahdida, Bozzato, et~al.]{fluka1}
Ahdida, C.; Bozzato, D.; Calzolari, D.; Cerutti, F.; Charitonidis, N.; Cimmino, A.; Coronetti, A.; D’aLessandro, G.L.; Servelle, A.D.; Esposito, L.S.;  et al. New Capabilities of the FLUKA Multi-Purpose Code. \emph{Front. Phys.} \textbf{2022}, \emph{9}, 788253. \url{https://doi.org/10.3389/fphy.2021.788253}.

\bibitem[Battistoni et~al.(2015)Battistoni, Boehlen, et~al.]{fluka2}
Battistoni, G.; Boehlen, T.; Cerutti, F.; Chin, P.W.; Esposito, L.S.; Fassò, A.; Ferrari, A.; Lechner, A.; Empl, A.; Mairani, A.;  et al. Overview of the FLUKA code. \emph{Ann. Nucl. Energy} \textbf{2015}, \emph{82}, 10--18.

\bibitem[Ikram et~al.(2025)Ikram, Basnet, et~al.]{RPC-Ikram2025}
Ikram, S.; Basnet, S.; Gil, E.C.; Demin, P.; Gamage, R.M.I.D.; Giammanco, A.; Karnam, R.; Kashyap, V.K.S.; Kumar, V.; Mohanty, B.;  et al. Development and performance analysis of glass-based gas-tight RPCs for muography applications. \emph{J. Appl. Phys.} \textbf{2025}, \emph{138}, 174502. \url{https://doi.org/10.1063/5.0275200}.

\bibitem[Kumar et~al.(2025)]{RPC-Kumar2025}
Kumar, V.; Basnet, S.; Gil, E.C.; Gamage, R.; Giammanco, A.; Moussawi, M.; Samalan, A.; Tytgat, M.; Karnam, R.  Characterization and stability tests of gas-tight RPC for muography application. \emph{Nucl. Instrum. Meth. A} \textbf{2025}, \emph{1070}, 170025.

\bibitem[Chaulagain et~al.(2022)Chaulagain, Lamač, et~al.]{bib:gammatron}
Chaulagain, U.; Lamač, M.; Raclavský, M.; Khakurel, K.P.; Rao, K.H.; Ta-Phuoc, K.; Bulanov, S.V.; Nejdl, J. ELI Gammatron Beamline: A Dawn of Ultrafast Hard X-ray Science. \emph{Photonics} \textbf{2022}, \emph{9}, 853.  \url{https://doi.org/10.3390/photonics9110853}.

\bibitem[Feldman and Cousins(1998)]{FeldmanCousins1998}
Feldman, G.; Cousins, R. Unified approach to the classical statistical analysis of small signals. \emph{Phys. Rev. D} \textbf{1998}, \emph{57}, 3873--3889. \url{https://doi.org/10.1103/PhysRevD.57.3873}.

\bibitem[Abbrescia et~al.(2000)Abbrescia, Colaleo, et~al.]{Abbrescia2000}
Abbrescia, M.; Colaleo, A.; Iaselli, G.; Loddo, F.; Maggi, M.; Marangelli, B.; Natali, S.; Nuzzo, S.; Pugliese, G.; Ranieri, A.;  et al.  New developments on front-end electronics for the CMS Resistive Plate Chambers.  \emph{Nucl. Instruments Methods Phys. Res. Sect. A Accel. Spectrometers Detect. Assoc. Equip.} \textbf{2000}, \emph{456}, 143--149.

\bibitem[Jolliffe and Cadima(2016)]{pca_review}
Jolliffe, I.; Cadima, J. Principal component analysis: A review and recent developments. \emph{Philos. Trans. R. Soc. A Math. Phys. Eng. Sci.} \textbf{2016}, \emph{374}, 20150202. \url{https://doi.org/10.1098/rsta.2015.0202}.

\end{thebibliography}
\end{document}